\documentclass{aa}  

\usepackage{graphicx}
\usepackage{txfonts}

\usepackage{graphicx,color}
\usepackage{graphicx}
\usepackage{amssymb,amsmath}
\usepackage{amsmath} 
\usepackage{multicol}
\usepackage{stmaryrd}
\usepackage{latexsym}
\usepackage{amsfonts}
\usepackage{amssymb}
\usepackage{wasysym}
\usepackage{mathcomp}
\usepackage{textcomp}
\usepackage{longtable,lscape}
\usepackage{mathtools}
\usepackage{morefloats} 
\usepackage{rotating}

\begin{document} 

  \authorrunning{M. Angelo, J. Santos Jr., W. Corradi, F. Maia}
  \titlerunning{Investigating dynamically evolved open clusters}

   \title{Investigating dynamical properties of evolved Galactic open clusters\thanks{The appendix tables are only available in electronic form at the CDS via anonymous ftp to cdsarc.u-strasbg.fr (130.79.128.5) or via http://cdsweb.u-strasbg.fr/cgi-bin/qcat?J/A+A/}}


   \author{M. S. Angelo
          \inst{1},\,
          J. F. C. Santos Jr.\inst{2},\,
          W. J. B. Corradi\inst{2}\,
          \and
          F. F. S. Maia\inst{3}
          }

   \institute{Centro Federal de Educaç\~ao Tecnol\'ogica de Minas Gerais, Av. Monsenhor Luiz de Gonzaga 103, 37250-000 Nepomuceno, MG, Brazil\\ 
              \email{mateusangelo@cefetmg.br}
         \and
             Departamento de F\'isica, ICEx, Universidade Federal de Minas Gerais, Av. Ant\^onio Carlos 6627, 31270-901 Belo Horizonte, MG, Brazil
         \and
             Universidade de S\~ao Paulo, Instituto de Astronomia, Geof\'isica e Ci\^encias Atmosf\'ericas, Rua do Mat\~ao 1226, 05508-090 S\~ao Paulo, SP, Brazil    
             }

   \date{Received 26 October 2018 / Accepted 08 February 2019}

 
  \abstract
   {The stellar content of Galactic open clusters  is gradually depleted during their evolution as a result of internal relaxation and external interactions. The final residues of the long-term evolution of open clusters are called open cluster remnants. These are sparsely populated structures that can barely be distinguished from the field.}
   {We aimed to characterise and compare the dynamical states of a set of 16 objects catalogued as remnants or remnant candidates. We employed parameters that are intimately associated with the dynamical evolution: age, limiting radius, stellar mass, and velocity dispersion. The sample also includes 7 objects that are catalogued as dynamically evolved open clusters for comparison purposes.}
   {We used photometric data from the 2MASS catalogue, proper motions and parallaxes from the GAIA DR2 catalogue, and a decontamination algorithm that was applied to the three-dimensional astrometric space of proper motions and parallaxes ($\mu_{\alpha},\mu_{\delta},\varpi$) for stars in the objects' areas. The luminosity and mass functions and total masses for most open cluster remnants are derived here for the first time. Our analysis used predictions of $N$-body simulations to estimate the initial number of stars of the remnants from their dissolution timescales.}
   {The investigated open cluster remnants present masses ($M$) and velocity dispersions ($\sigma_{v}$) within well-defined ranges: $M$ between $\sim10-40\,M_{\odot}$ and $\sigma_{v}$ between $\sim1-7\,$km\,s$^{-1}$. Some objects in the remnant sample have a limiting radius $R_{\textrm{lim}}\lesssim2\,$pc, which means that they are more compact than the investigated open clusters; other remnants have $R_{\textrm{lim}}$ between $\sim2-7\,$pc, which is comparable to the open clusters. We suggest that cluster NGC\,2180 (previously classified as an open cluster) is entering a remnant evolutionary stage. In general, our clusters show signals of depletion of low-mass stars. This confirms their dynamically evolved states.}
   {We conclude that the open cluster remnants we studied are in fact remnants of initially very populous open clusters ($N_{0}$$\,\sim\,$$10^3-10^4$ stars). The outcome of the long-term evolution is to bring the final residues of the open clusters to dynamical states that are similar to each other, thus masking out the memory of the initial formation conditions of star clusters.}


   \maketitle
%

\section{Introduction}

Galactic open clusters (OCs) gradually lose their stellar content and eventually dissolve. Their evolution can be split into three phases: (i) the first lasts for $\sim3\,$Myr, during which the cluster is embedded in its progenitor molecular cloud and stars are still forming; (ii) the clusters that survive the early gas-expulsion phase (e.g. supernova explosions and stellar winds) and are largely gas free and their overall dynamics is dominated by stellar mass loss (\citeauthor{Portegies-Zwart:2007}\,\,\citeyear{Portegies-Zwart:2007}; \citeauthor{Portegies-Zwart:2010}\,\,\citeyear{Portegies-Zwart:2010}); and finally, (iii) the long-term evolutionary phase ($t\gtrsim100\,$Myr), when timescales for stellar mass loss through stellar evolution are considerably longer than dynamical timescales and purely dynamical processes dominate the evolution of the cluster. Internal and external forces (two-body or higher-order interactions, interactions with the Galactic tidal field, collisions with molecular clouds, and disc shocking) contribute to the decrease of total mass in the cluster (\citeauthor{Pavani:2011}\,\,\citeyear{Pavani:2011}; \citeauthor{Pavani:2001}\,\,\citeyear{Pavani:2001}).  



The final residue of an OC evolution is often called open cluster remnant (OCR). \cite{Bica:2001} assumed that a cluster becomes significantly depopulated with a remnant appearence after losing two-thirds of its initial stellar content. They also suggested the acronym POCR (possible open cluster remnant) for remnant candidates, that is, objects presenting significant number density contrast with respect to the general Galactic field, but with dubious evolutionary sequences in colour-magnitude diagrams (CMDs) due to large contamination by field stars. As stated by \cite{Carraro:2007}, the application of the criterion established by \cite{Bica:2001} is quite straightforward in the case of simulated clusters. On the other hand, this criterion is difficult to use with real clusters, since a reliable estimate of the initial number of stars is inaccessible from observations. In this context, how can we objectively distinguish  OCRs from well-known OCs? The present work is a contribution in this sense. In addition to being heavily underpopulated in relation to the OCs, the studied OCRs present compatible masses between them, they tend to present larger velocity dispersions as a result of dynamical internal heating, and they define a distinctive locus in the plot of density versus limiting radius (Section \ref{analysing_OCRs_evolut_stages}). Additionally, there is evidence that the currently observed OCRs are remnants of initially very populous OCs ($N_{0}\sim10^3-10^4\,$stars; Sect. \ref{comparison_with_simulations}).

Because of their physical nature, these objects are intrinsically sparsely populated. They typically consist of a few tens of stars, but they have  enough members to show evolutionary sequences in CMDs. This was noted by \cite{Pavani:2011} in the cases of ESO\,435-SC48 and ESO\,324-SC15. They developed a diagnostic tool to analyse CMDs of sparsely populated stellar systems and verified that the evolutive sequences defined by stars in the central regions of ESO\,435-SC48 and ESO\,324-SC15 are statistically distinguishable from their fields, thus classifying them as true OCRs. Regarding star counts in the remnants' inner areas, \cite{Bica:2001} analysed a list of 34 POCRs and verified that they exhibit significant overdensities compared to the surrounding field and to Galactic model predictions. This is the first step towards establishing the physical nature of an OCR. 

\cite{Pavani:2007} analysed a sample of 18 POCRs and devised a systematic method for characterizing such objects as physical systems or field fluctuactions based on three criteria: (a) radial density profile, (b) probability for the object CMD to be representative of the offset field CMD, and (c) homogeneity in the distribution of stars along CMDs sequences and number of stars compatible with isochrones for single and binary stars. Thirteen objects were classified by \cite{Pavani:2007} as genuine OCRs. The remaining studied objects retained their status as POCRs or were dismissed as coeval structures. The asymmetries and peaks observed in the OCRs proper motion histograms revealed the presence of binary and/or multiple systems. 


Some OCRs have been analysed from spectroscopic, photometric, and proper motions data. This is the case of the sparsely populated star cluster NGC\,1901, projected against the Large Magellanic Cloud NGC\,1901. \cite{Carraro:2007} studied its stellar content and kinematics by employing photometry in the $UBVI$ pass-bands, proper motions from the UCAC2 catalogue \citep{Zacharias:2004}, and multi-epoch high-resolution spectroscopic data. Based on the coherence of the kinematical data and considering the positions of each star in various photometric diagrams, they considered NGC\,1901 a prototype of an OCR. The physical nature of this cluster has also recently been confirmed by \cite{Kos:2018} by means of photometry and proper motions from the GAIA DR2 catalogue \citep{Gaia-Collaboration:2018} and radial velocities from the Galactic Archeology with HERMES (GALAH) survey \citep{De-Silva:2015}. \cite{Pavani:2003} investigated Ruprecht\,3 by employing 2MASS $JH$ photometry, proper motions from the Tycho-2 catalogue \citep{Hog:2000}, and low-resolution spectroscopy. After applying a decontamination algorithm on the object CMD in order to statistically remove field stars, they selected probable member stars and obtained basic parameters from isochrone fitting. The coherence between the proper motions of the four brightest stars and the spread of their derived spectral types along the red giant branch were used as additional membership criteria. The authors proposed that Ruprecht\,3 is a genuine OCR. 



$N$-body simulations of star clusters in an external potential show typical dissolution times in the range 500$-$2\,500\,Myr (\citeauthor{Portegies-Zwart:2001}\,\,\citeyear{Portegies-Zwart:2001} and heferences therein). \citeauthor{de-La-Fuente-Marcos:1998}\,\,(\citeyear{de-La-Fuente-Marcos:1998}, hereafter M98) showed that the initial number of stars in an OC is the main parameter determining its lifetime for a given Galactocentric distance. The disruption times predicted by M98 are comparable to those of \cite{Portegies-Zwart:2001}, whose simulations, differently from those of M98, used a fixed \cite{Scalo:1986} initial mass function (IMF) and included binary evolutionary effects. These numerical simulations suggest that currently observable OCRs can be descendants of initially rich OCs, containing as many as $N_{0}\sim10^3-10^4$ stars when they were born. 

Based on the $N-$body simulations of \cite{Baumgardt:2003} for clusters in an external tidal field, \cite{Lamers:2005} found that the cluster disruption time also depends on the local ambient density of the host galaxy (their section 2). 
The final stellar content of the remnant depends on the initial mass function and on the fraction of primordial binaries as well. As a consequence of dynamical interactions (internal relaxation and the action of the Galactic tidal field), remnants are expected to be biased towards stable dynamical configurations (binaries and long-lived triples) and deficient in low-mass stars because of their preferential evaporation and tidal stripping.

In this study we select a sample of objects catalogued as OCRs or OCR candidates. Our main goal is to compare their evolutionary stages by employing parameters that are intimately associated with the dynamical evolution: age, limiting radius, mass, and velocity dispersion. The masses of most of the OCRs are estimated here for the first time. We also discuss our results in the light of the expected properties predicted by $N-$body simulations. This study is a contribution towards clarifying the topic of OCRs and providing observational constraints to evolutionary models.

This paper is organized as follows: in Section \ref{sample} we present our sample. In Section \ref{method} we describe our method for deriving cluster parameters. In Section \ref{results} we present the results for the whole sample. The results are discussed in Sections \ref{discussion} and \ref{comparison_OCR_dyn_properties}. In Sect. \ref{conclusions} we summarise our conclusions.

 \section{Sample}
\label{sample}

\begin{figure*}
   \centering
   
   \parbox[c]{0.84\textwidth}
  {

    \includegraphics[width=0.42\textwidth]{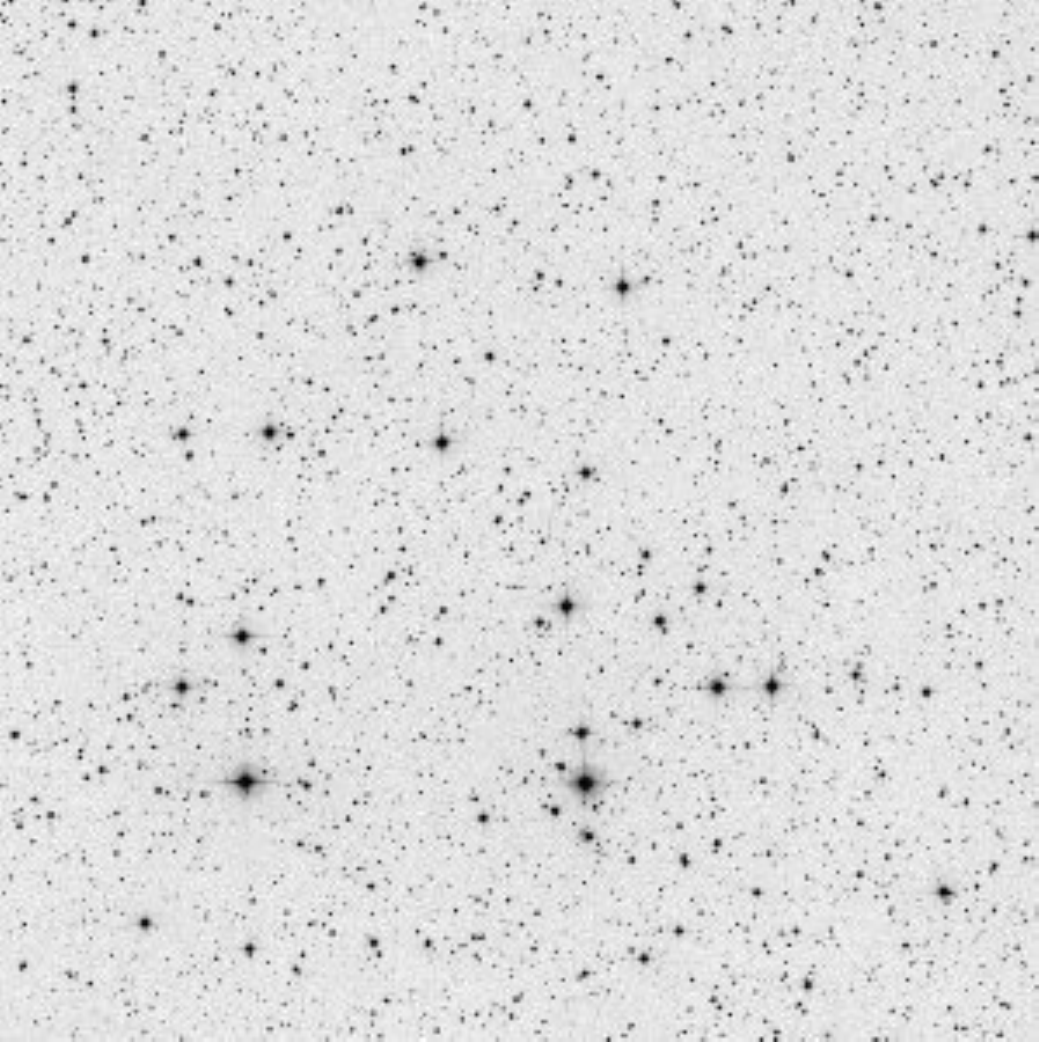}
    \includegraphics[width=0.42\textwidth]{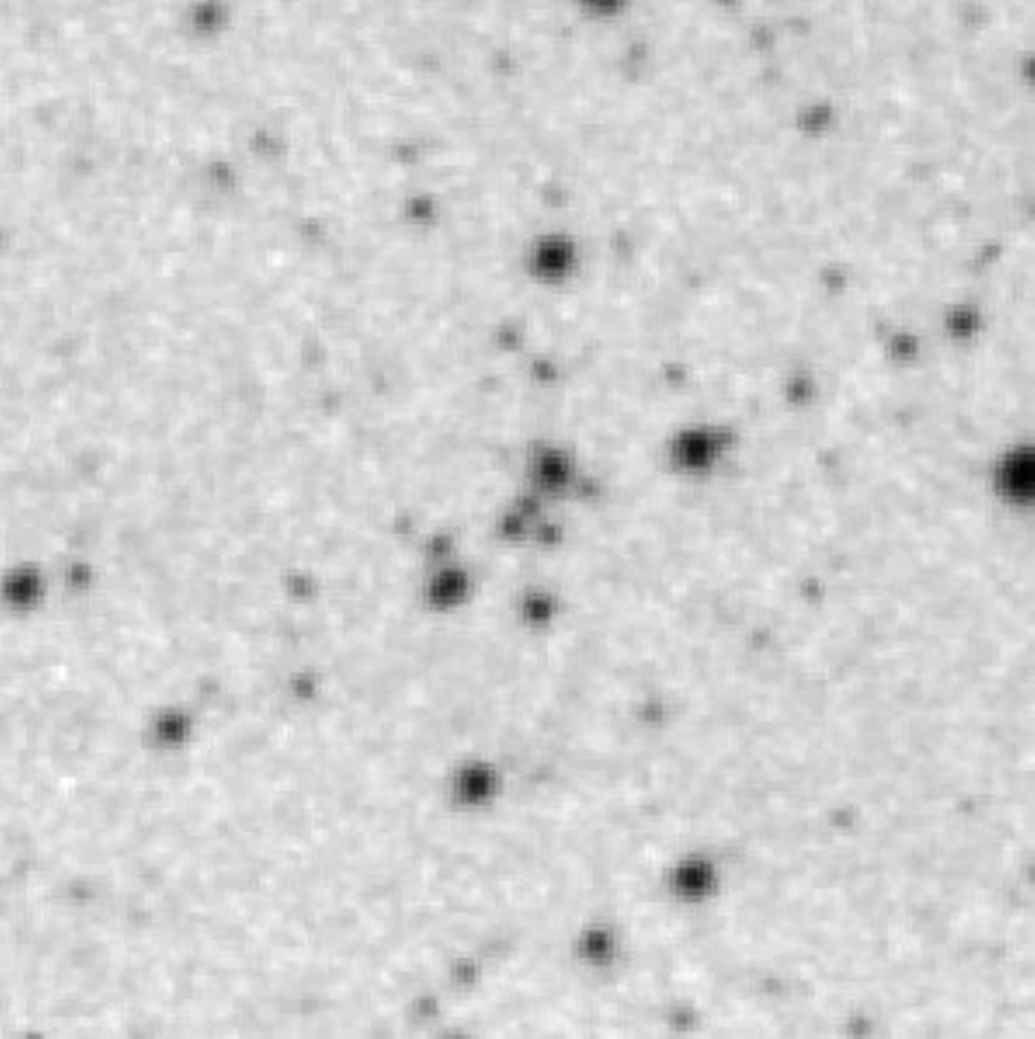}
    \includegraphics[width=0.42\textwidth]{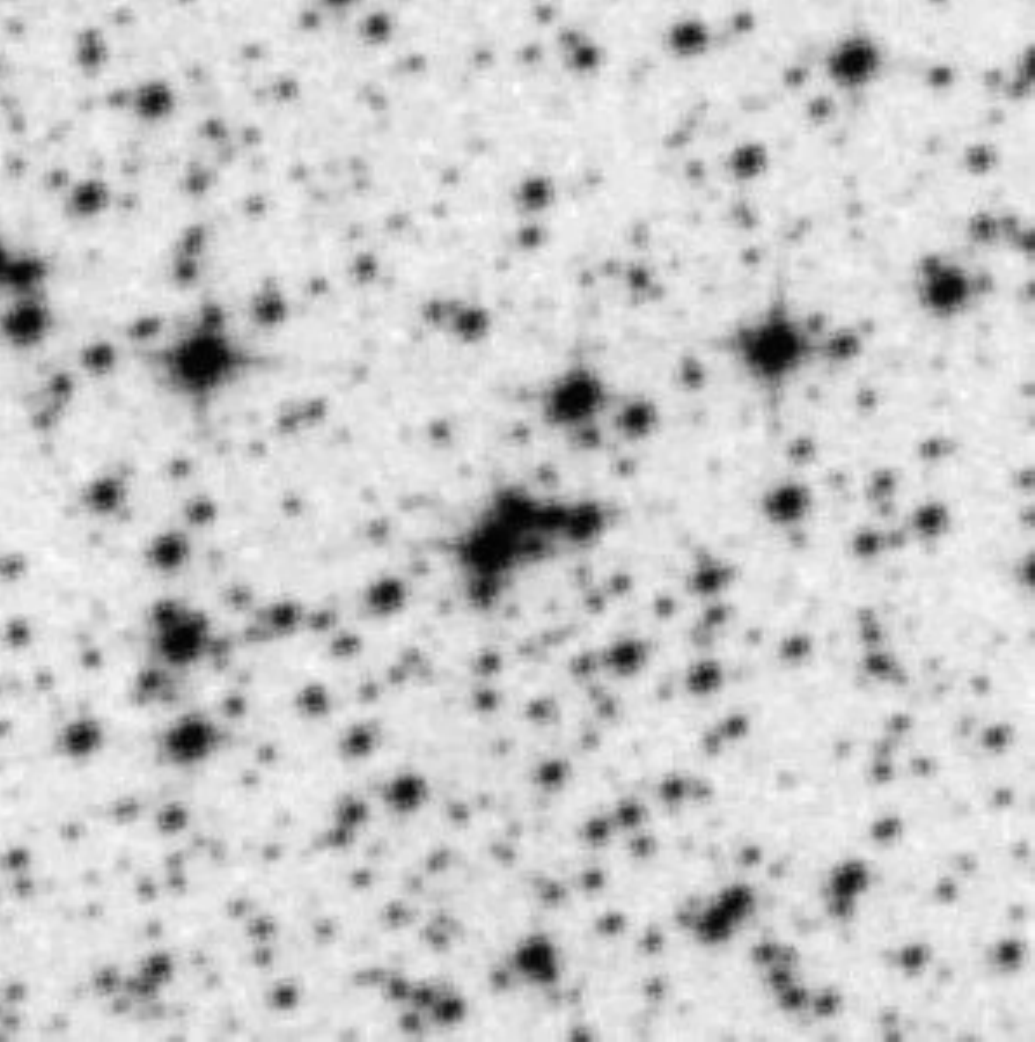}    
    \includegraphics[width=0.42\textwidth]{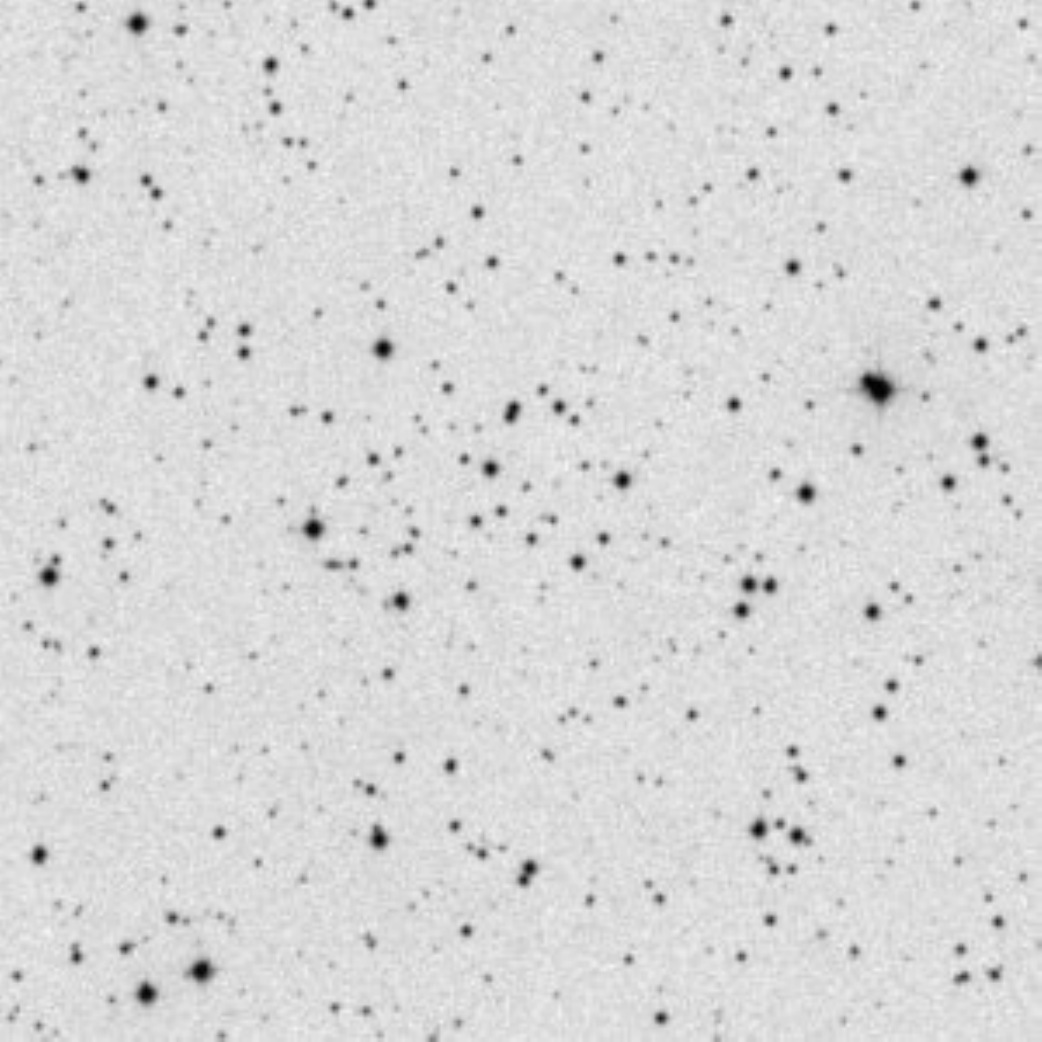}
    \includegraphics[width=0.42\textwidth]{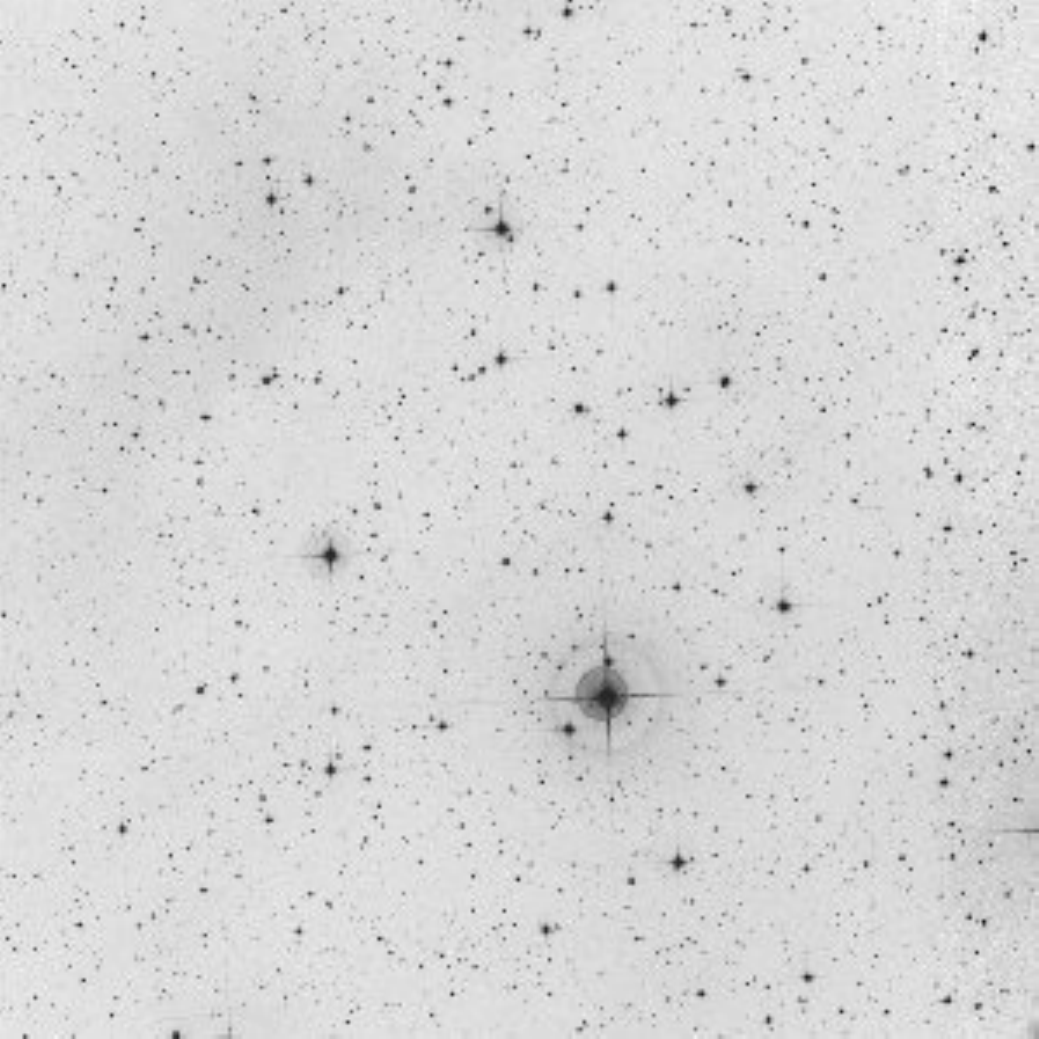}        
    \includegraphics[width=0.42\textwidth]{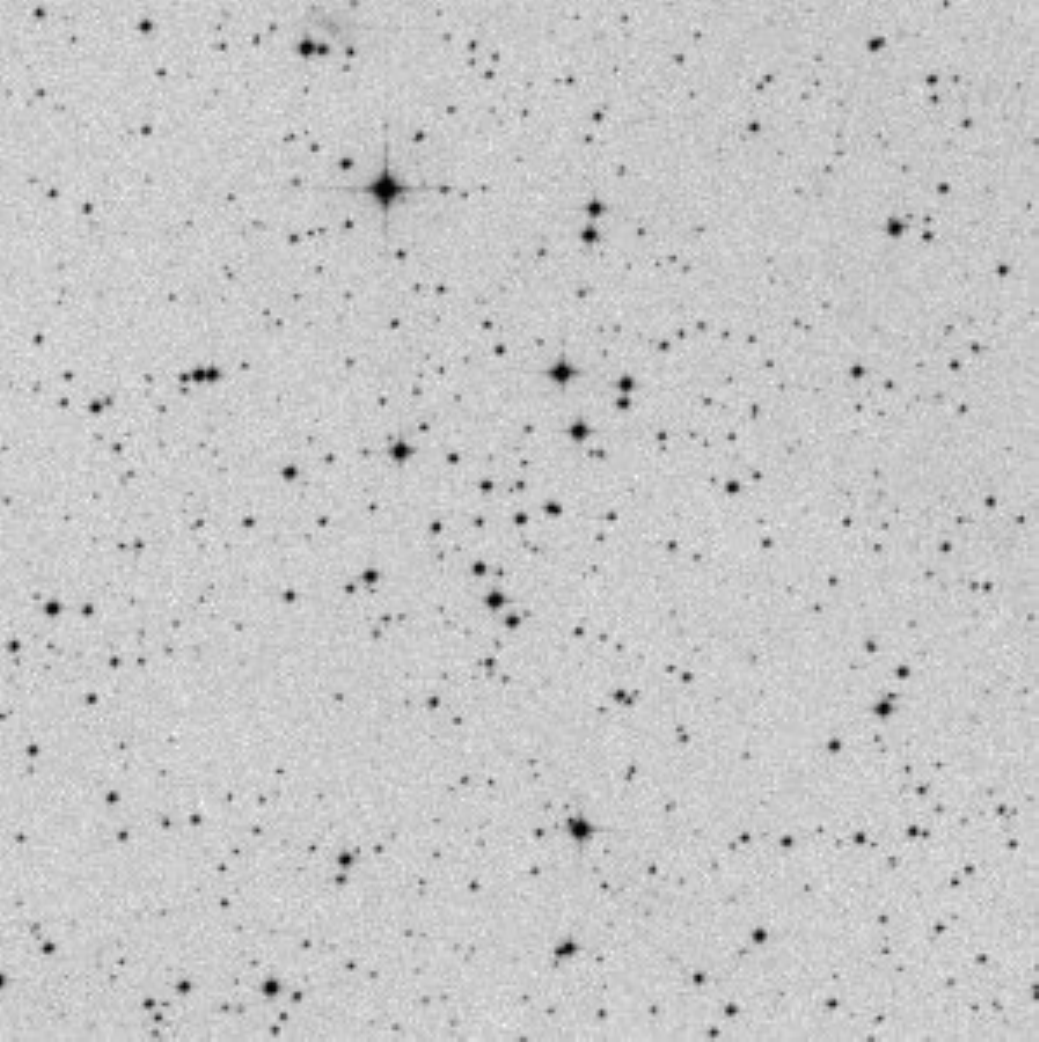}
 
  }
   
   \caption{DSS2 R images of the OCRs (from top left to lower right) NGC\,2180, Lynga\,8, Bica\,5, NGC\,7036, Alessi\,3, and ESO\,426-SC26. Image sizes are $25\arcmin\times25\arcmin$, $3\arcmin\times3\arcmin$, $4\arcmin\times4\arcmin$, $10\arcmin\times10\arcmin$, $35\arcmin\times35\arcmin$ and $14\arcmin\times14\arcmin$, respectively. North is up and east to the left.}             
 \label{fig_OCRs_parte1}
 \end{figure*}

\begin{sidewaystable*}
\caption{Coordinates and parameters of the studied open clusters and remnants.}\label{info_sample_OCs_OCRs}
\centering
\begin{tabular}{lccccrrrrrrr}

 \hline
  Cluster  & $\ell$  & $b$   &  RA$_{J2000}$  & DEC$_{J2000}$   &  $d_{\odot}$  &$R_{\textrm{G}}$   &$Z$    & Age     &$R_{\textrm{lim}}$  & $M_{\textrm{phot}}$  & $\sigma_v$   \\
        
         & $({\degr})$  & $({\degr})$  & $(h:m:s)$   & $(\degr:\arcmin:\arcsec)$   & (kpc)      & (kpc)     &(kpc)   & (Gyr)    & (pc)   & $(M_{\odot})$  & (km s$^{-1}$)  \\
               
 \hline
\vspace{0.005cm}\\
 \multicolumn{12}{c}{Open clusters (OCs)} \\
\vspace{0.005cm}\\ 
 \hline

NGC\,4337$^a$   &299.3    &  4.6    & 12:24:05  & -58:07:19  &   $ 2.75\pm0.38 $   &  $7.1\pm0.5 $   &  $ 0.22\pm0.03$     &  $1.58\pm0.18$     & $ 8.01\pm1.60$  & $186\pm18$  &$0.9\pm0.1$ \\ 
NGC\,3680       &286.8    & 16.9    & 11:25:40  & -43:14:11  &   $ 0.95\pm0.13 $   &  $7.8\pm0.5 $   &  $ 0.28\pm0.04$     &  $1.58\pm0.36$     & $ 6.11\pm0.28$  & $102\pm12$  &$0.9\pm0.1$ \\
NGC\,752$^a$    &137.1    &-23.3    & 01:57:38  &  37:49:01  &   $ 0.43\pm0.04 $   &  $8.3\pm0.5 $   &  $-0.17\pm0.02$     &  $1.41\pm0.16$     & $ 3.72\pm0.25$  & $158\pm15$  &$0.7\pm0.1$ \\ 
M\,67           &215.7    & 31.9    & 08:51:21  &  11:48:47  &   $ 0.85\pm0.04 $   &  $8.6\pm0.5 $   &  $ 0.45\pm0.02$     &  $3.55\pm0.41$     & $ 6.93\pm0.49$  & $772\pm29$  &$1.0\pm0.0$ \\
NGC\,1582       &159.3    & -2.9    & 04:31:50  &  43:47:46  &   $ 0.95\pm0.09 $   &  $8.9\pm0.5 $   &  $-0.05\pm0.00$     &  $0.20\pm0.09$     & $ 5.56\pm0.56$  & $197\pm19$  &$1.2\pm0.1$ \\
NGC\,2180$^b$   &203.8    & -7.0    & 06:09:44  &  04:49:24  &   $ 1.05\pm0.14 $   &  $9.0\pm0.5 $   &  $-0.13\pm0.02$     &  $0.71\pm0.33$     & $ 3.35\pm0.61$  & $ 27\pm 7$  &$5.2\pm0.9$ \\
NGC\,188        &122.9    & 22.4    & 00:48:01  &  85:14:45  &   $ 1.82\pm0.17 $   &  $9.1\pm0.5 $   &  $ 0.69\pm0.06$     &  $7.08\pm1.63$     & $13.23\pm1.06$  & $323\pm19$  &$1.0\pm0.1$ \\

\hline
\vspace{0.005cm}\\ 
\multicolumn{12}{c}{Open cluster remnants (OCRs)} \\
\vspace{0.005cm}\\ 
\hline

Lynga\,8        &333.3  & -0.1    &16:20:04  &-50:13:59  &   $2.40\pm0.33$   &  $ 6.0\pm0.6$   &   $ 0.00\pm0.00$   &  $0.40\pm0.14$   & $1.40\pm0.35$   & $38\pm10$   & $6.8\pm1.4$  \\      
NGC\,6481       & 29.9  & 14.9    &17:52:48  & 04:10:03  &   $1.51\pm0.21$   &  $ 6.8\pm0.5$   &   $ 0.39\pm0.05$   &  $2.24\pm0.77$   & $1.32\pm0.22$   & $ 8\pm 3$   & $7.2\pm2.1$  \\
ESO\,324-SC15   &309.3  & 20.6    &13:23:33  &-41:53:33  &   $1.32\pm0.18$   &  $ 7.3\pm0.5$   &   $ 0.46\pm0.06$   &  $3.16\pm0.36$   & $3.83\pm0.77$   & $13\pm 4$   & $6.4\pm1.9$  \\
Bica\,5         &294.9  & -0.6    &11:41:35  &-62:25:04  &   $1.91\pm0.26$   &  $ 7.4\pm0.5$   &   $-0.02\pm0.00$   &  $0.50\pm0.17$   & $1.66\pm0.28$   & $25\pm 7$   & $6.1\pm1.2$  \\
NGC\,7036       & 64.5  &-21.4    &21:09:58  & 15:30:52  &   $1.46\pm0.20$   &  $ 7.5\pm0.5$   &   $-0.53\pm0.07$   &  $2.24\pm0.52$   & $3.40\pm0.42$   & $15\pm 4$   & $6.3\pm1.4$  \\  
NGC\,7193       & 70.1  &-34.3    &22:03:08  & 10:48:14  &   $1.25\pm0.17$   &  $ 7.7\pm0.5$   &   $-0.70\pm0.10$   &  $4.47\pm1.54$   & $3.64\pm0.36$   & $11\pm 3$   & $5.7\pm1.5$  \\ 
NGC\,1901       &279.0  &-33.6    &05:18:14  &-68:26:38  &   $0.42\pm0.06$   &  $ 8.0\pm0.5$   &   $-0.23\pm0.03$   &  $0.50\pm0.23$   & $1.21\pm0.24$   & $22\pm 6$   & $0.7\pm0.1$  \\  
Alessi\,3       &257.9  &-15.4    &07:16:33  &-46:40:31  &   $0.26\pm0.04$   &  $ 8.1\pm0.5$   &   $-0.07\pm0.01$   &  $0.63\pm0.22$   & $1.53\pm0.15$   & $18\pm 5$   & $0.9\pm0.2$  \\ 
ESO\,435-SC48   &264.8  & 22.3    &10:09:32  &-28:21:43  &   $2.69\pm0.37$   &  $ 8.7\pm0.5$   &   $ 1.02\pm0.14$   &  $1.58\pm0.73$   & $6.25\pm0.78$   & $16\pm 5$   & $9.2\pm2.5$  \\
NGC\,3231       &142.0  & 44.6    &10:26:58  & 66:47:54  &   $1.08\pm0.17$   &  $ 8.7\pm0.5$   &   $ 0.76\pm0.12$   &  $2.00\pm0.92$   & $2.82\pm0.31$   & $11\pm 4$   & $3.1\pm1.0$  \\ 
Ruprecht\,31    &250.1  & -6.0    &07:42:60  &-35:36:00  &   $2.29\pm0.32$   &  $ 9.0\pm0.5$   &   $-0.24\pm0.03$   &  $0.79\pm0.18$   & $4.00\pm1.00$   & $27\pm 7$   & $8.1\pm1.8$  \\
ESO\,425-SC15   &236.4  &-20.4    &06:14:33  &-29:21:54  &   $1.91\pm0.26$   &  $ 9.1\pm0.5$   &   $-0.66\pm0.09$   &  $1.12\pm0.52$   & $3.88\pm0.55$   & $16\pm 5$   & $7.1\pm1.8$  \\
ESO\,426-SC26   &239.6  &-16.5    &06:36:18  &-30:51:57  &   $2.00\pm0.28$   &  $ 9.1\pm0.5$   &   $-0.57\pm0.08$   &  $0.63\pm0.14$   & $5.80\pm1.16$   & $29\pm 7$   & $5.3\pm1.2$  \\ 
ESO\,425-SC06   &235.4  &-22.3    &06:04:52  &-29:11:04  &   $2.00\pm0.18$   &  $ 9.2\pm0.5$   &   $-0.76\pm0.07$   &  $3.16\pm0.36$   & $3.48\pm0.58$   & $14\pm 4$   & $6.9\pm2.0$  \\
Ruprecht\,3     &238.8  &-14.8    &06:42:08  &-29:27:07  &   $2.29\pm0.32$   &  $ 9.4\pm0.5$   &   $-0.59\pm0.08$   &  $1.12\pm0.26$   & $3.33\pm0.67$   & $24\pm 7$   & $7.1\pm1.7$  \\  
NGC\,1663       &185.9  &-19.7    &04:49:24  & 13:08:25  &   $2.40\pm0.33$   &  $10.3\pm0.6$   &   $-0.81\pm0.11$   &  $5.01\pm2.31$   & $6.98\pm1.40$   & $15\pm 4$   & $3.9\pm1.0$  \\

\hline

\multicolumn{12}{l}{    }  \\
\multicolumn{12}{l}{  $^{a}$ [Fe/H]=0.12 \cite{Carraro:2014}.  }  \\
\multicolumn{12}{l}{ $^{b}$ In this paper, we advocate that this cluster (previously classified as an OC) is entering a remnant evolutionary stage. } \\

\end{tabular}
\end{sidewaystable*}

The selected sample is composed of 16 objects catalogued as OCRs or OCR candidates. Images for 6 of them (including NGC\,2180) are shown in Fig. \ref{fig_OCRs_parte1}. In this paper (see Section \ref{discussion}), we propose that NGC\,2180 is better classified as a remnant. Images of the remaining sample are shown in the appendix. We adopt the same procedure throughout for other figures. Six objects, catalogued as OCs in advanced dynamical states, were also included for comparison purposes. The complete set of parameters for the studied OCs is shown in Table \ref{info_sample_OCs_OCRs}. The columns list the cluster name, Galactic ($\ell,b$) and equatorial (RA,DEC) coordinates, distance from the Sun ($d_{\odot}$), Galactocentric distance ($R_{G}$), distance from the Galactic plane ($Z_{G}$), age, limiting radius ($R_{\textrm{lim}}$), photometric mass ($M_{\textrm{phot}}$), and velocity dispersion ($\sigma_v$).

For each target in Table \ref{info_sample_OCs_OCRs}, near-infrared ($J,\,H,\,$ and $K_{s}$\,band) photometric data from the 2MASS catalogue \citep{Skrutskie:2006} were extracted for stars inside a circular region, centered in the cluster coordinates, with radius larger than five times the literature limiting radius, when available. Otherwise, a visual estimate of the limiting radius was used to establish the field size extraction. Magnitudes in $J,\,H,\,$ and $K_{s}$ were restricted to cutoff values following the catalogue completeness limit according to the Galactic coordinates\footnote[1]{http://www.astro.caltech.edu/$\sim$jmc/2mass/v3/gp/analysis.html}. Proper motions and parallaxes for stars in these same regions were extracted from the GAIA DR2 catalogue. Data from both catalogues were extracted by means of the Vizier Service \footnote[2]{http://vizier.u-strasbg.fr/viz-bin/VizieR}.

\section{Method}
\label{method}
\subsection{Determining the radial density profile and limiting radius}
\label{rdp_limiting_radius}

\begin{figure*}
\begin{center}
\parbox[c]{0.90\textwidth}
  {

    \includegraphics[width=0.45\textwidth]{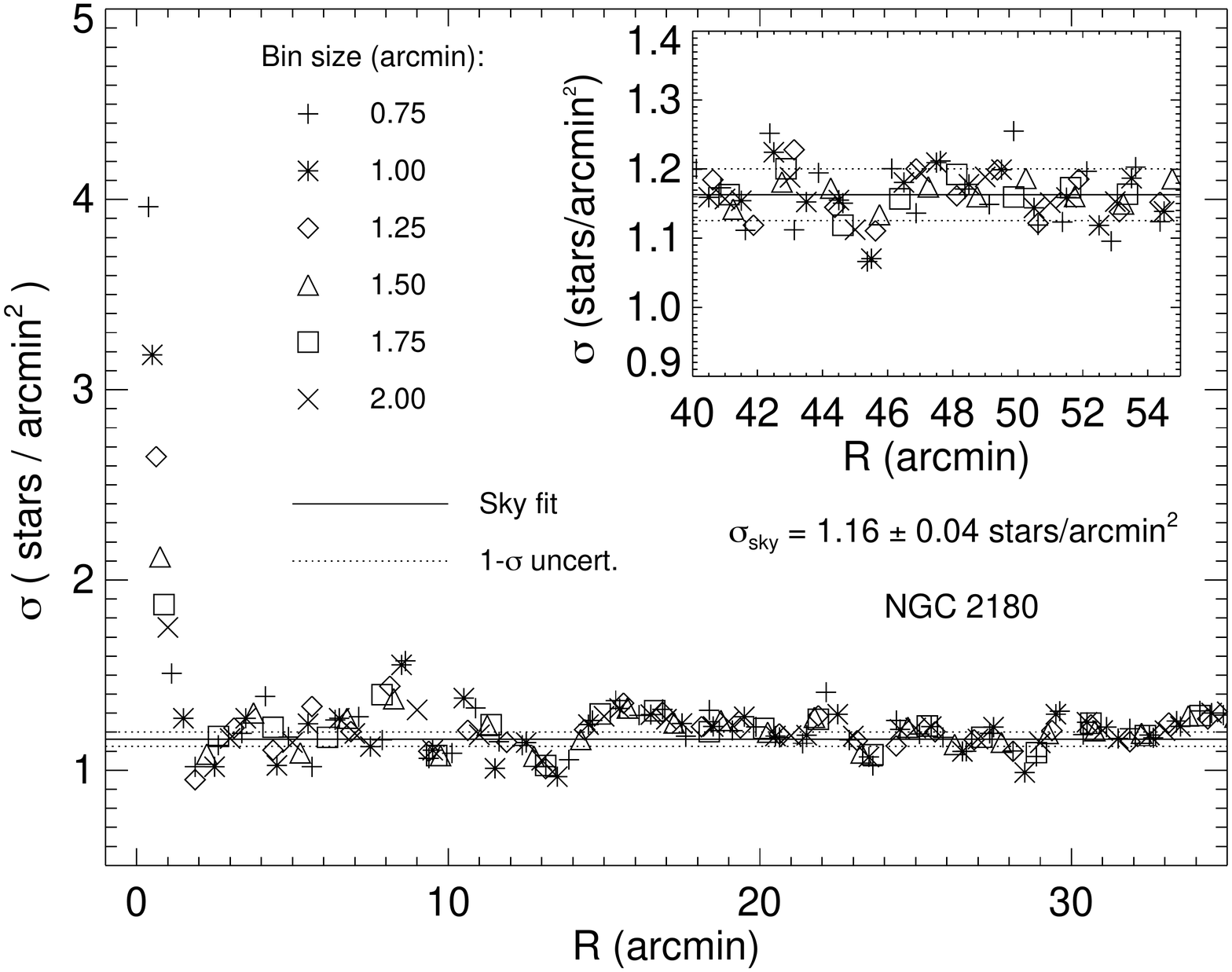}
    \includegraphics[width=0.45\textwidth]{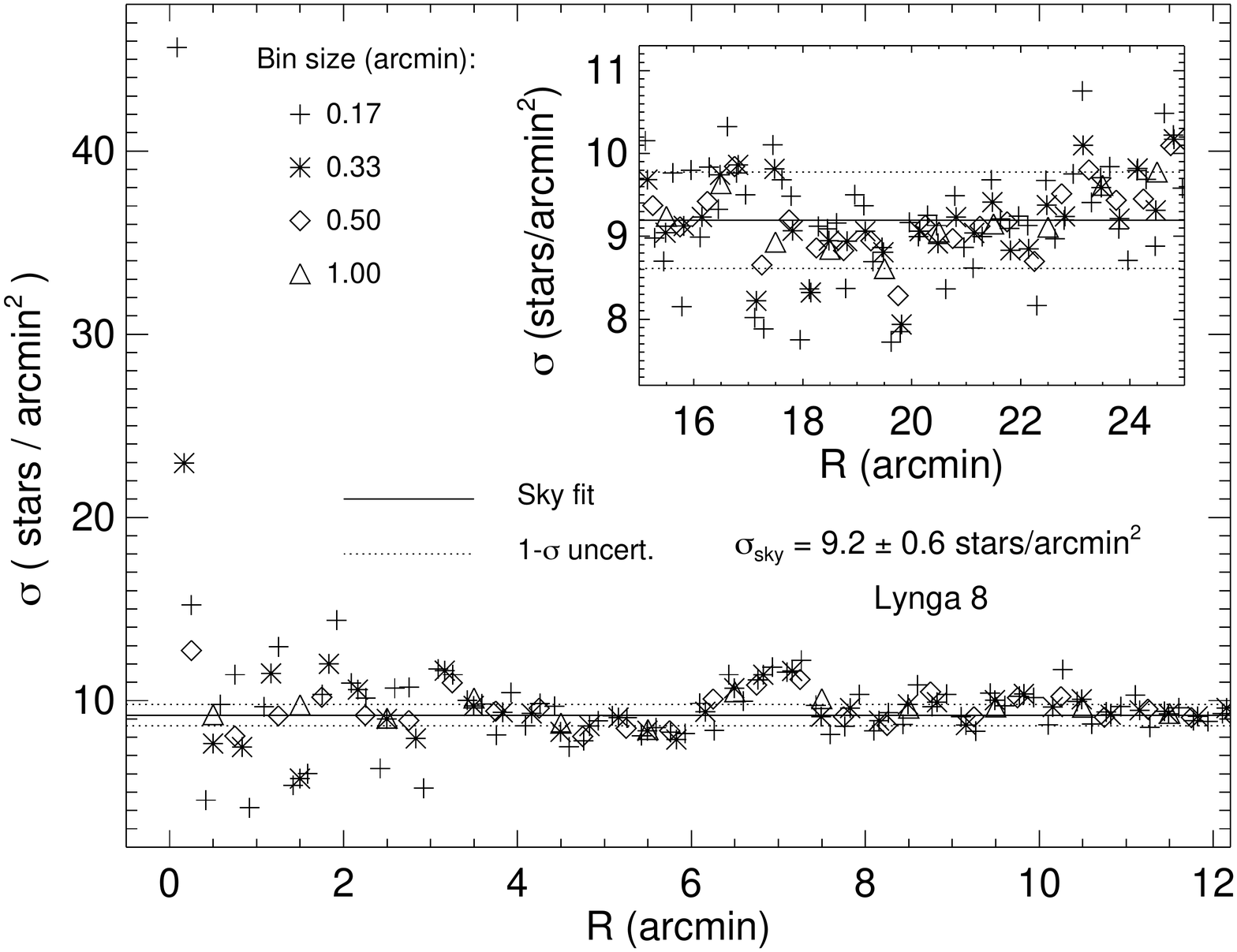}
    \includegraphics[width=0.45\textwidth]{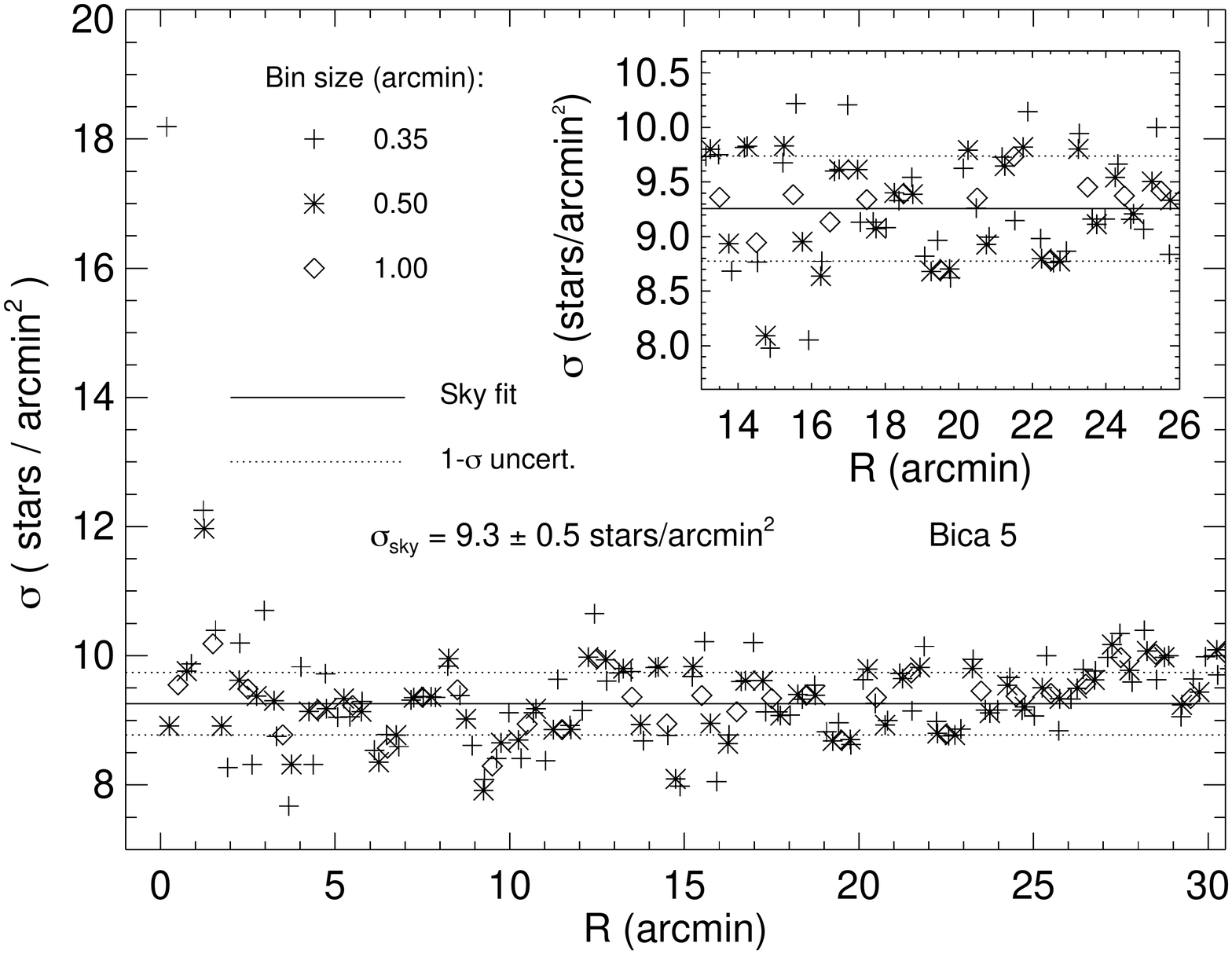}
    \includegraphics[width=0.45\textwidth]{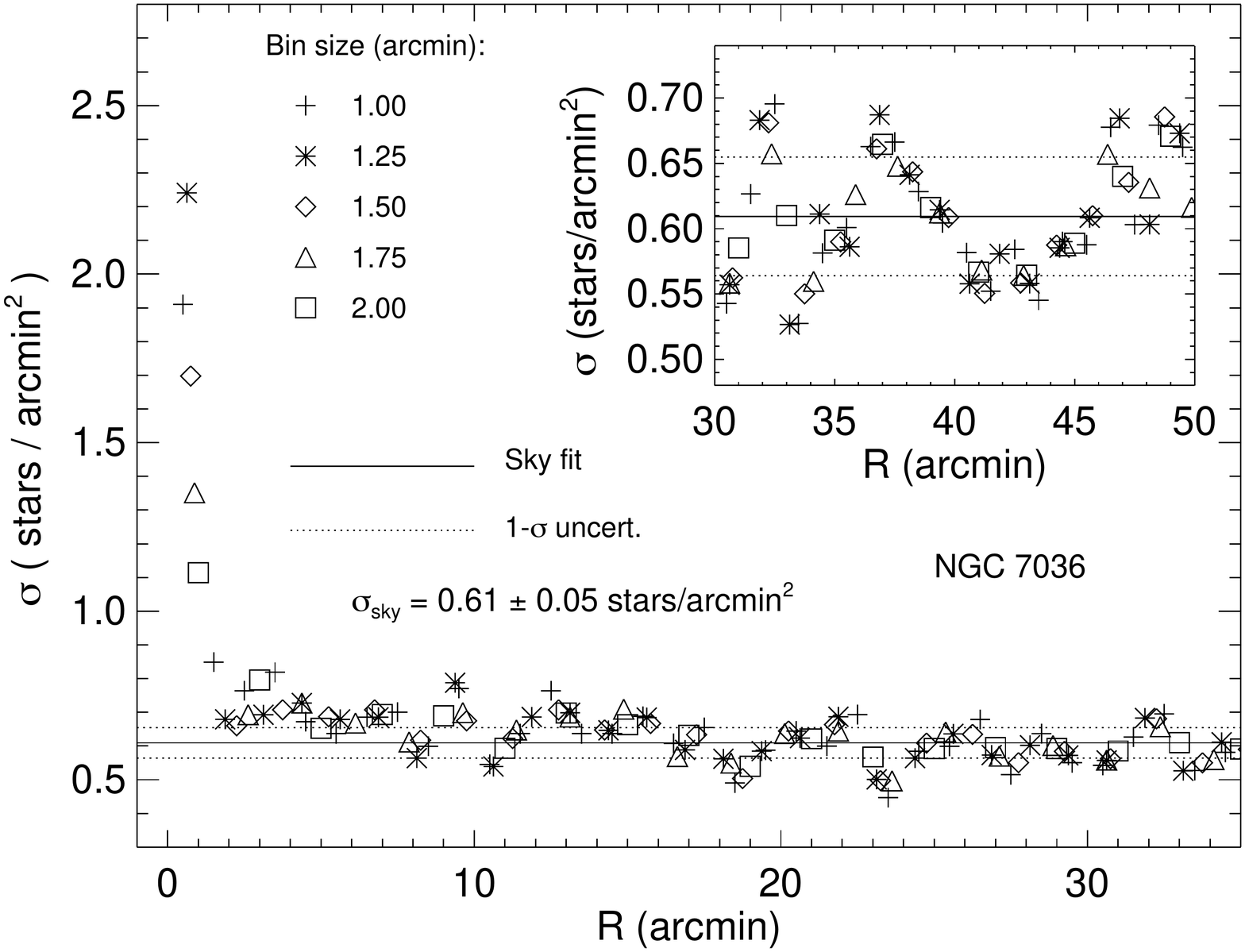}
    \includegraphics[width=0.45\textwidth]{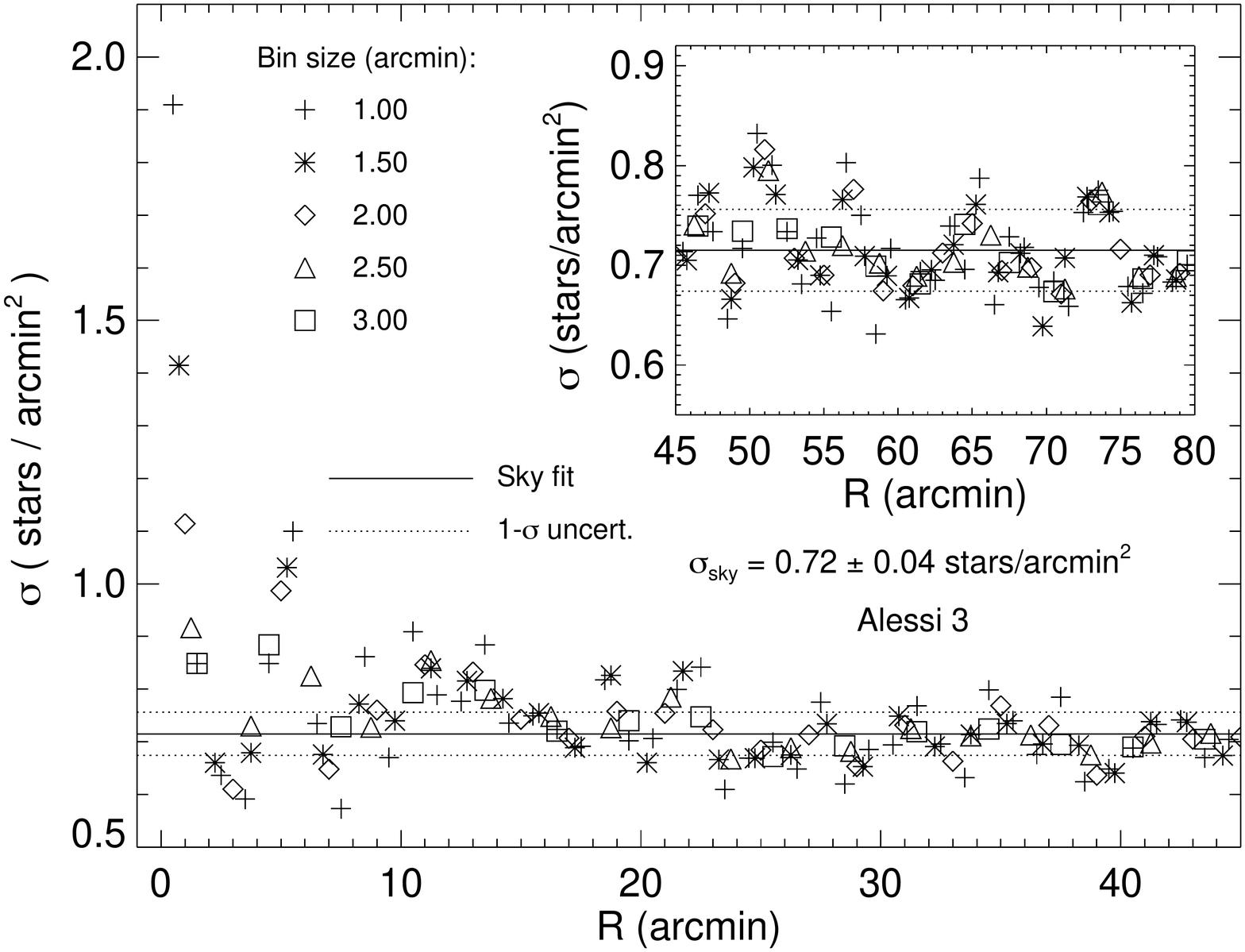}
    \includegraphics[width=0.45\textwidth]{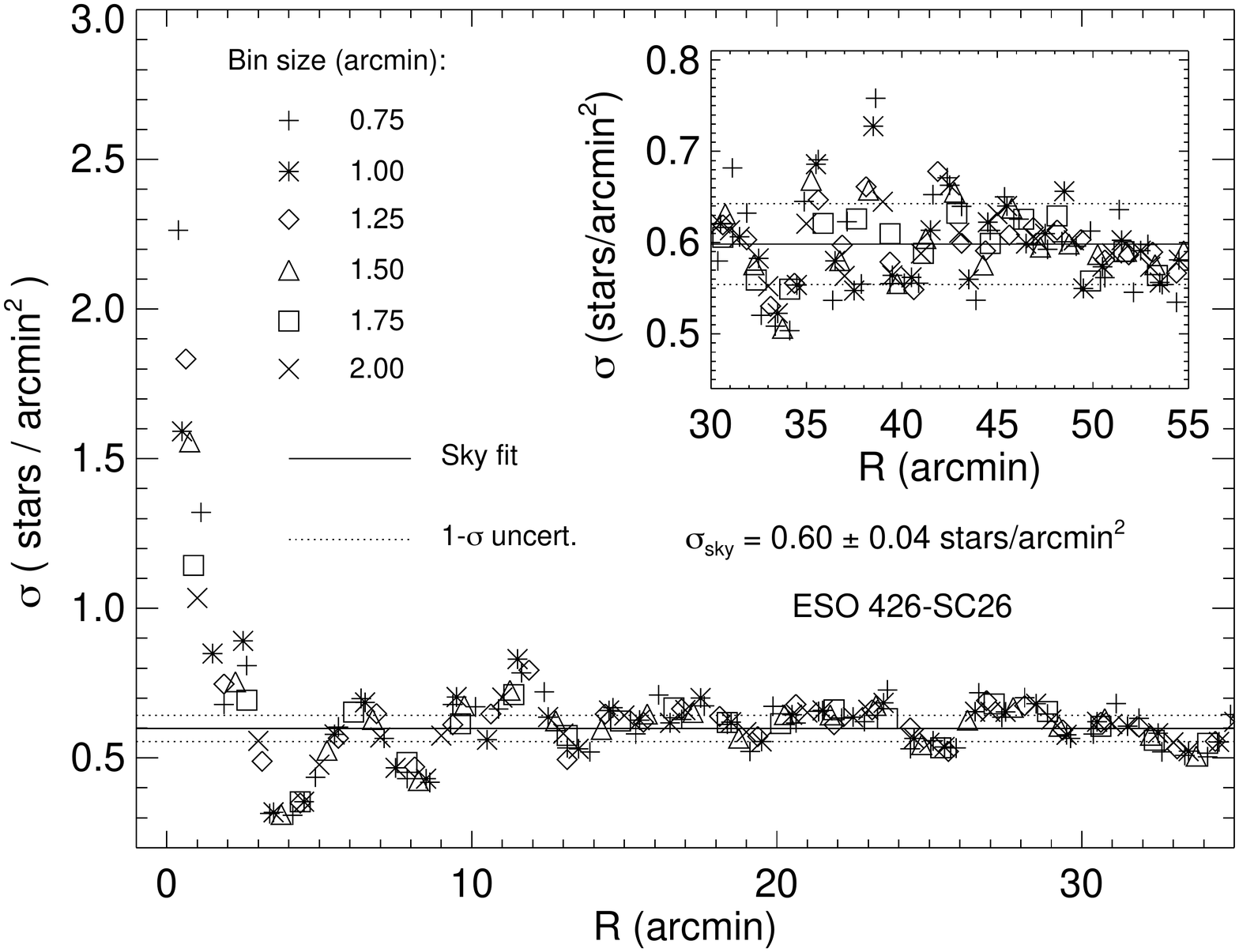}

  }

\caption{Radial density profiles for the OCRs NGC\,2180, Lynga\,8, Bica\,5, NGC\,7036, Alessi\,3, and ESO\,426-SC26. Symbols representing different bin sizes are overplotted. For each cluster, the inset shows the region (dashed rectangle) selected for estimation of the mean background density ($\sigma_{\textrm{sky}}$, continuous lines) and its 1$\sigma$ uncertainty (dotted lines).}
\label{RDPs_parte1}
\end{center}
\end{figure*}

\begin{figure}
\begin{center}
 \includegraphics[width=6.0cm]{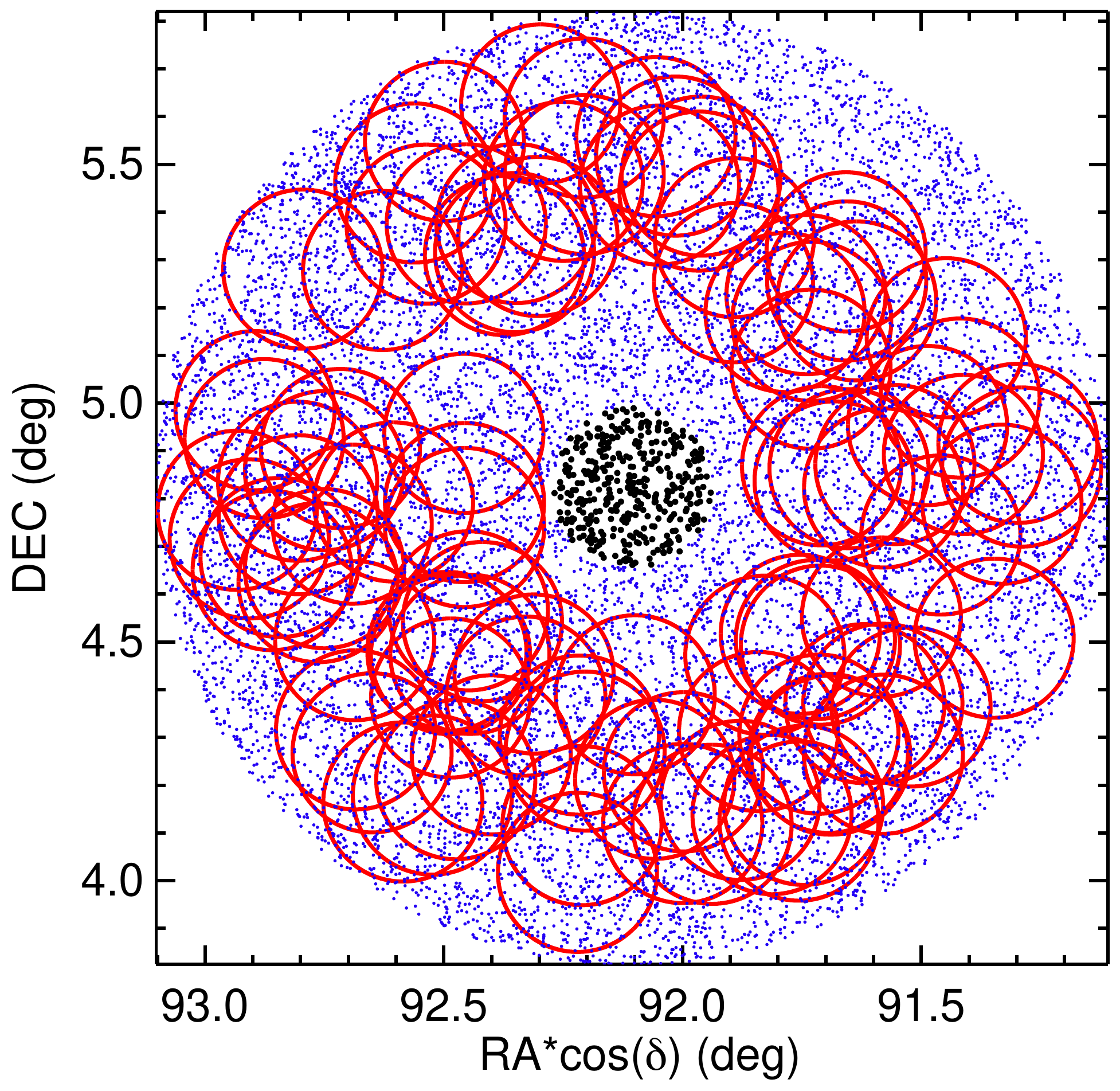}
 \caption{Sky map of NGC\,2180. Stars with $J$$\,\leq\,$15.8, $H$$\,\leq\,$15.1, and $K_{s}$$\,\leq\,$14.3\,mag that are located within 1\,degree of the central coordinates are shown. Black circles represent stars within 10\,arcmin. Other stars are plotted as blue dots. Red circles illustrate a set of 100 randomly selected field samples with radii equal to 10\,arcmin.}
   \label{skymap_randomfieldsamples_Rcut_NGC2180}
\end{center}

\end{figure}

\begin{figure*}
\begin{center}
\parbox[c]{1.00\textwidth}
  {

    \includegraphics[width=0.33333\textwidth]{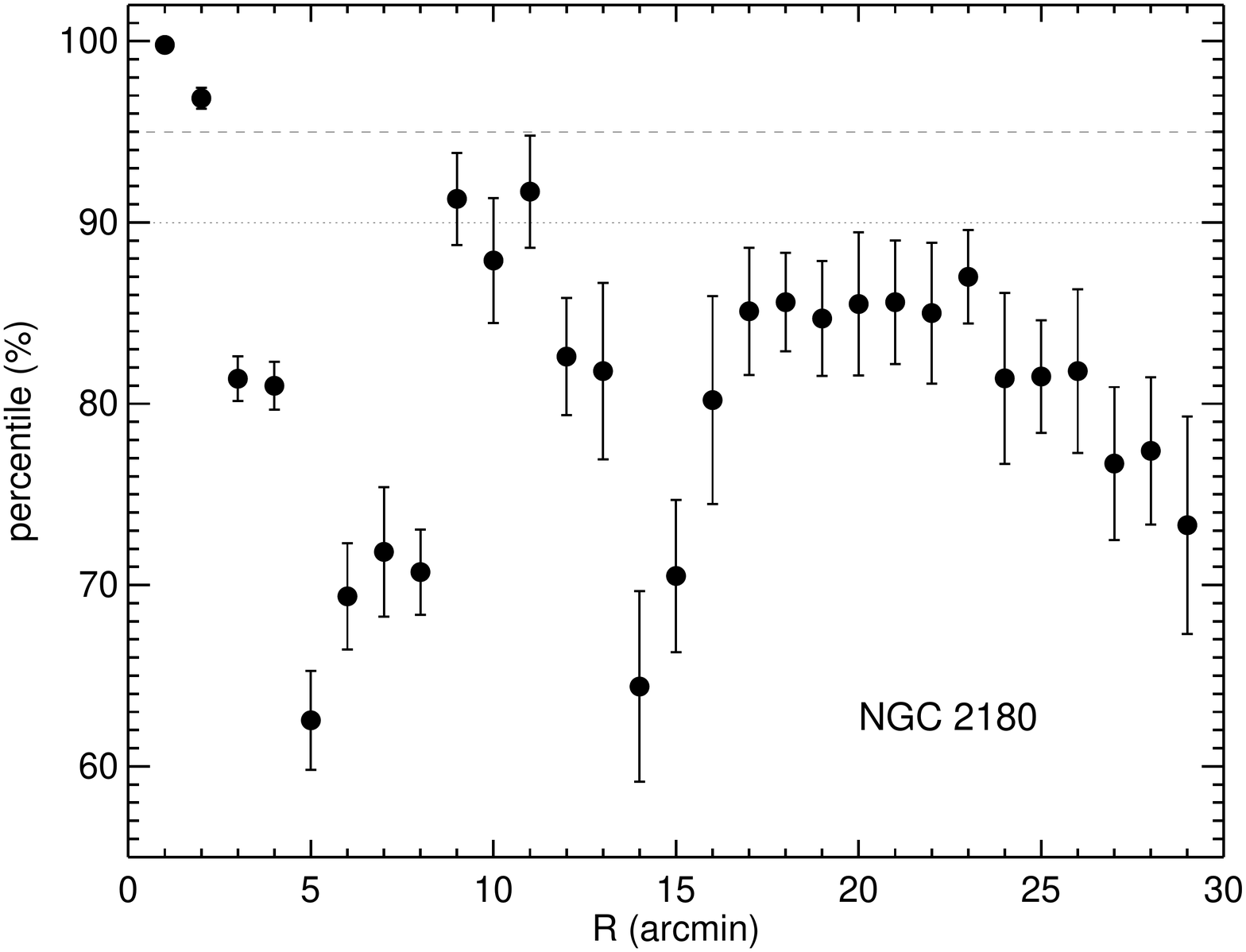}
    \includegraphics[width=0.33333\textwidth]{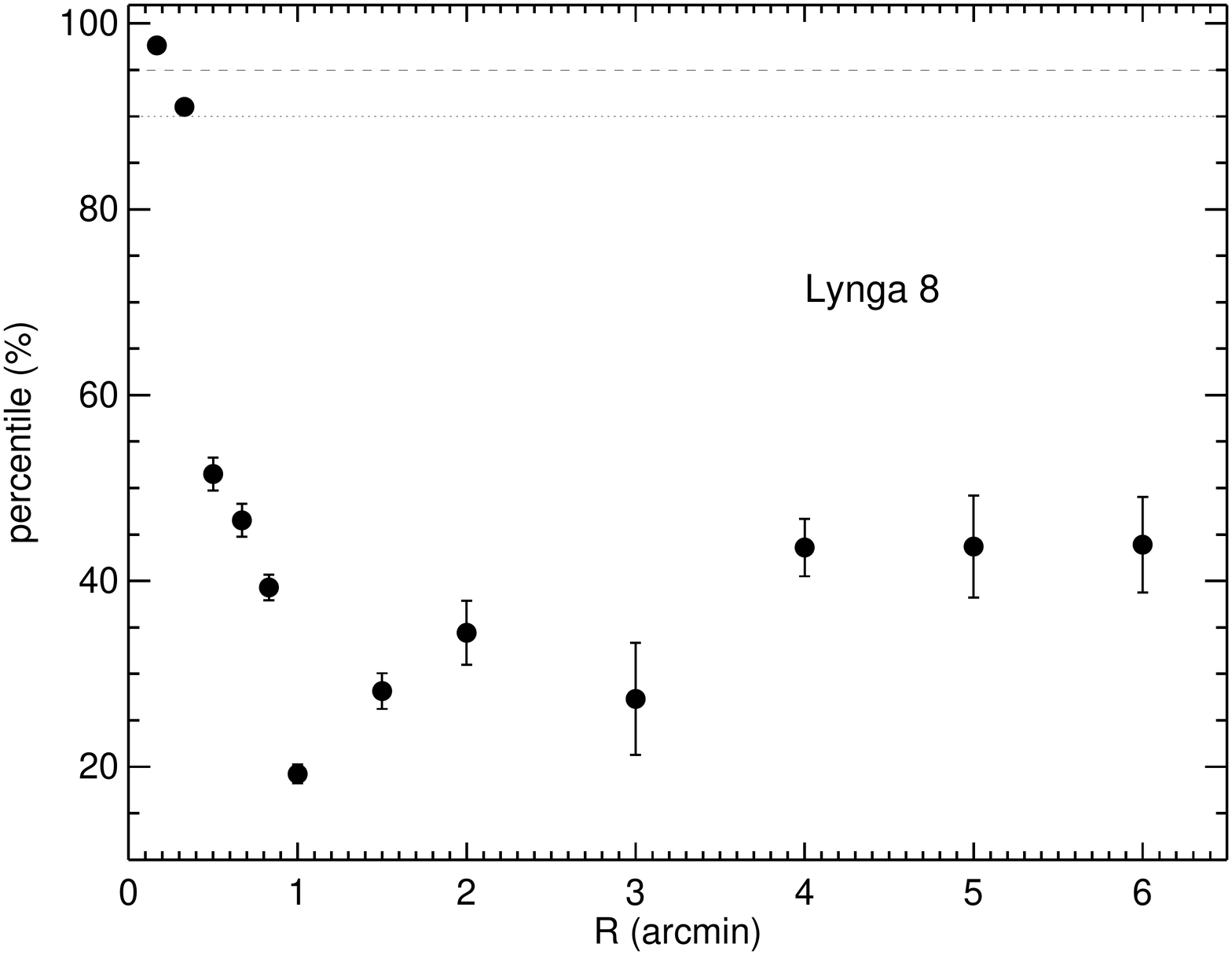}
    \includegraphics[width=0.33333\textwidth]{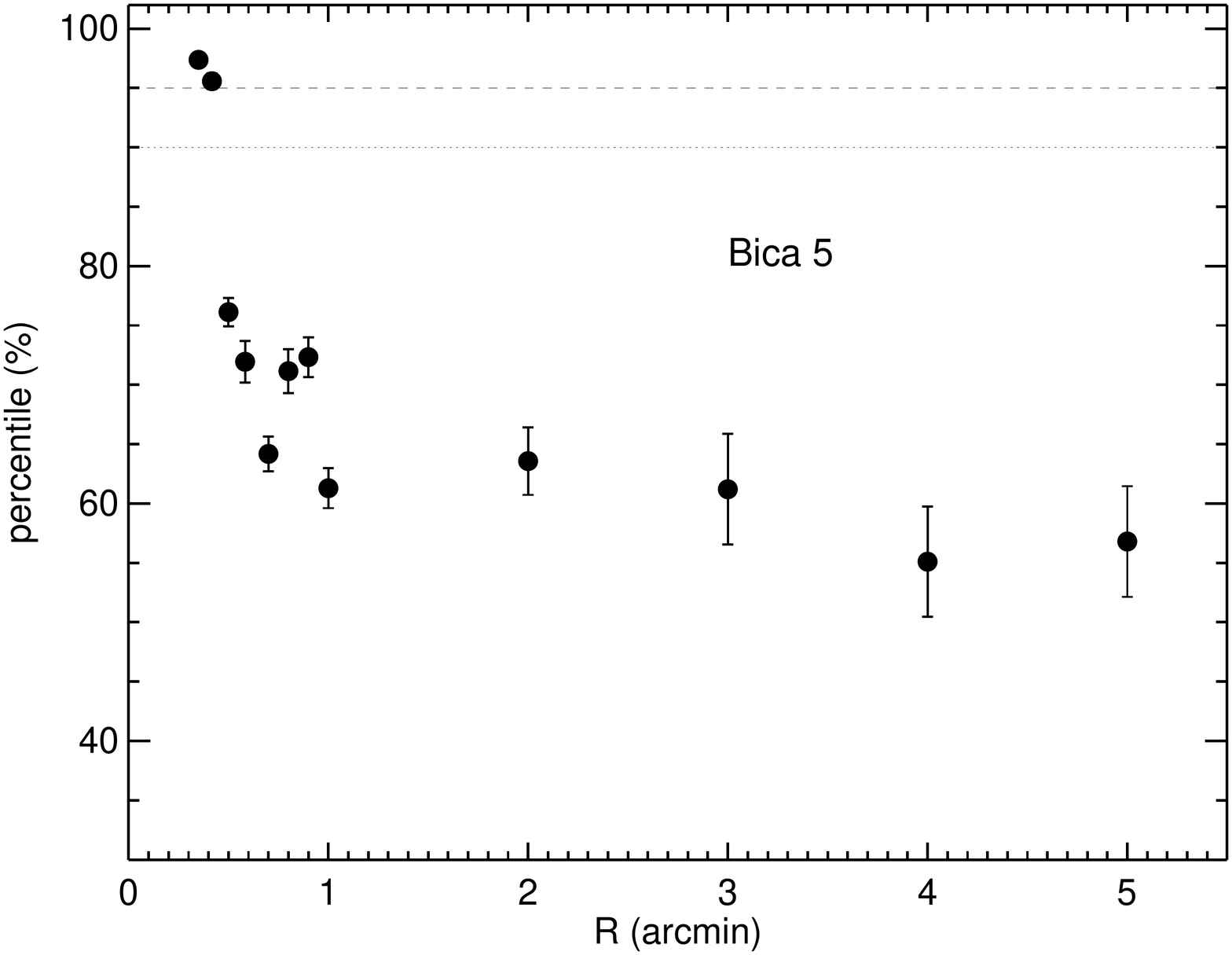}
    \includegraphics[width=0.33333\textwidth]{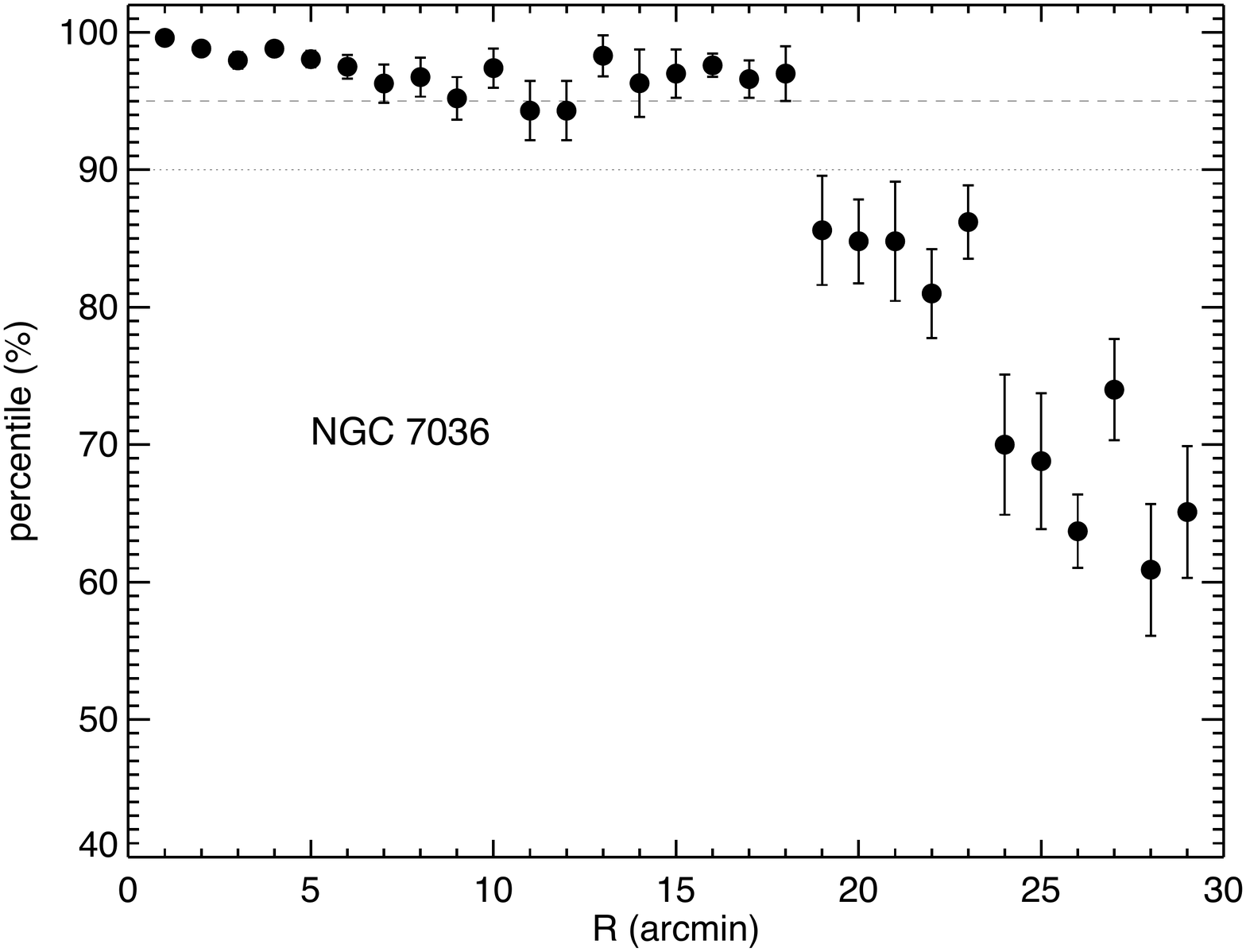}
    \includegraphics[width=0.33333\textwidth]{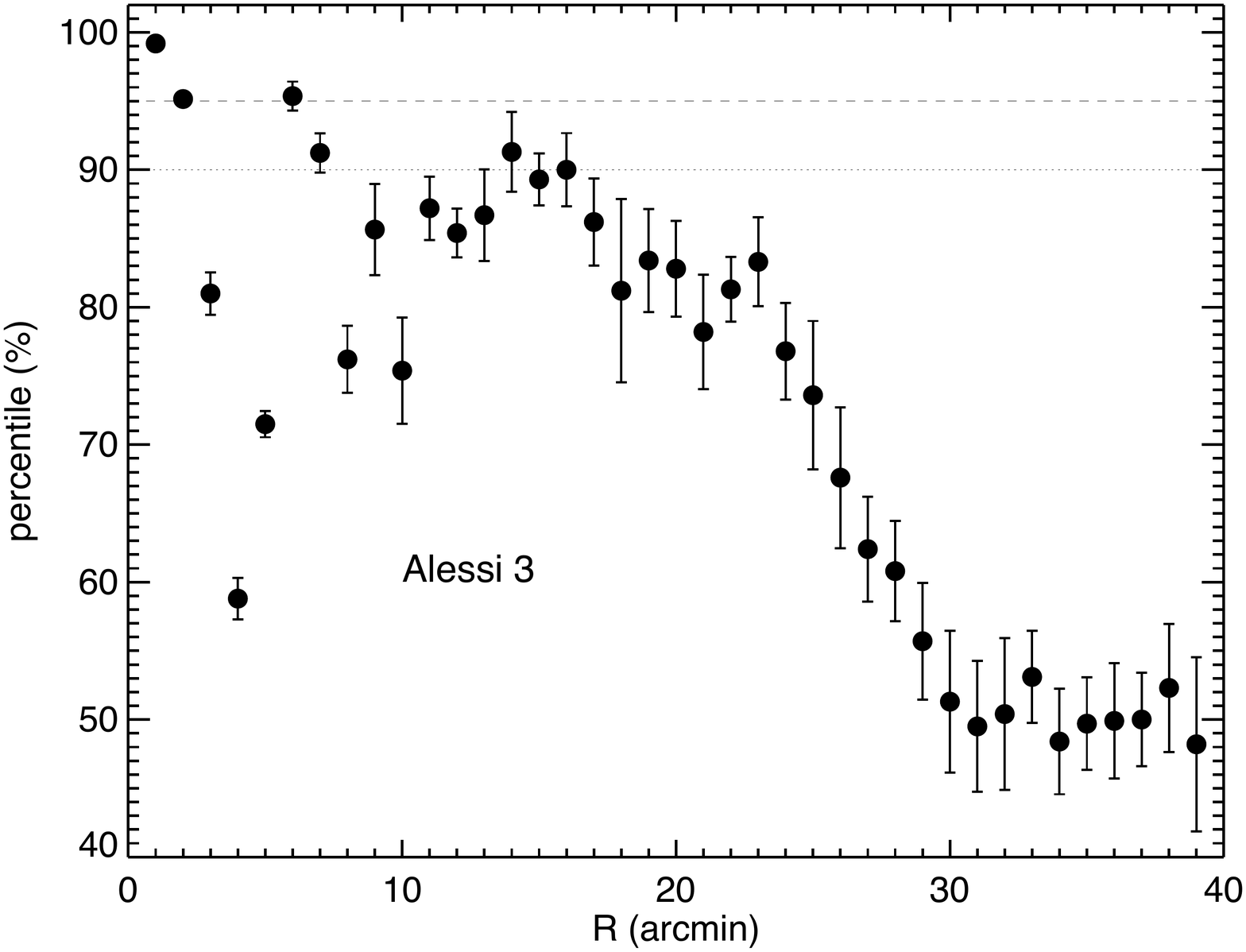}
    \includegraphics[width=0.33333\textwidth]{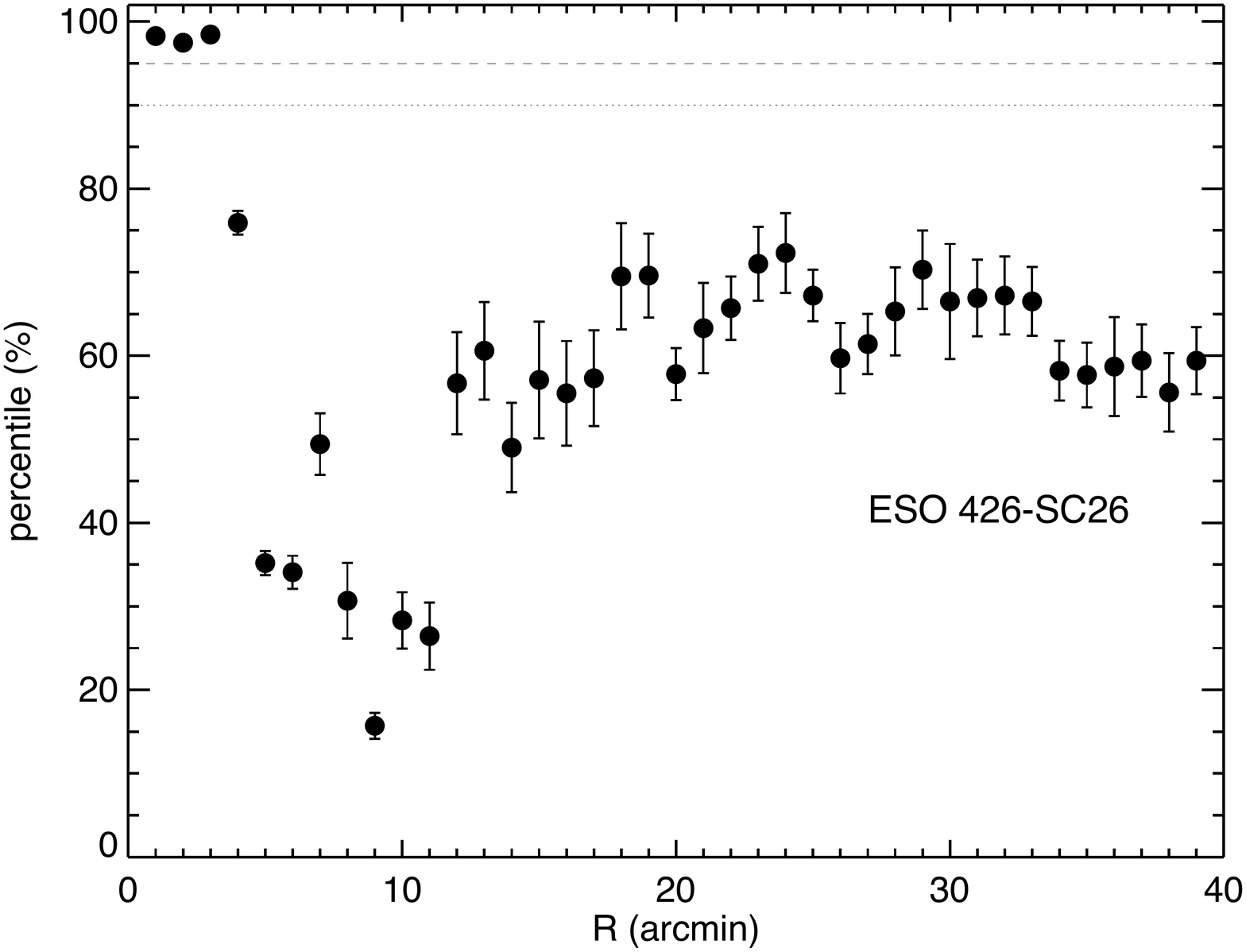}
  }

\caption{Average percentiles and associated $1\sigma$ uncertainties corresponding to the ratio $N_{\textrm{cluster}}/\langle N_{\textrm{field}}\rangle$ within the $N_{\textrm{field}}/\langle N_{\textrm{field}}\rangle$ distribution for each test radius. Dotted and dashed lines mark the 90th and 95th percentiles, respectively. The results are shown for the OCRs (from top left to lower right) NGC\,2180, Lynga\,8, Bica\,5, NGC\,7036, Alessi\,3, and ESO\,426-SC26.}
\label{percentile_vs_radius_parte1}
\end{center}
\end{figure*}

For each object listed in Table \ref{info_sample_OCs_OCRs}, a tentative central coordinates redetermination was performed by employing spatial density profiles along RA and DEC. Peaks in these profiles were adopted as first guesses for the central coordinates. Next, we manually varied these coordinates looking for a radial density profile (RDP) with a clear central stellar overdensity compared to the mean background density. The RDPs were constructed by counting the number of stars inside circular rings with variable widths and dividing this number by the respective ring area. The narrower rings are ideal for probing the central regions because their stellar densities are higher, while the wider rings are better suited to probing the external regions. This prevents either region of the cluster from being undersampled (e.g. \citeauthor{Maia:2010}\,\citeyear{Maia:2010}).

The RDPs for six of our clusters are shown in Fig. \ref{RDPs_parte1}. See the appendix for the full sample plots. The insets in these figures show the region selected for estimation of the mean background densities ($\sigma_{\textrm{sky}}$, continuous lines) and their 1\,$\sigma$ uncertainties (dotted lines). In each case, the limiting radius ($R_{\textrm{lim}}$) is considered as the radius from which the stellar densities  tend to fluctuate around $\sigma_{\textrm{sky}}$. The angular $R_{\textrm{lim}}$ radii were converted into parsec using the distances (Table \ref{info_sample_OCs_OCRs}) derived from isochrone fitting (Sect. \ref{results}).

Although the RDPs show a central stellar overdensity compared to the background, we evaluated the statistical significance of these density peaks compared to the number of stars counted in a set of field samples randomly chosen around each object. A real stellar aggregate is expected to exhibit a contrast compared to the field, which is the first step to establishing its physical nature \citep{Bica:2001}.

To accomplish this, we devised an algorithm that is based on the prescriptions detailed in \cite{Pavani:2011}. The procedure consists of comparing the number of stars that are counted in the central region of a given object (within $R_{\textrm{lim}}$) with that counted in a set of field samples of the same area. Fig. \ref{skymap_randomfieldsamples_Rcut_NGC2180} illustrates the procedure for the OC NGC\,2180. For a given $R_{\textrm{lim}}$, the distribution of $N_{\textrm{field}}/\langle N_{\textrm{field}}\rangle$ is built, where $N_{\textrm{field}}$ is the number of stars counted within each of the red circles in Fig. \ref{skymap_randomfieldsamples_Rcut_NGC2180} and $\langle N_{\textrm{field}}\rangle$ is the average value taken over the whole ensemble. Then the ratio $N_{\textrm{cluster}}/\langle N_{\textrm{field}}\rangle$ is evaluated, where $N_{\textrm{cluster}}$ is the number of stars counted in the cluster centre (black dots in Fig. \ref{skymap_randomfieldsamples_Rcut_NGC2180}), and the percentile corresponding to this ratio within the $N_{\textrm{field}}/\langle N_{\textrm{field}}\rangle$ distribution is determined. This procedure is performed for a set of radius values, thus allowing us to estimate the percentile as a function of $R_{\textrm{lim}}$. Fluctuations around each test radius are taken into account by repeating the algorithm ten times.

Fig. \ref{percentile_vs_radius_parte1} shows the percentiles as a function of $R_{\textrm{lim}}$ for six objects of our sample (Appendix A contains the plots for the full sample). In all cases there is at least one percentile value above 90\%, which reveals significant contrasts between the targets and the field with respect to star counts in the neighbouring regions. The percentiles tend to decrease as we take increasingly larger $R_{\textrm{lim}}$ , which is a result of the increasing field contamination.

\subsection{Membership determination}
\label{memberships}

Photometric and astrometric data from the 2MASS and GAIA DR2 catalogues, respectively, were extracted for stars in circular regions, centred on each cluster, with radius $r>5\times R_{\textrm{lim}}$. Then we cross-matched the information from both catalogues in these areas and executed a routine that evaluates the clustering of stars in each part of the three-dimensional (3D) astrometric space of proper motions and parallaxes ($\mu_{\alpha}, \mu_\delta, \varpi$), assigning membership likelihoods. The main assumption is to consider that cluster members must be more tightly distributed than field stars because stars within a cluster are expected to be located at almost the same distance and to share the same global movement, regardless of their colours and luminosities \citep{Cantat-Gaudin:2018}.

To ensure the quality of the astrometric information, we applied cuts to the GAIA DR2 data following equations 1 and 2 of \cite{Arenou:2018}. Only stars restricted to these relations were employed in the astrometric decontamination procedure. Additionally, we restricted our sample to  stars with $J\leq15.8\,$mag, $H\leq15.1\,$mag, and $K_s\leq14.3\,$mag. These limits correspond to the 99\% completeness limit of the 2MASS catalogue.
The procedure follows four steps, as described below.

\section*{Step 1}
We divide the astrometric space in cells with widths proportional to the sample mean uncertainties ($\langle\Delta\mu_{\alpha}\rangle, \langle\Delta\mu_{\delta}\rangle, \langle\Delta\varpi\rangle$): cell widths are $\sim10\times\langle\Delta\mu_{\alpha}\rangle, \sim10\times\langle\Delta\mu_{\delta}\rangle$ and $\sim1\times\langle\Delta\varpi\rangle$. These values translate into $\sim$1.0\,mas.yr$^{-1}$ and $\sim$0.1\,mas for proper motion components and parallaxes, respectively. These cell dimensions are large enough to accommodate a significant number of stars, but small enough to detect local fluctuations in the stellar density through the whole data domain. For all clusters, the typical number of cells employed in the decontamination procedure was greater than 100. The search for member stars was restricted to the limiting radius of each cluster. Stars for a comparison field were selected in a large external ring with an internal radius $r>3\times R_{\textrm{lim}}$.

\section*{Step 2}
In the second step, the algorithm verifies whether the group of stars contained within a given 3D cell can be statistically distinghuished from a sample of field stars in the same data domain. To perform this verification, we followed a procedure analogous to that adopted by \cite{Dias:2012} and \cite{Dias:2018}, and also employed by \cite{Angelo:2017}, in order to establish membership likelihoods. 

For each star selected in the cluster sample contained within a given 3D cell, a membership likelihood ($l_{\textrm{star}}$) is computed according to the relation:

\begin{equation}
\begin{aligned}
        l_{\textrm{star}} = \frac{\textrm{exp}\left[-\frac{1}{2}(\boldsymbol{X}-\boldsymbol{\mu})^{\textrm{T}}\boldsymbol{\sum}^{-1}(\boldsymbol{X}-\boldsymbol{\mu})\right]}{\sqrt{(2\pi)^3\vert\boldsymbol{\sum}\vert}}                                   
\end{aligned}   
\label{likelihood_formula}
,\end{equation}

\noindent
where we employed the same notation as \cite{Dias:2018}: $\boldsymbol{X}$ is the column vector ($\mu_{\alpha}\textrm{cos}\,\delta$,\,$\mu_{\delta}$,\,$\varpi$) containing the astrometric information for a given star; $\sigma_{\mu_{\alpha}}$, $\sigma_{\mu_{\delta}}$ and $\sigma_{\varpi}$ are calculated through the quadrature sum of the individual errors with the intrinsic dispersion of each parameter for the sample of cluster stars; $\boldsymbol{\mu}$ is the mean column vector for the group of cluster stars \textit{\textup{within the cell}}, that is, these means are defined locally\textup{} in the parameter space. Consequently, this calculation takes into account the three astrometric parameters simultaneously and provides a combined membership likelihood, considering the measurement uncertainties. $\boldsymbol{\sum}$ is the full covariance matrix, which incorporates the uncertainties ($\sigma$) and their correlations ($\rho$) according to the equation below:

\begin{equation}
\boldsymbol{\sum} = 
 \begin{bmatrix}
  \sigma_{\mu_{\alpha}}^2   & \sigma_{\mu_{\alpha}}\sigma_{\mu_{\delta}}\rho_{\mu_{\alpha}\mu_{\delta}} & \sigma_{\mu_{\alpha}}\sigma_{\varpi}\rho_{\mu_{\alpha}\varpi} \\
  
  \sigma_{\mu_{\alpha}}\sigma_{\mu_{\delta}}\rho_{\mu_{\alpha}\mu_{\delta}} & \sigma_{\mu_{\delta}}^2   & \sigma_{\mu_{\delta}}\sigma_{\varpi}\rho_{\mu_{\delta}\varpi} \\
  \sigma_{\mu_{\alpha}}\sigma_{\varpi}\rho_{\mu_{\alpha}\varpi} & \sigma_{\mu_{\delta}}\sigma_{\varpi}\rho_{\mu_{\delta}\varpi} & \sigma_{\varpi}^2
 \end{bmatrix}
.\end{equation}

The same calculation was then performed for the group of field stars within the same cell in the parameter space, keeping the dispersions as defined above, that is, relative to the mean values for the group of cluster stars. The total likelihood for a group of stars  is given by $\mathcal{L}=\prod_{i}^{} l_i$, from which we defined the objective function

\begin{equation}
   S = -\textrm{log}\,\mathcal{L}
   \label{func_entropia}
.\end{equation}

\noindent
We then verified the degree of similarity between the two groups (cluster and field stars). Cluster stars within cells for which $S_{\textrm{clu}}<S_{\textrm{fld}}$ were flagged (``1") as possible cluster members (non-member stars according to this criterion were flagged as ``0"). The $S$ function could be interpreted as analogous to the entropy of this parameter space, representing the degree of spreadedness of the data. Thus, lower $S$ values indicate more clustered data.


\normalsize

\section*{Step 3}
The third step of the procedure consists of obtaining the final membership likelihood for each star ($L_{\textrm{star}}$). We took cells that contained stars flagged as ``1" (i.e. member candidates), as described in step 2 above, and evaluated an exponential factor of the form

\begin{equation}
   L_{\textrm{star}} \propto\,\textrm{exp}\left(-\frac{\langle N_{\textrm{clu}}\rangle}{N_{\textrm{clu}}}\right),     
   \label{termo_exponencial} 
\end{equation}

\noindent
where $\langle N_{\textrm{clu}}\rangle$ is the weighted average number of stars taken over the whole set of cells for a grid configuration. This exponential factor ensures that only stars contained in cells with $N_{\textrm{clu}}$ considerably greater than $\langle N_{\textrm{clu}}\rangle$ will receive appreciable membership likelihoods, thus reducing the presence of outliers (i.e. small groups of field stars with non-zero membership likelihoods) in the final list of member stars.

\section*{Step 4}

Cell sizes are then increased and decreased by one-third of their mean sizes in each of the three axes and the complete procedure (steps 1 to 3) is repeated. Therefore we have a total of 27 different grid configurations. Finally, the algorithm registers the median of the 27 likelihood values (equation \ref{termo_exponencial}) computed for a given star.  The proportionality constant is obtained by taking the complete sample of cluster stars and normalising the maximum value of $L_{\textrm{star}}$ to 1.

The procedure outlined in this section helps in the search for significant local overdensities in the ($\mu_{\alpha}, \mu_\delta, \varpi$) space that are statistically distinguishable from the distribution of field stars, taking uncertainties into account. For verification purposes, we employed this method to investigate two well-known OCs: NGC\,752 and NGC\,188.

\subsection{Method verification}
\label{method_verification}

\begin{figure*}
\begin{center}
\parbox[c]{0.8\textwidth}
  {
    
    \includegraphics[width=0.40\textwidth]{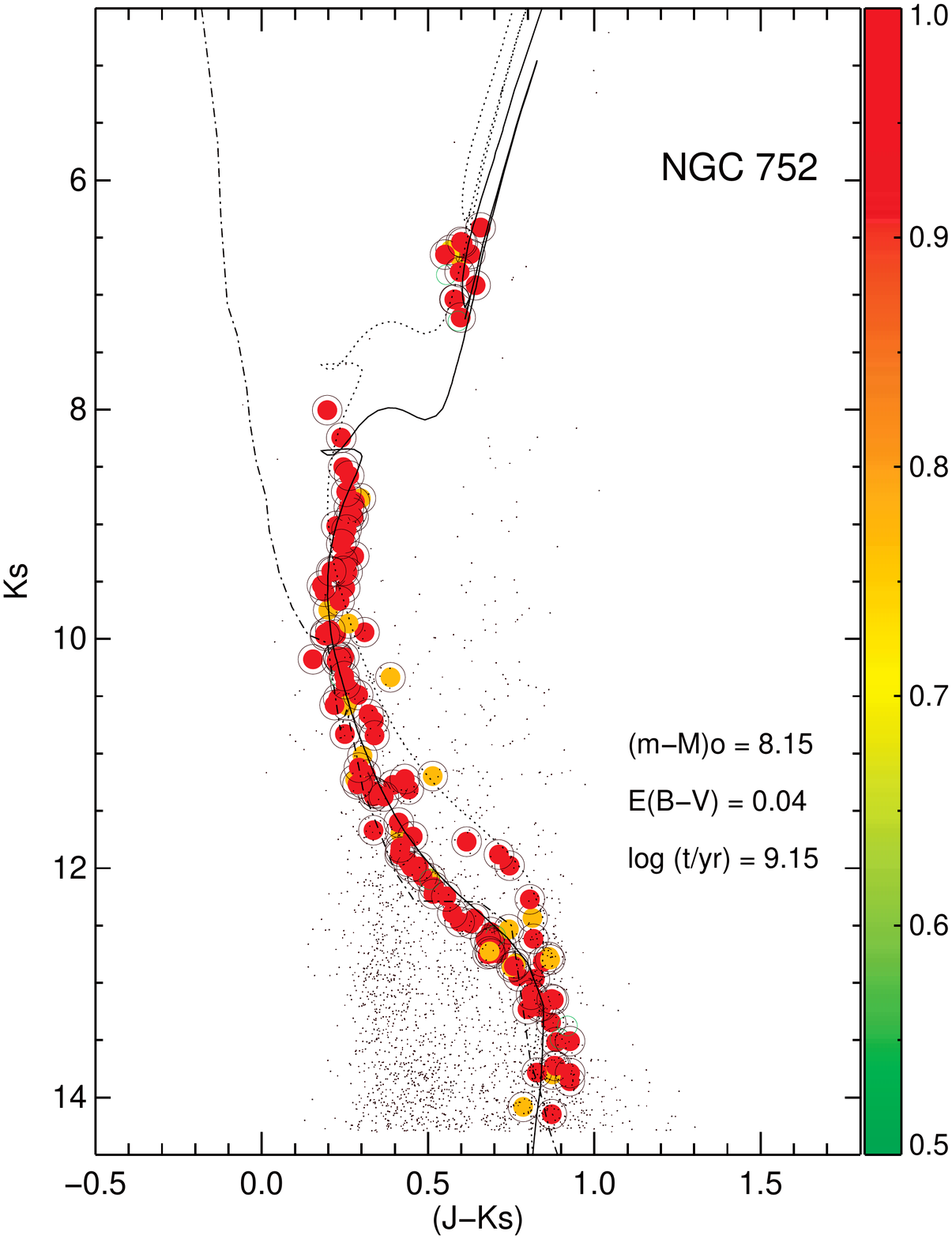}
    \includegraphics[width=0.40\textwidth]{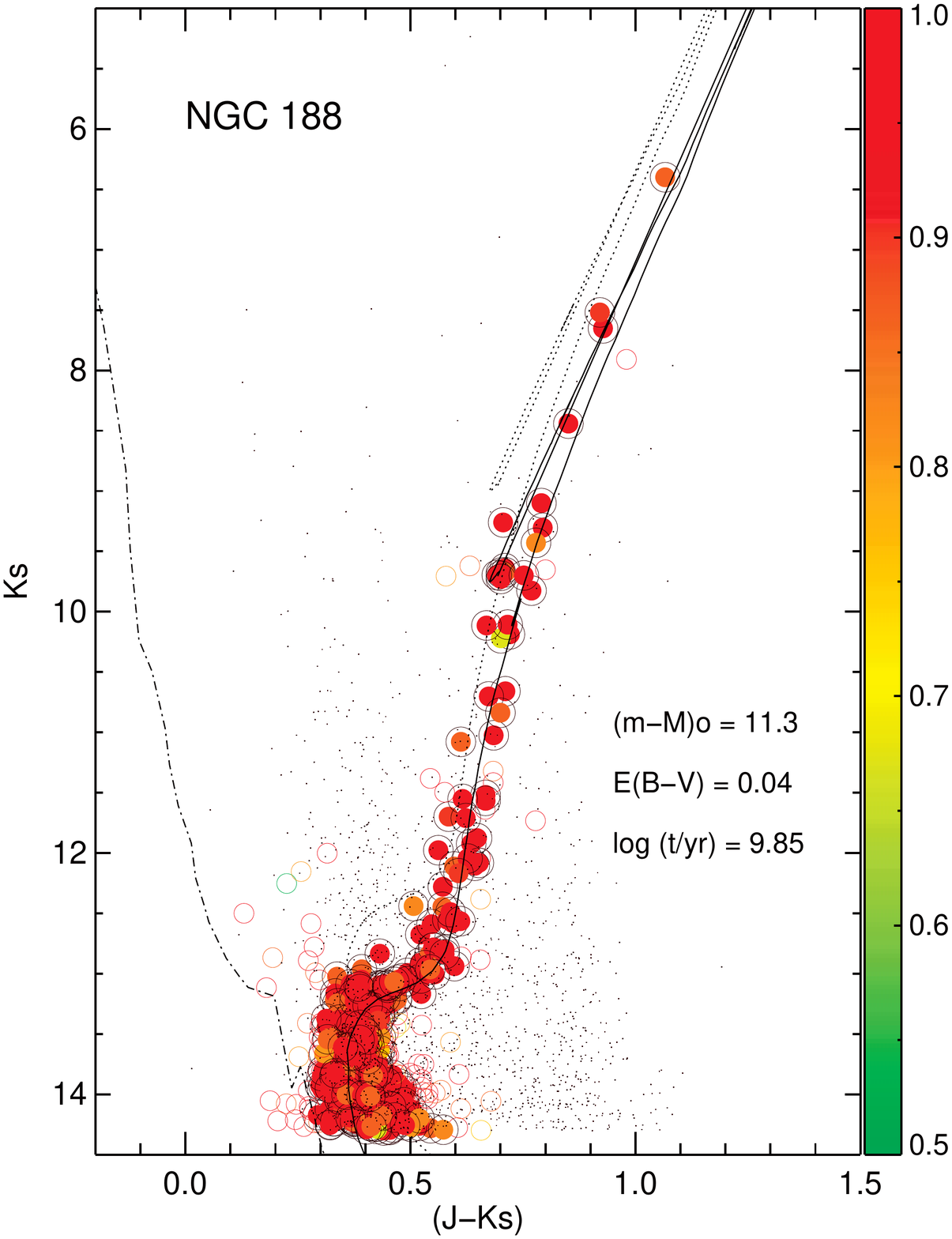}     
  }

\caption{ $K_\textrm{s}\times(J-K_\textrm{s})$ CMDs of NGC\,752 (left panel) and NGC\,188 (right panel) after the decontamination procedure outlined in this section. The colour bar indicates membership likelihoods according to eq. \ref{termo_exponencial}. Filled (open) symbols represent (non-) member stars, according to our method, and small black dots are stars in a comparison field. Circled filled symbols are stars with \textit{\textup{photometric}} membership likelihoods $L_{\textrm{phot}}\ge0.1$, according to the Maia et al. (2010) decontamination algorithm. The continuous lines are PARSEC isochrones fitted to the data of both clusters, and the dotted lines represent the locus of unresolved binaries of equal-mass components. Dot-dashed lines represent the empirical main sequence of \cite{Koornneef:1983} converted into 2MASS according to the relations of \cite{Carpenter:2001}. Fundamental astrophysical parameters are indicated (see also Table \ref{info_sample_OCs_OCRs}). }
\label{Ks_JKs_decontaminated_NGC752_NGC188}
\end{center}
\end{figure*}

\begin{figure*}
\begin{center}
\parbox[c]{0.9\textwidth}
  {
   
        \includegraphics[width=0.45\textwidth]{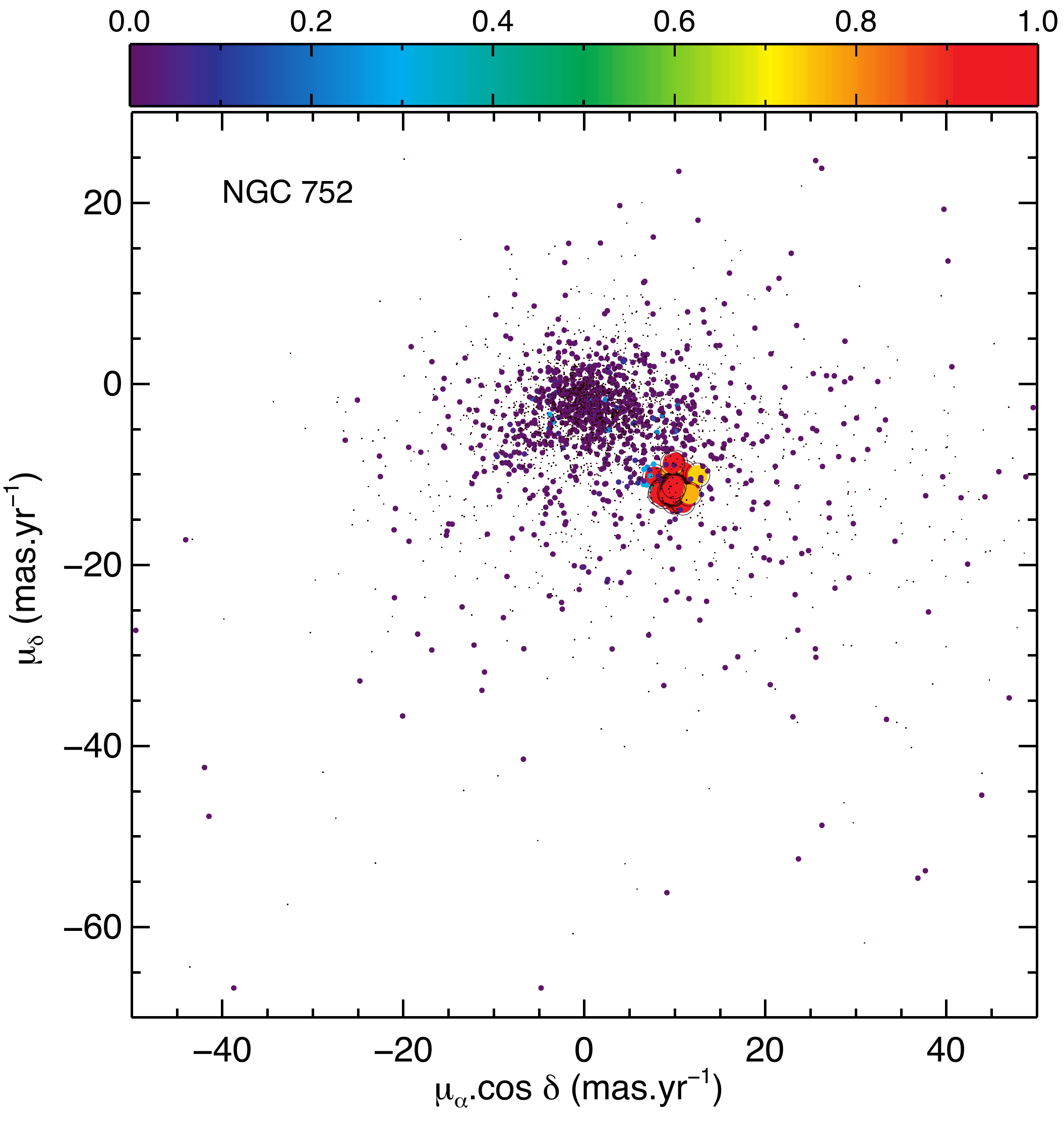}
        \includegraphics[width=0.45\textwidth]{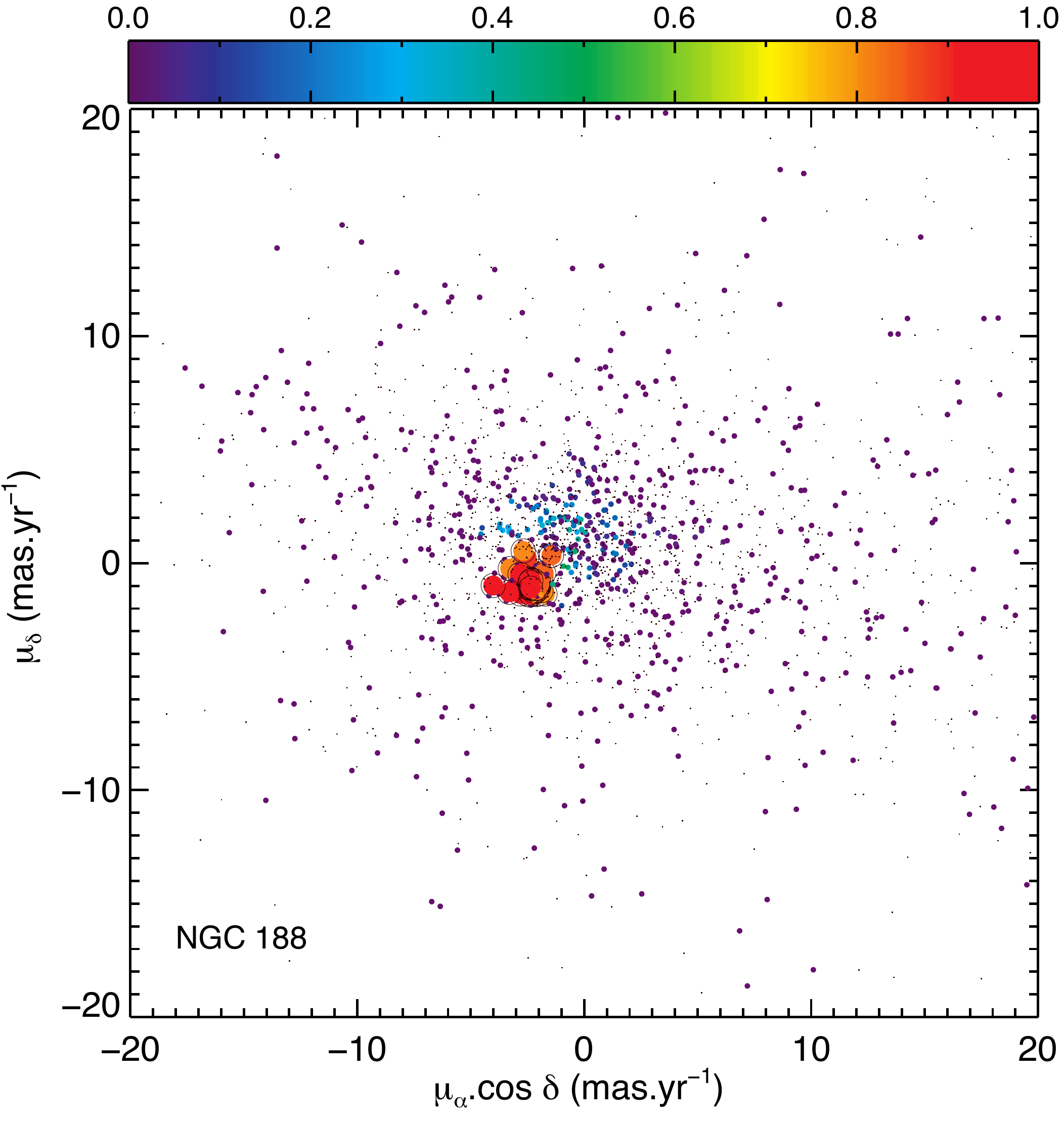}
        \includegraphics[width=0.45\textwidth]{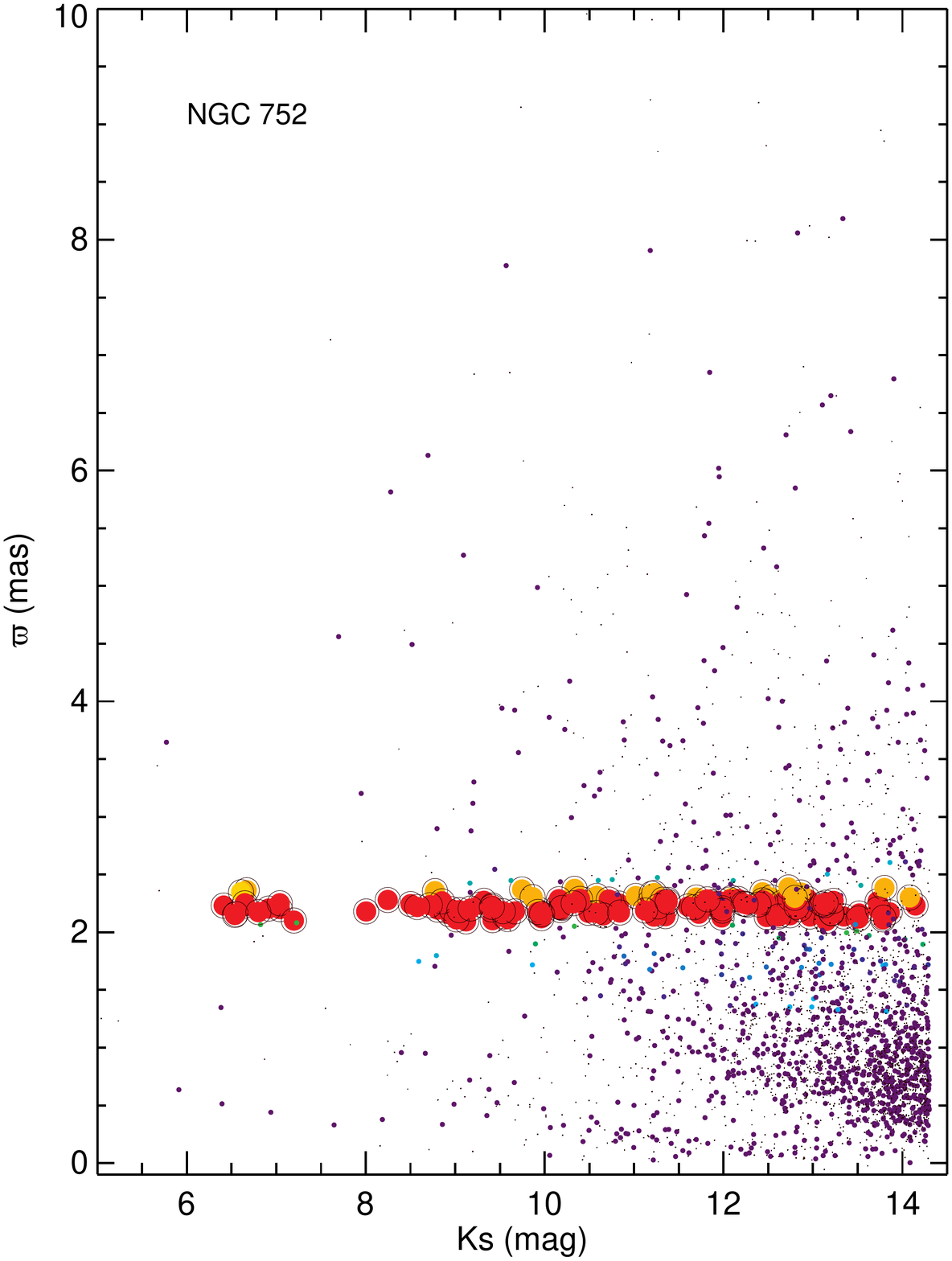}
        \includegraphics[width=0.45\textwidth]{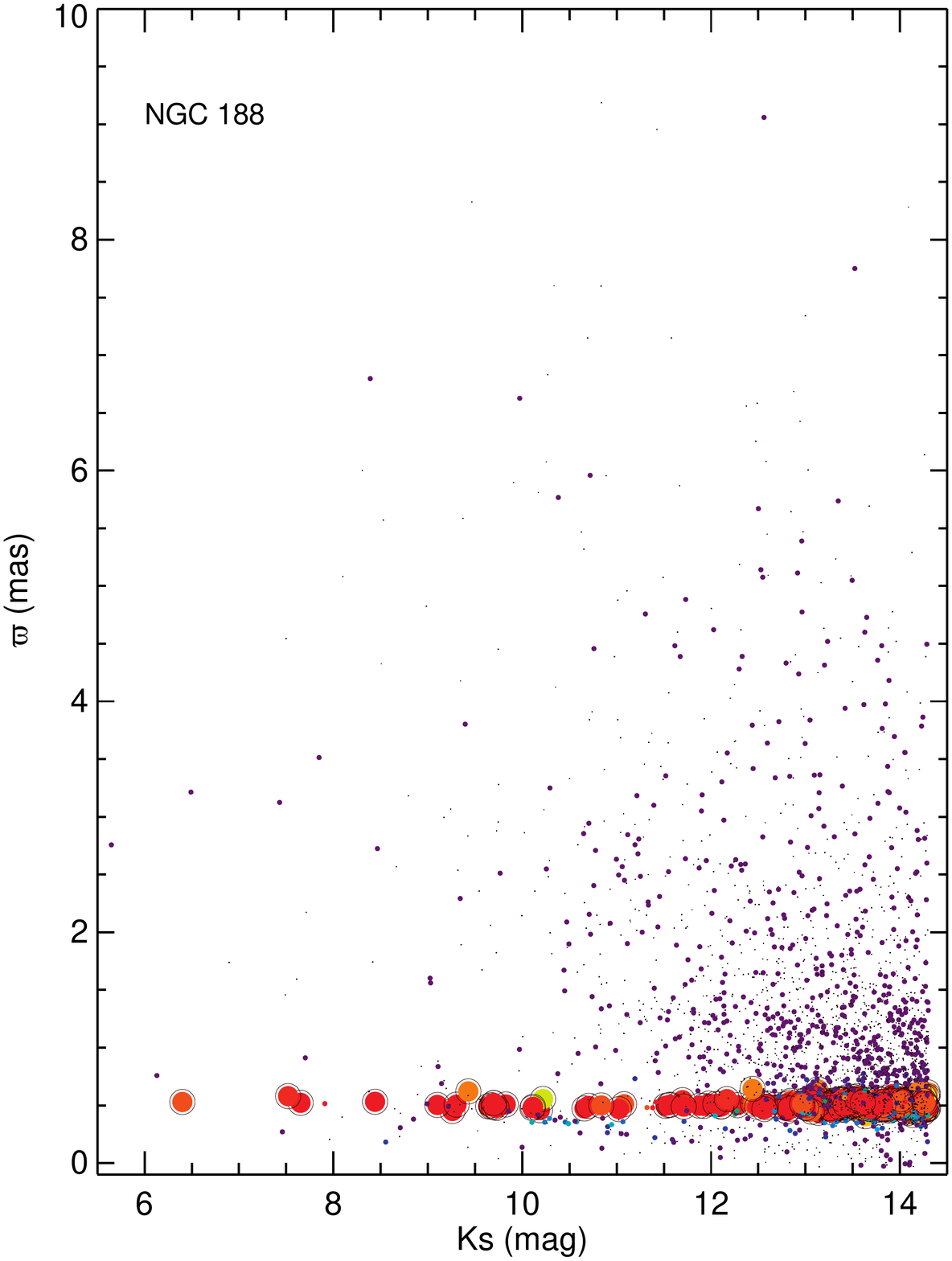}
        
  }
\caption{ Top panels: VPD for stars in the inner areas ($r\leq R_{\textrm{lim}}$) of NGC\,752 (left panel) and NGC\,188 (right) panel. Stars in the comparison field are plotted as small black dots. Symbols follow the same scheme as in Figure \ref{Ks_JKs_decontaminated_NGC752_NGC188}. The colour bars indicate membership likelihoods. Bottom panels: $\varpi\times K_\textrm{s}$ plots for NGC\,752 (left panel) and NGC\,188 (right panel). }

\label{VPDs_plx_Ks_NGC752_NGC188}
\end{center}
\end{figure*}

The decontaminated CMDs are shown in Figure \ref{Ks_JKs_decontaminated_NGC752_NGC188} for the OCs NGC\,752 (left panel) and NGC\,188 (right panel). We fitted a PARSEC isochrone (\citeauthor{Bressan:2012}\,\,\citeyear{Bressan:2012}; $[Fe/H]=0.0$ for NGC\,188 and $[Fe/H]=0.12$ for NGC\,752; see Table 1) to the more probable members $L_{\textrm{star}}\gtrsim0.8$ (equation \ref{termo_exponencial}) in order to derive fundamental astrophysical parameters. Groups of stars with high $L$ values significantly contrast with the field and provide useful constraints for isochrone fitting, although some interlopers remain.

As a first guess for the distance modulus, we employed $(m-M)_{0}$=5\,$\times$\,log(100/$\langle\varpi\rangle$), where $\langle\varpi(\textrm{mas})\rangle$ is the mean parallax for the high-probability stars. Each isochrone was then reddened and vertically shifted using the extinction relations \citep{Rieke:1985} $A_{K_\textrm{S}}=0.112\,A_V$, $A_{(J-K_S)}=0.17\,A_V$, and $A_V=3.09\,E(B-V)$ until the observed CMD features were matched. Ages were estimated by fitting the brightest stars close to the turnoff point, subgiant and giant stars. 


Member stars are plotted in Figure \ref{Ks_JKs_decontaminated_NGC752_NGC188} with filled symbols. Additionally, we ran a \textit{\textup{photometric}} decontamination procedure (fully described in \citeauthor{Maia:2010}\,\,\citeyear{Maia:2010}) using the same set of data as before, in order to check for possible member stars without astrometric information. For reference, the circled filled symbols in Figure \ref{Ks_JKs_decontaminated_NGC752_NGC188} represent members with \textit{\textup{photometric}} membership likelihoods $L_{\textrm{phot}}\ge0.1$. The same holds for the other CMDs presented in this paper (Figure \ref{Ks_JKs_clusters_decontaminated_parte1}). Open circles are non-member stars and small black dots represent stars in a comparison field. 

The top plots in Figure \ref{VPDs_plx_Ks_NGC752_NGC188} show the vector-point diagrams (VPDs) of NGC\,752 (left panel) and  NGC\,188 (right panel); the plots $\varpi\times K_{\textrm{s}}$ are shown in the bottom panels for the same OCs. Member stars (larger filled circles) form conspicuous clumps in the VPDs, present similar parallaxes, and define recognisable sequences in these OCs CMDs, as expected.

\section{Results}
\label{results}

\begin{figure*}
\begin{center}
\parbox[c]{1.0\textwidth}
  {

    \includegraphics[width=0.33333\textwidth]{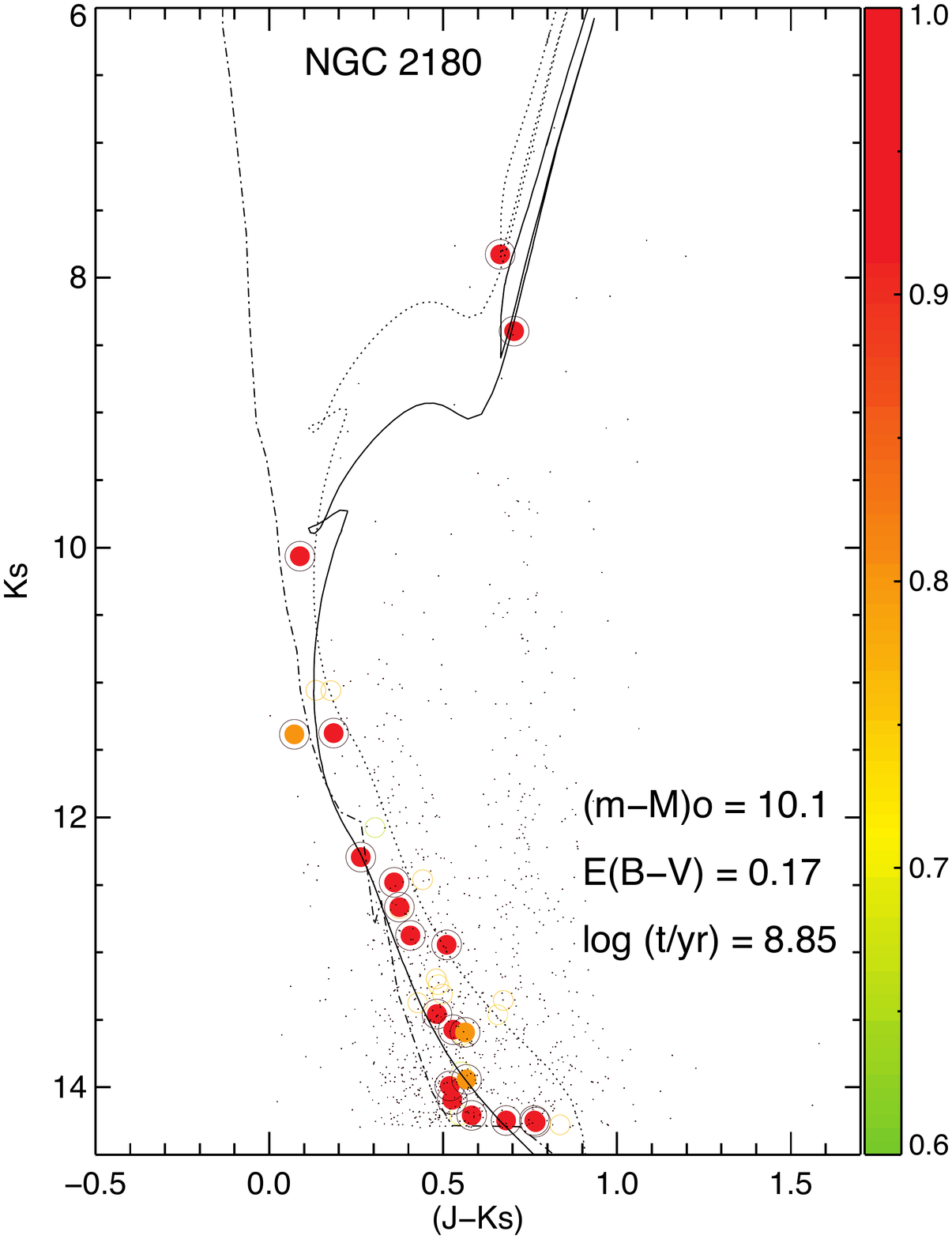}
    \includegraphics[width=0.33333\textwidth]{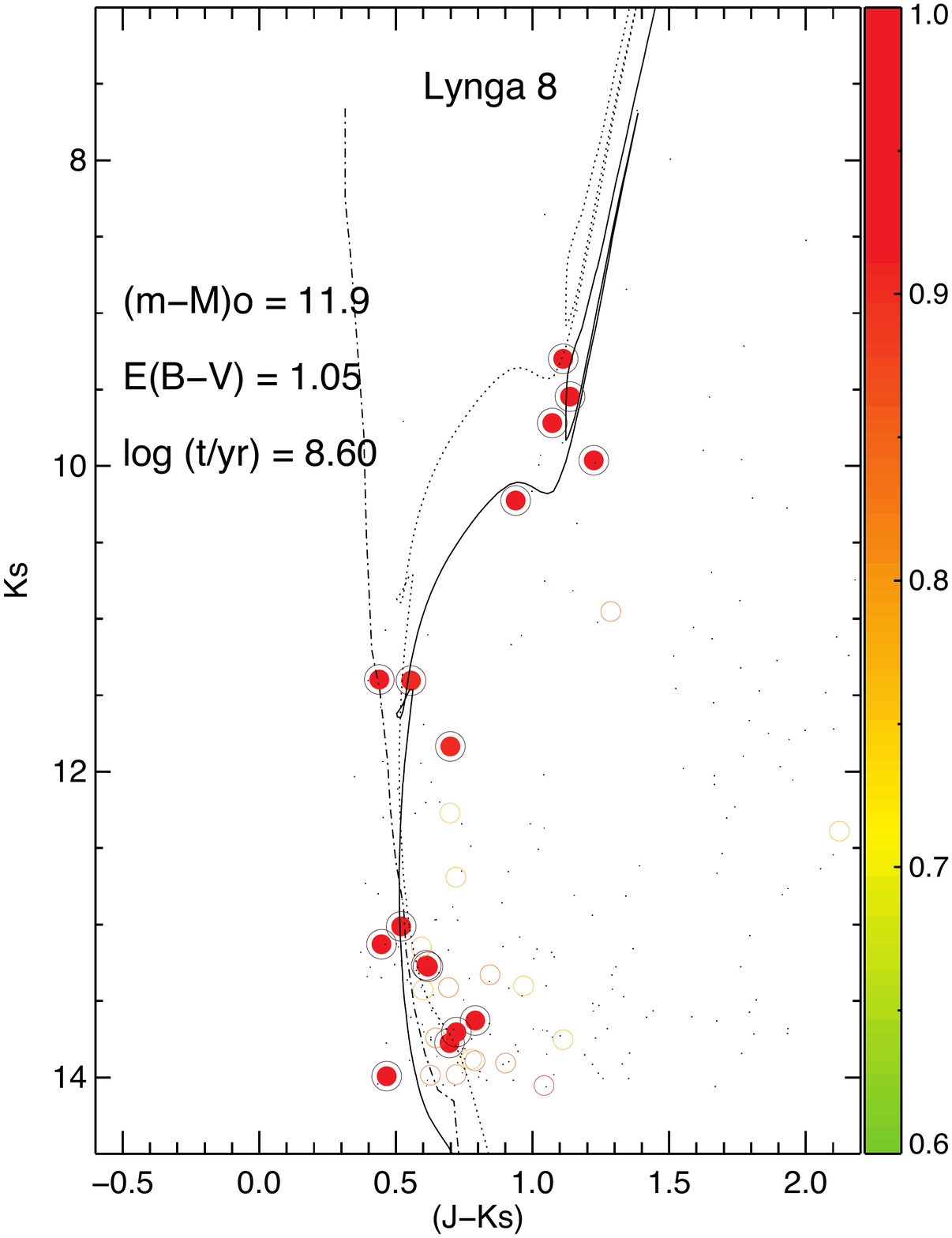}
    \includegraphics[width=0.33333\textwidth]{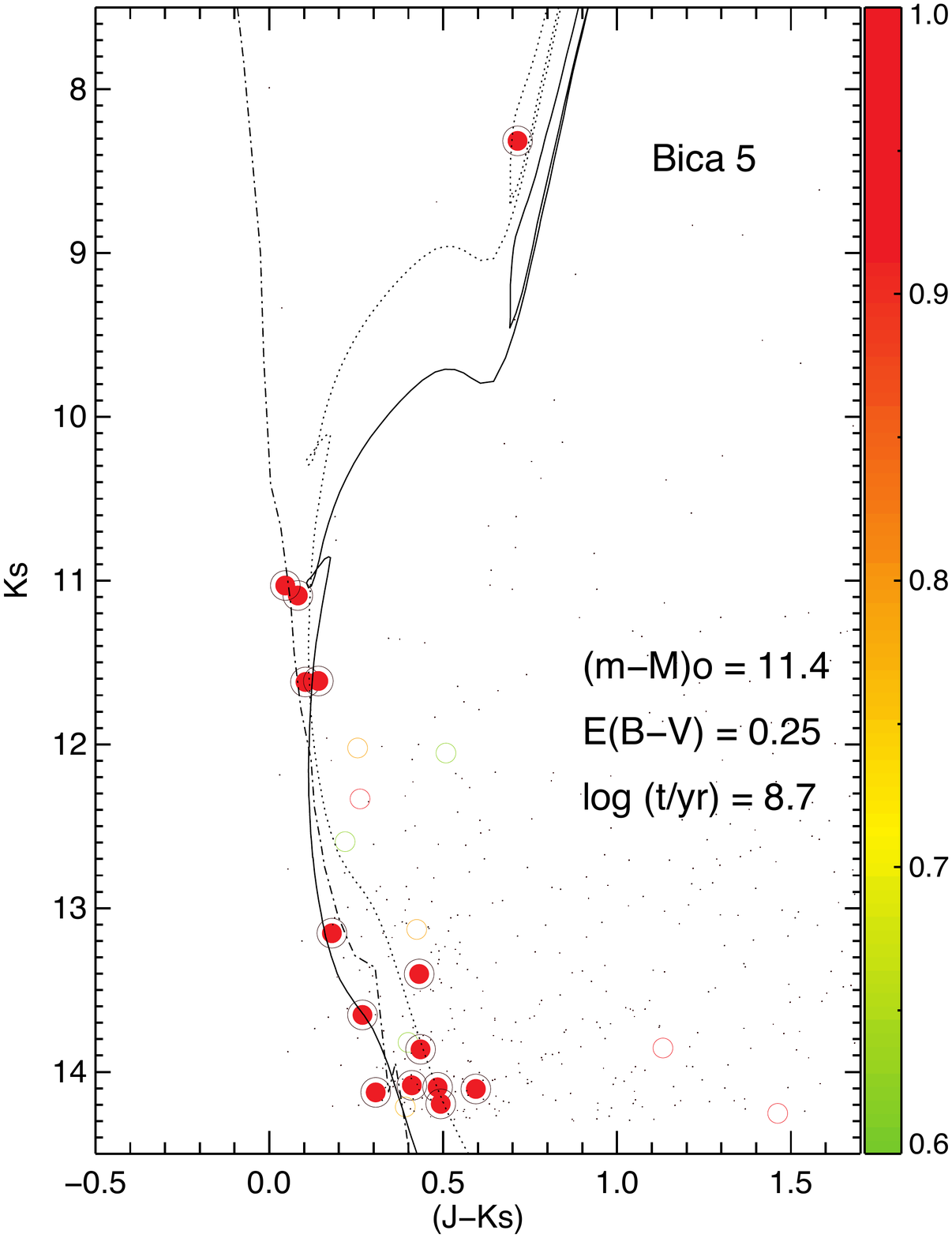}
    \includegraphics[width=0.33333\textwidth]{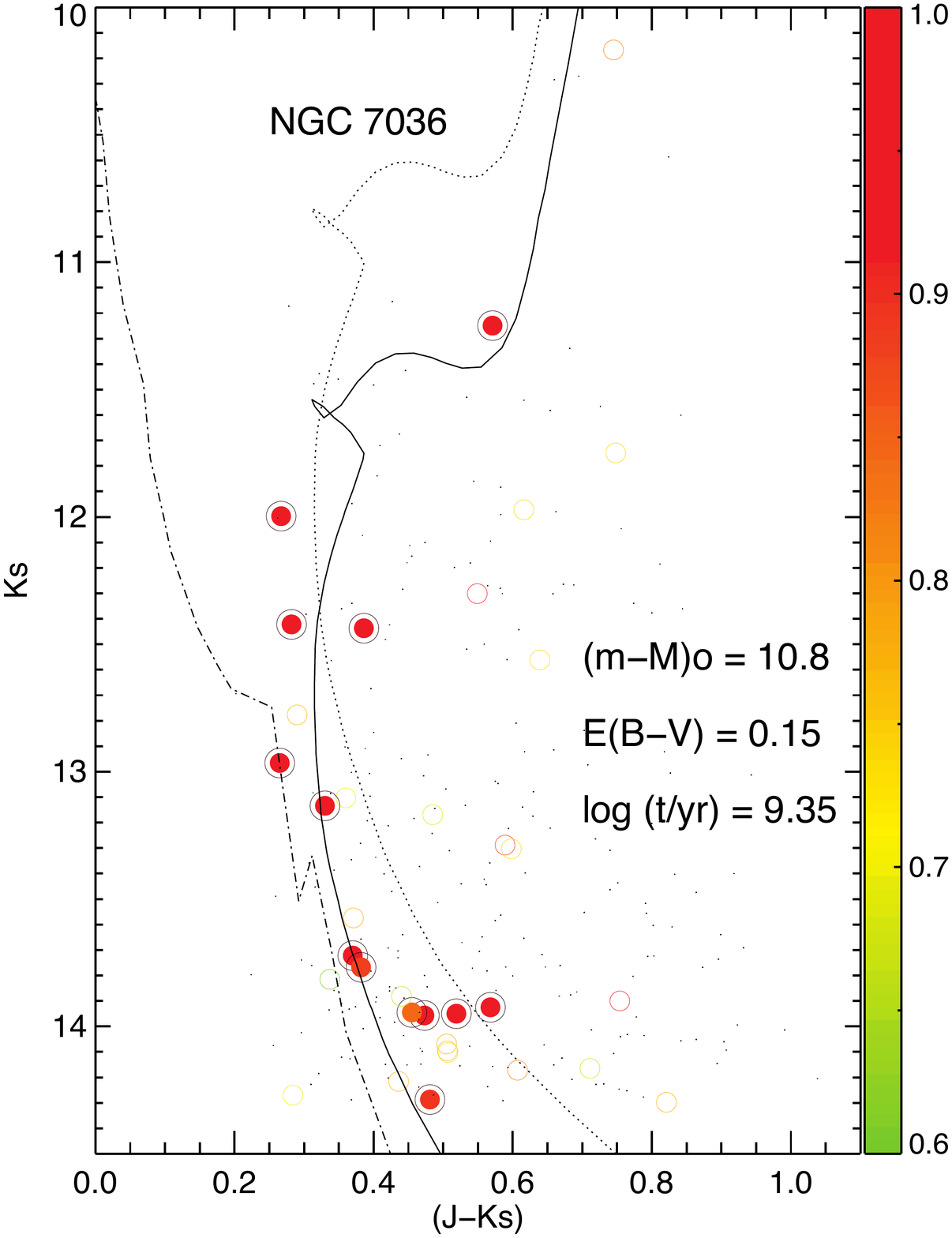}
    \includegraphics[width=0.33333\textwidth]{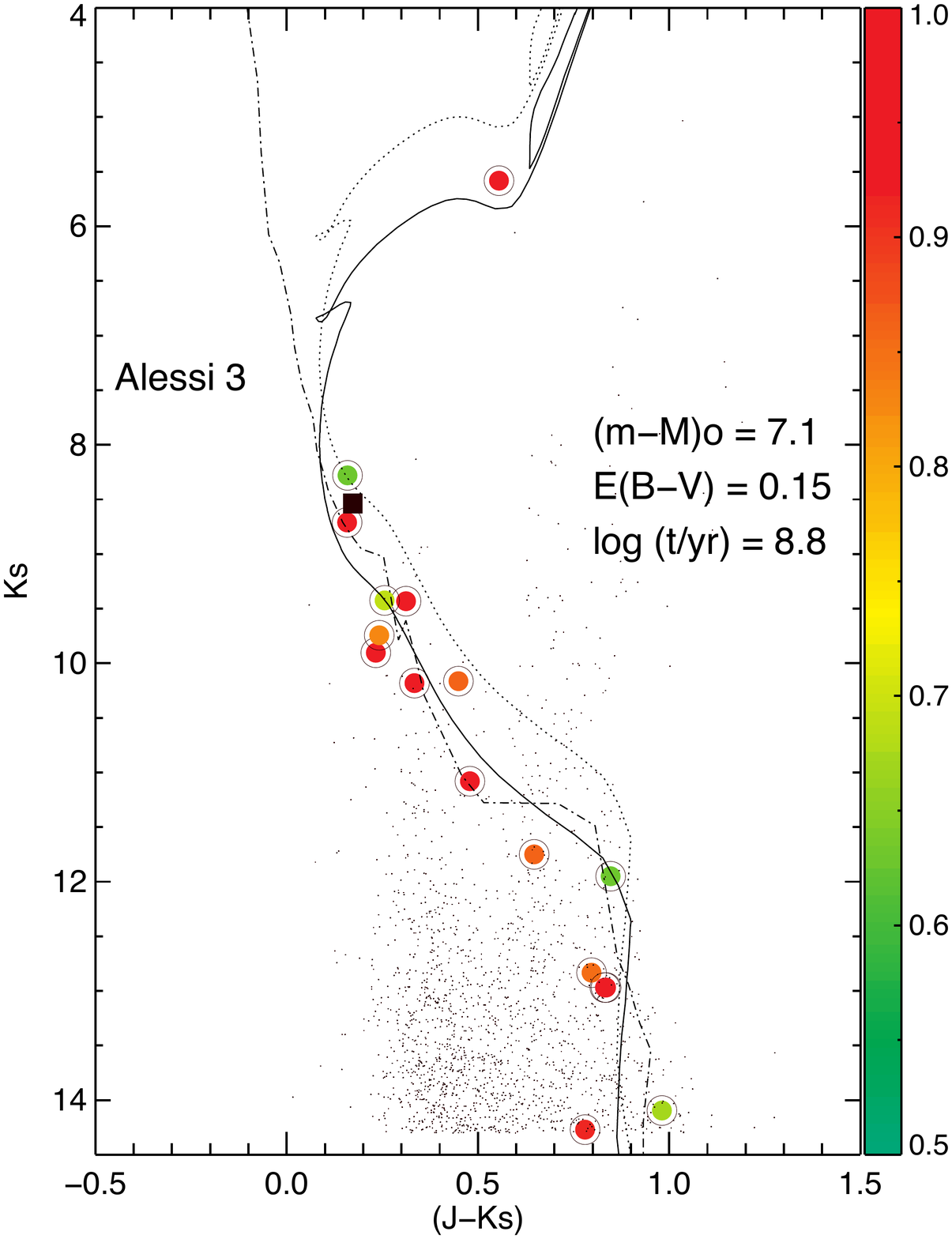}
    \includegraphics[width=0.33333\textwidth]{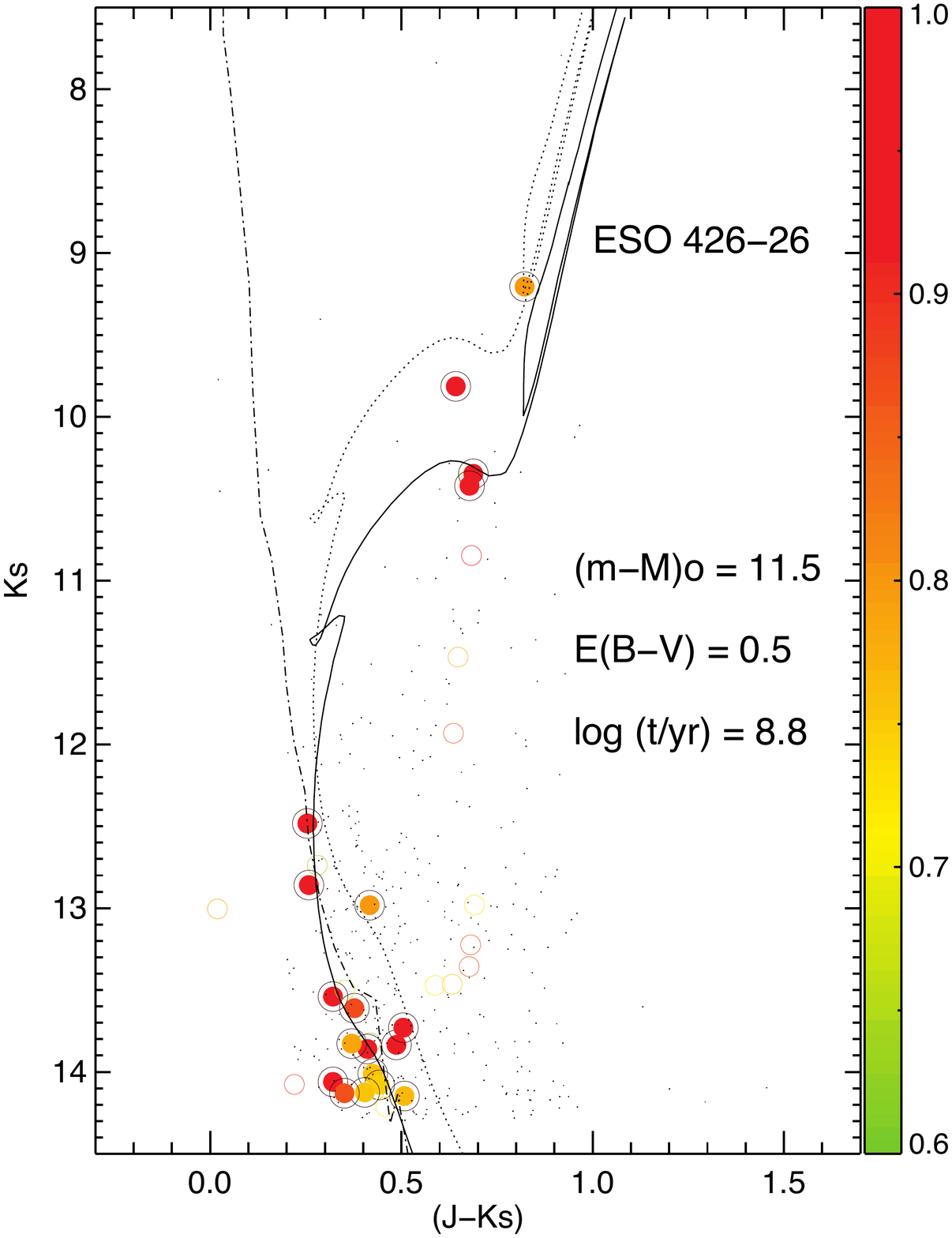}
  }

\caption{ Same as Figure \ref{Ks_JKs_decontaminated_NGC752_NGC188}, but for NGC\,2180, Lynga\,8, Bica\,5, NGC\,7036, Alessi\,3, and ESO\,426-SC26. The black filled rectangle in the CMD of Alessi\,3 represents a star with no astrometric data in GAIA DR2, but with \textit{\textup{photometric}} membership likelihood (see Section \ref{method_verification}) greater than 50\%. }
\label{Ks_JKs_clusters_decontaminated_parte1}
\end{center}
\end{figure*}

\begin{figure*}
\begin{center}
\parbox[c]{1.0\textwidth}
  {
   
        \includegraphics[width=0.33333\textwidth]{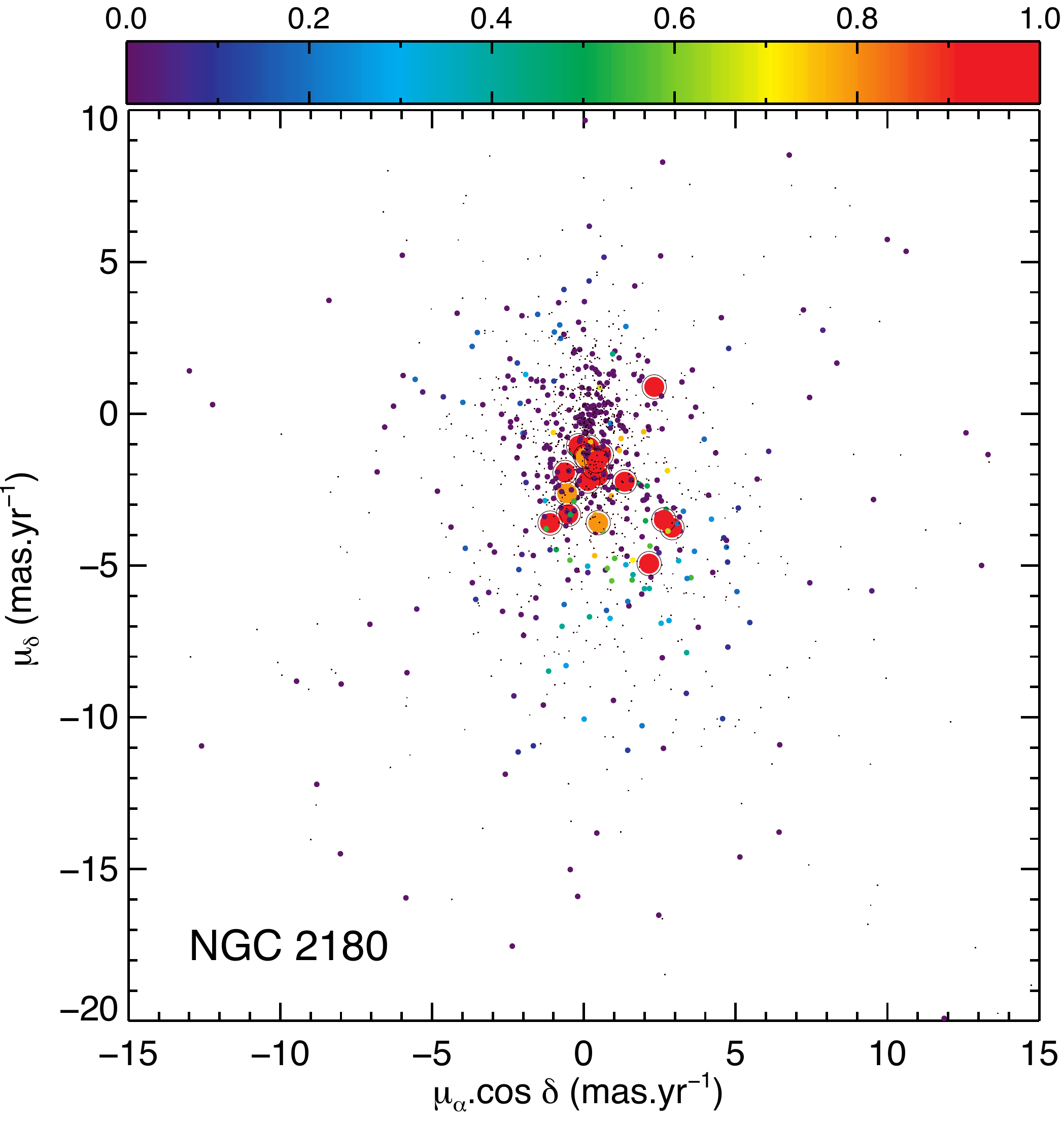}
        \includegraphics[width=0.33333\textwidth]{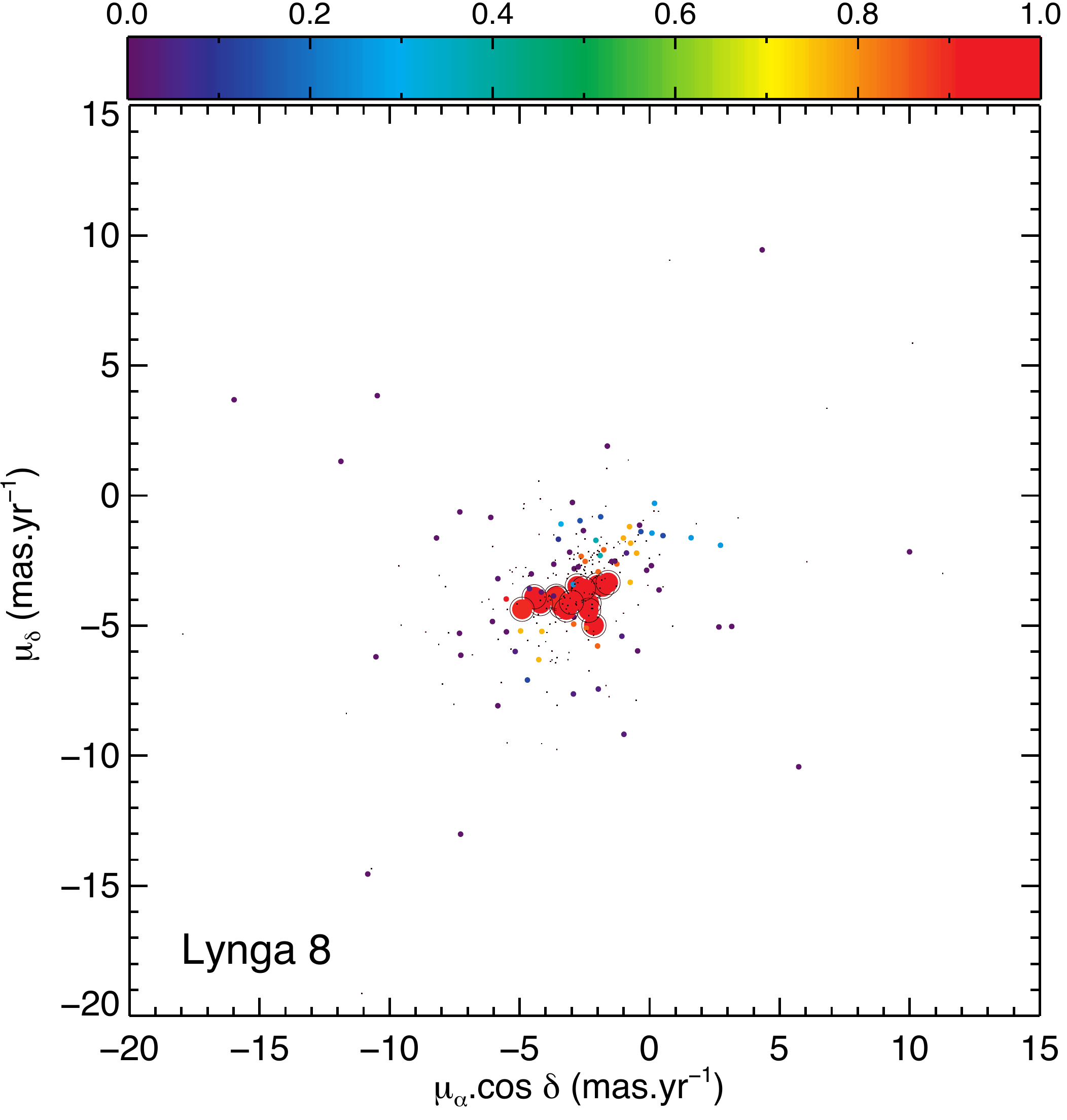}
        \includegraphics[width=0.33333\textwidth]{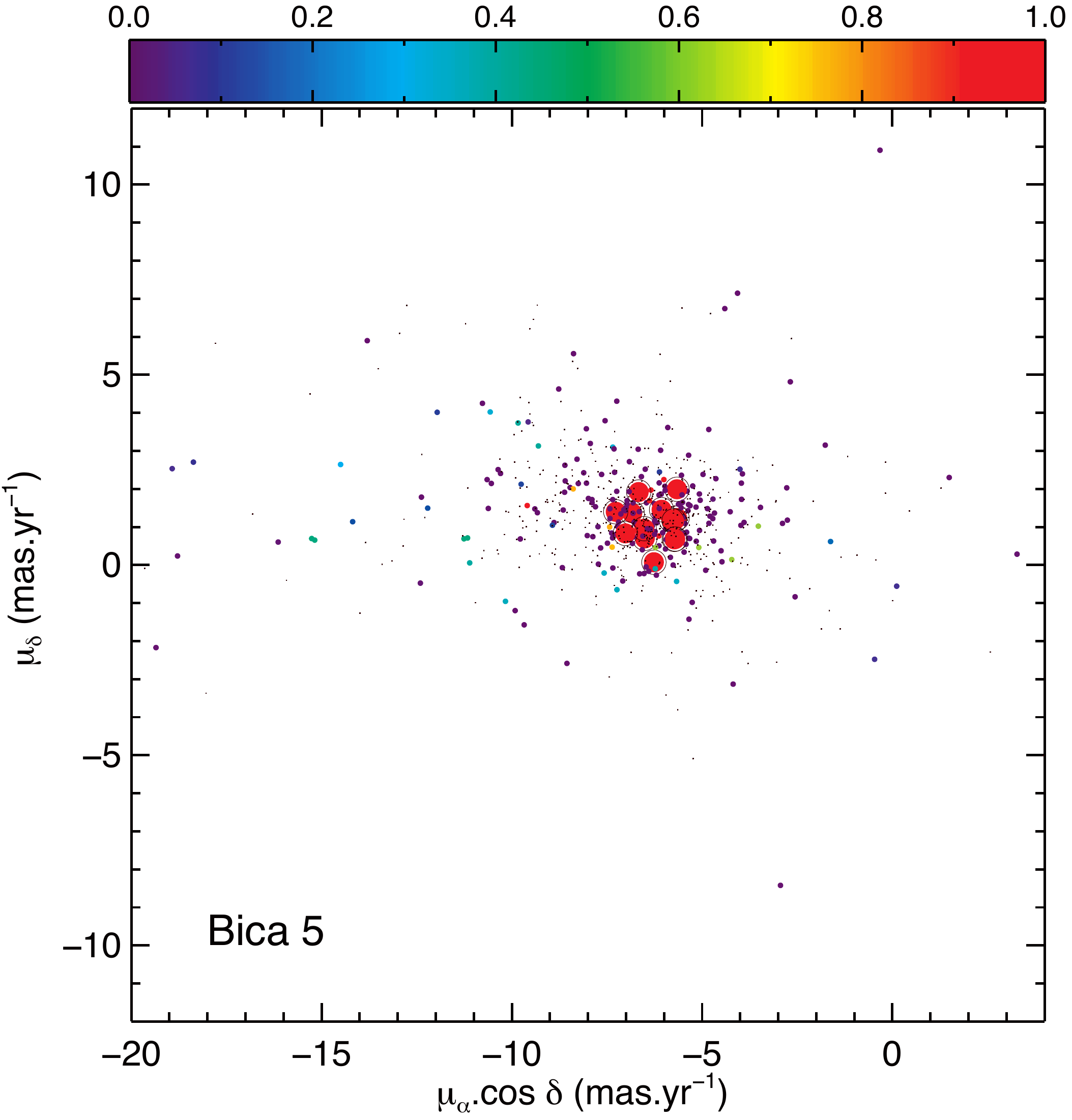}

        \includegraphics[width=0.33333\textwidth]{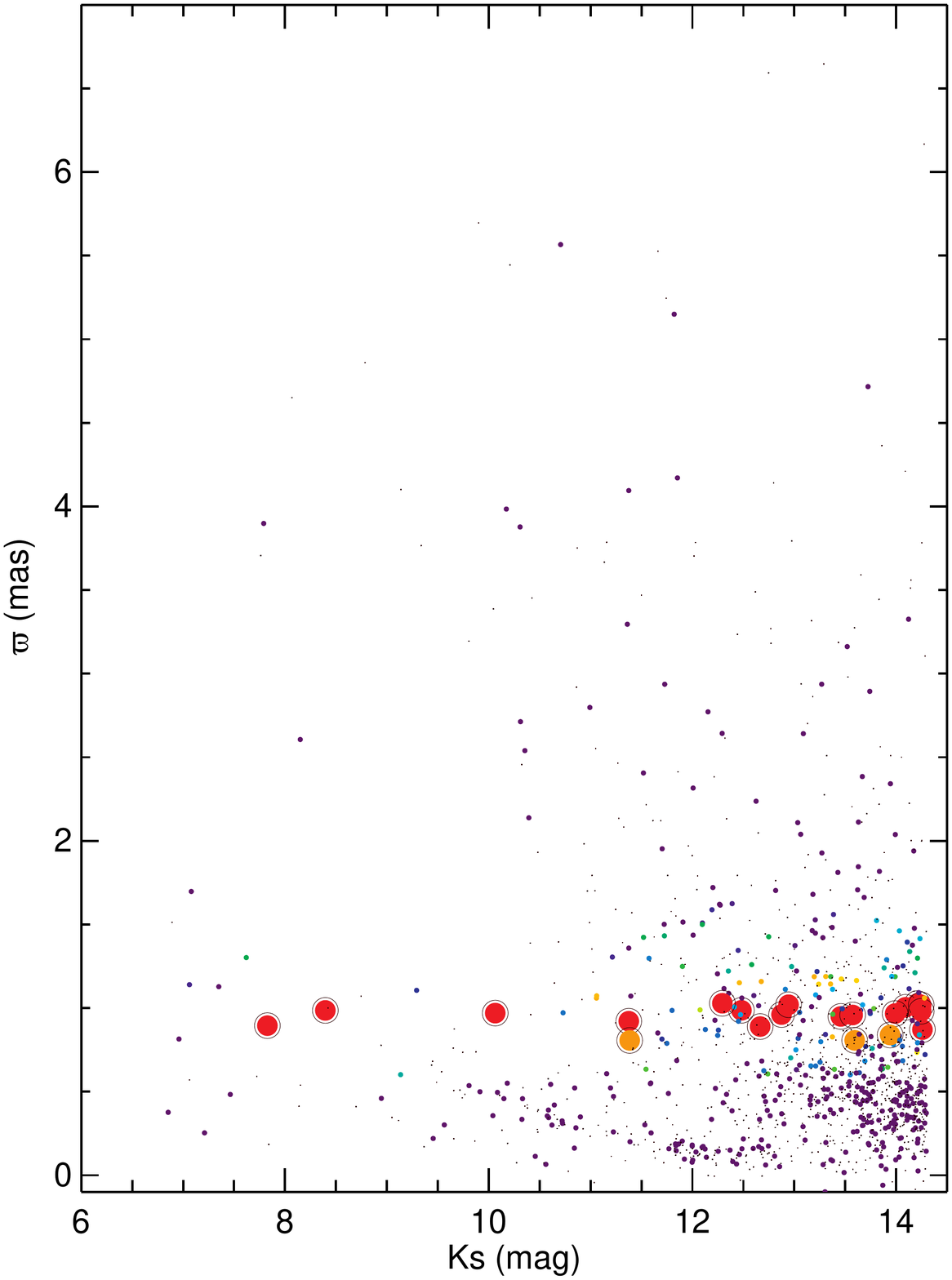}
        \includegraphics[width=0.33333\textwidth]{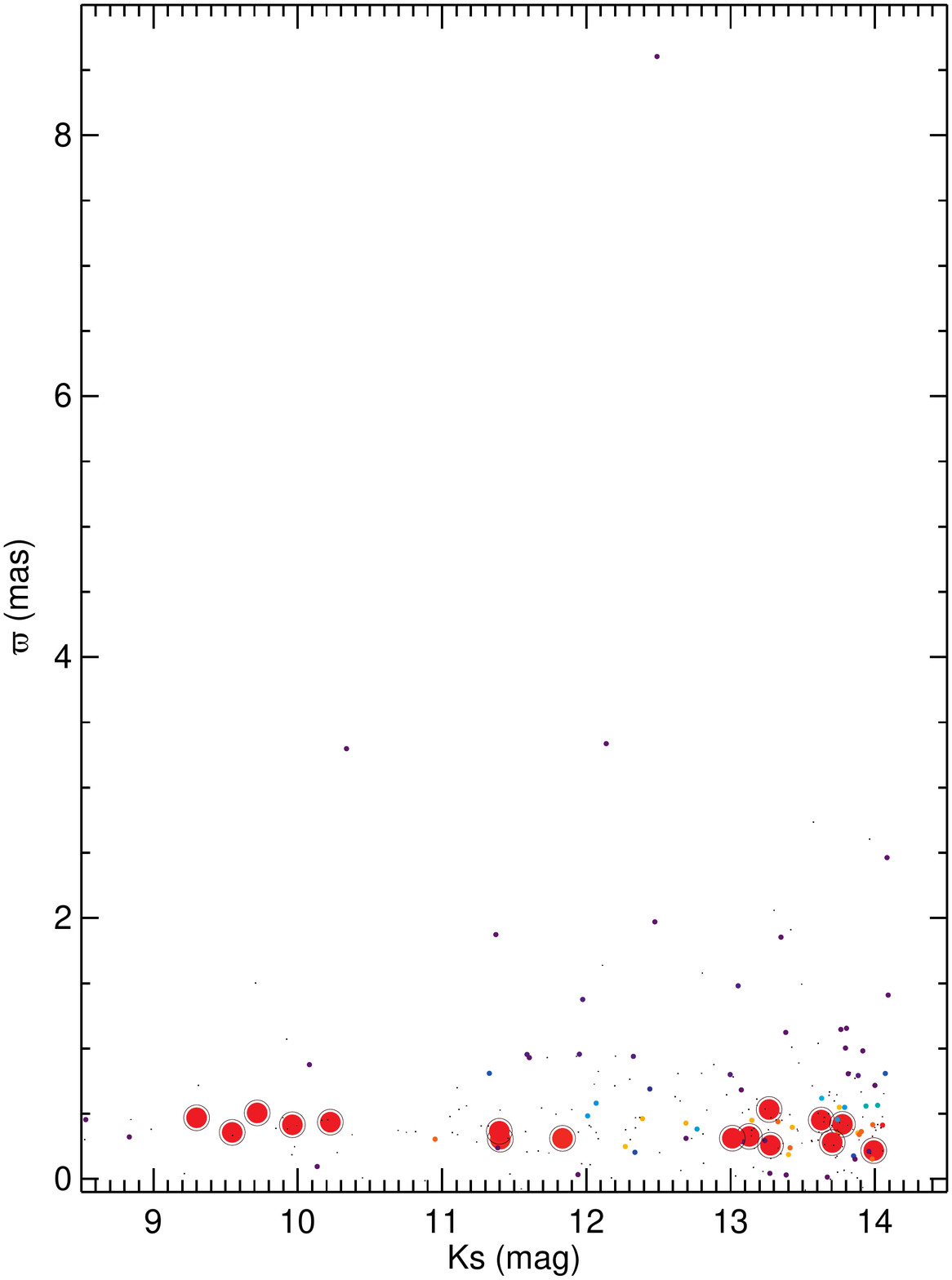}
        \includegraphics[width=0.33333\textwidth]{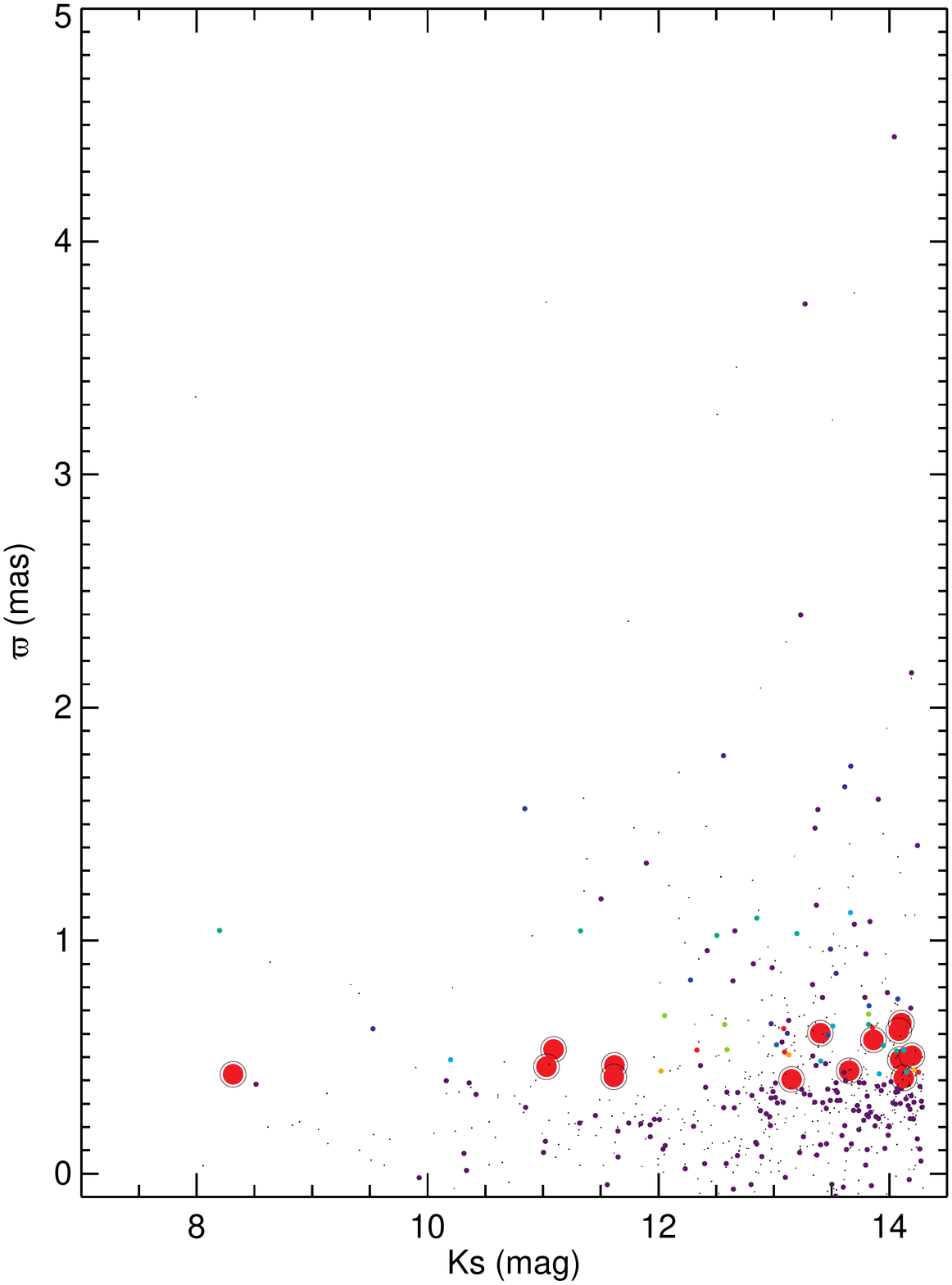}
        
  }
\caption{ Same as Figure \ref{VPDs_plx_Ks_NGC752_NGC188}, but for the OCRs NGC\,2180, Lynga\,8, and Bica\,5. }

\label{VPDs_plx_Ks_parte1}
\end{center}
\end{figure*}

The procedures described in Sections \ref{memberships} and \ref{method_verification} were applied to all clusters in our sample. The results are shown in Table \ref{info_sample_OCs_OCRs}. Galactocentric distances ($R_{\textrm{G}}$) were obtained assuming that the Sun is located at $8.0\pm0.5$\,kpc \citep{Reid:1993a} from the Galactic centre. The CMDs, VPDs, and $\varpi\times K_{\textrm{s}}$ plots for six clusters (NGC\,2180, Lynga\,8, Bica\,5, NGC\,7036, Alessi\,3, and ESO\,426-SC26) are shown in Figures \ref{Ks_JKs_clusters_decontaminated_parte1}, \ref{VPDs_plx_Ks_parte1}, and \ref{VPDs_plx_Ks_parte2}. Results for other clusters and the complete lists of members stars are shown in the appendix.

\begin{figure*}
\begin{center}
\parbox[c]{1.0\textwidth}
  {
   
        \includegraphics[width=0.33333\textwidth]{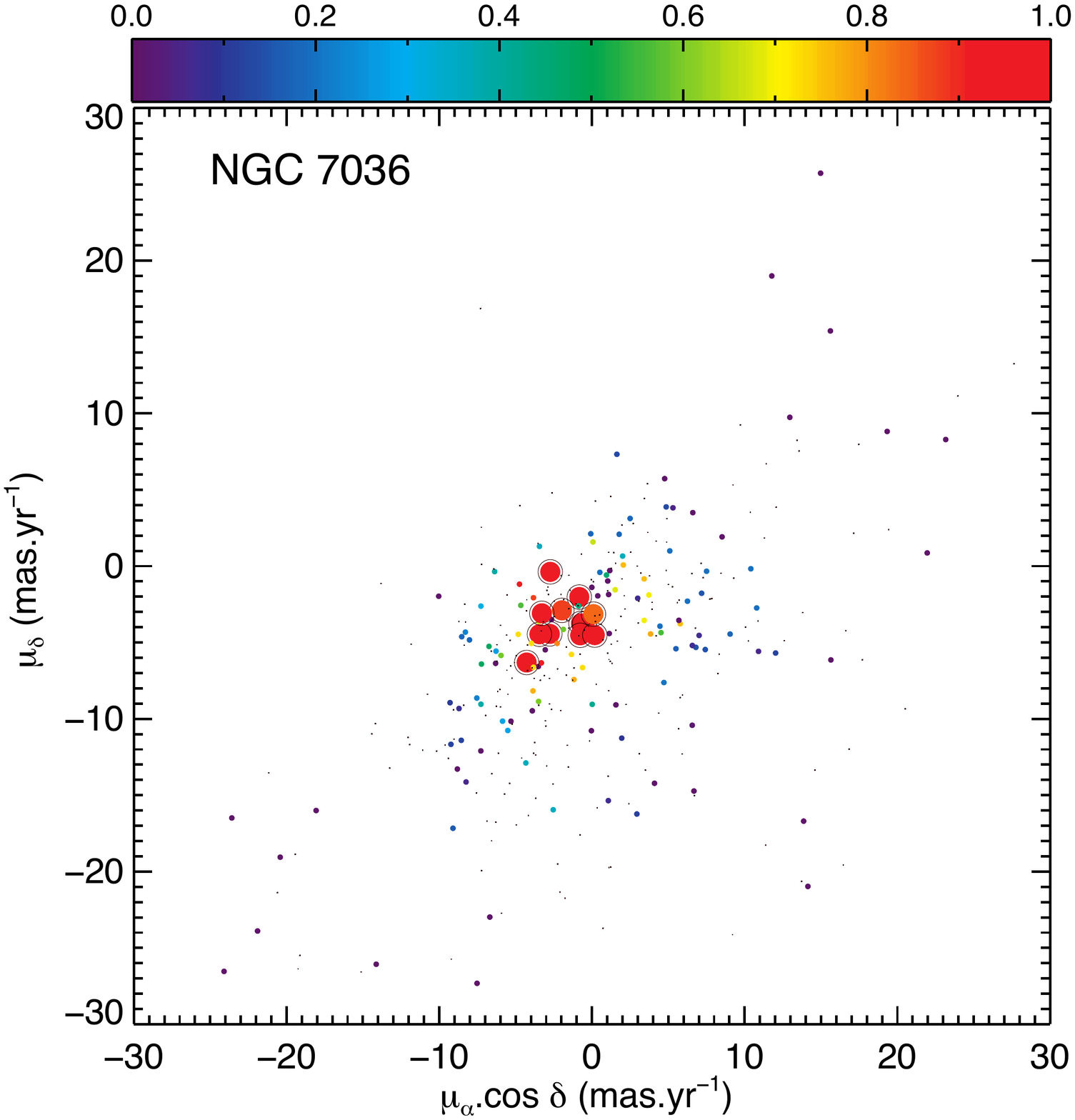}
        \includegraphics[width=0.33333\textwidth]{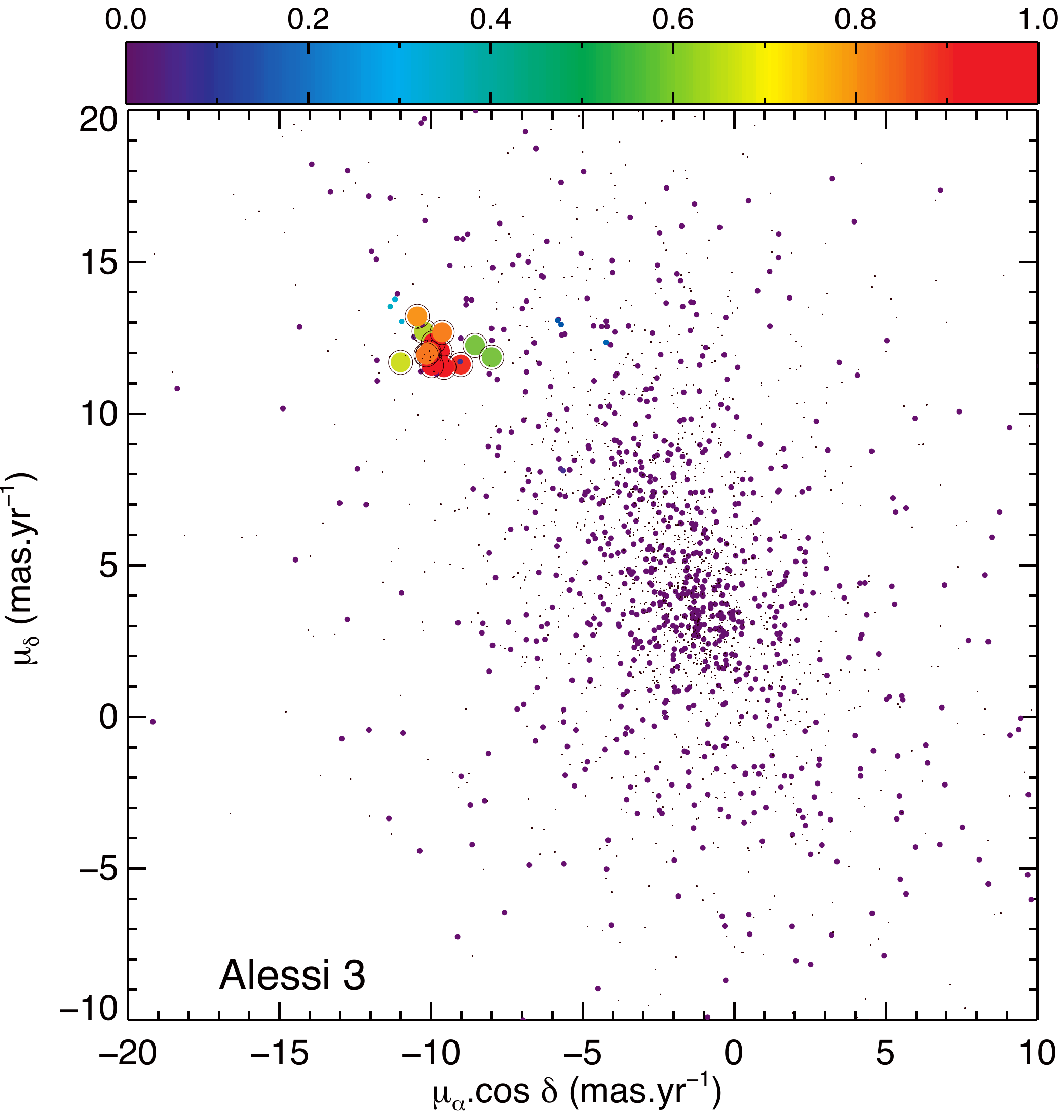}
        \includegraphics[width=0.33333\textwidth]{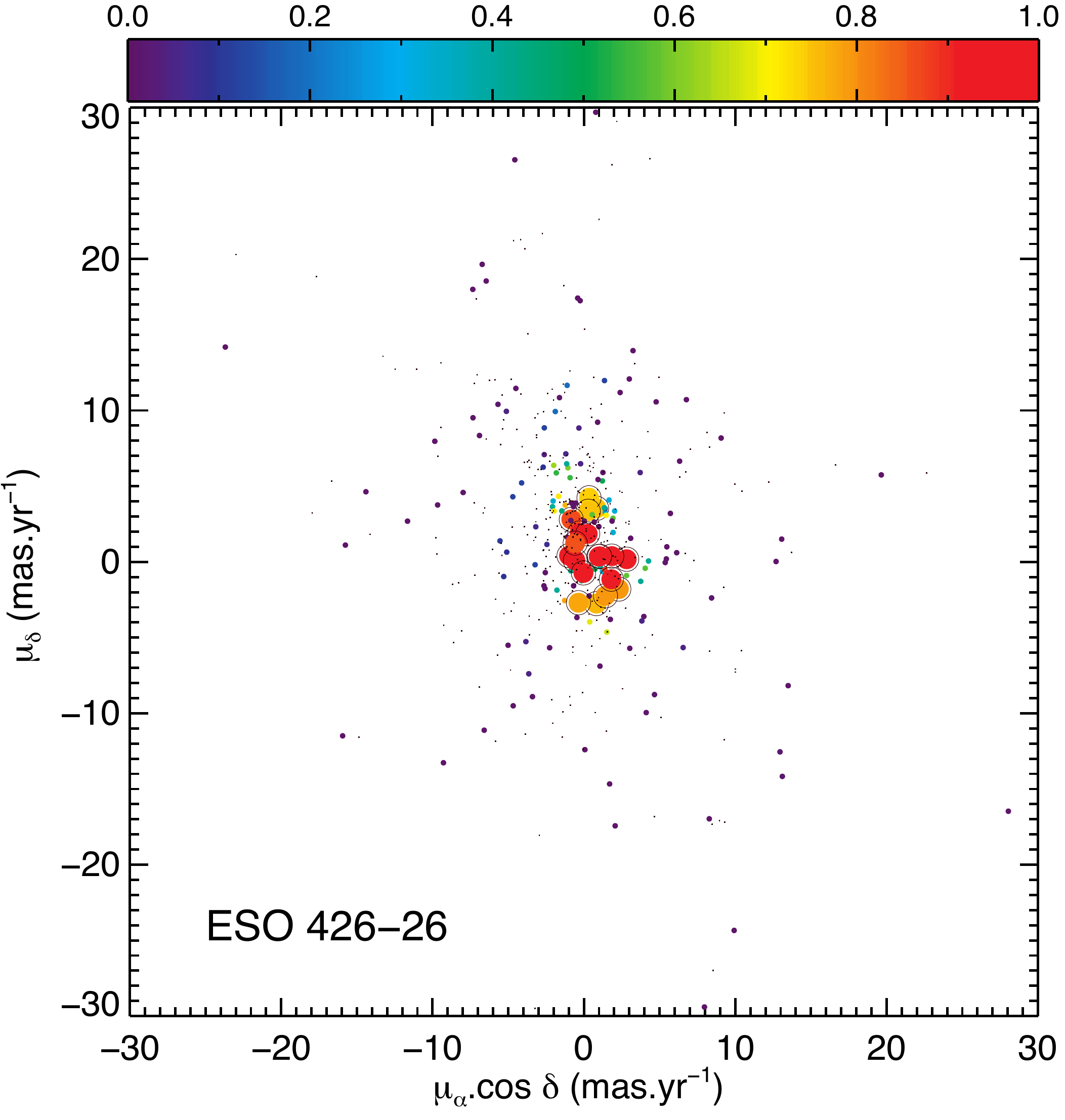}

        \includegraphics[width=0.33333\textwidth]{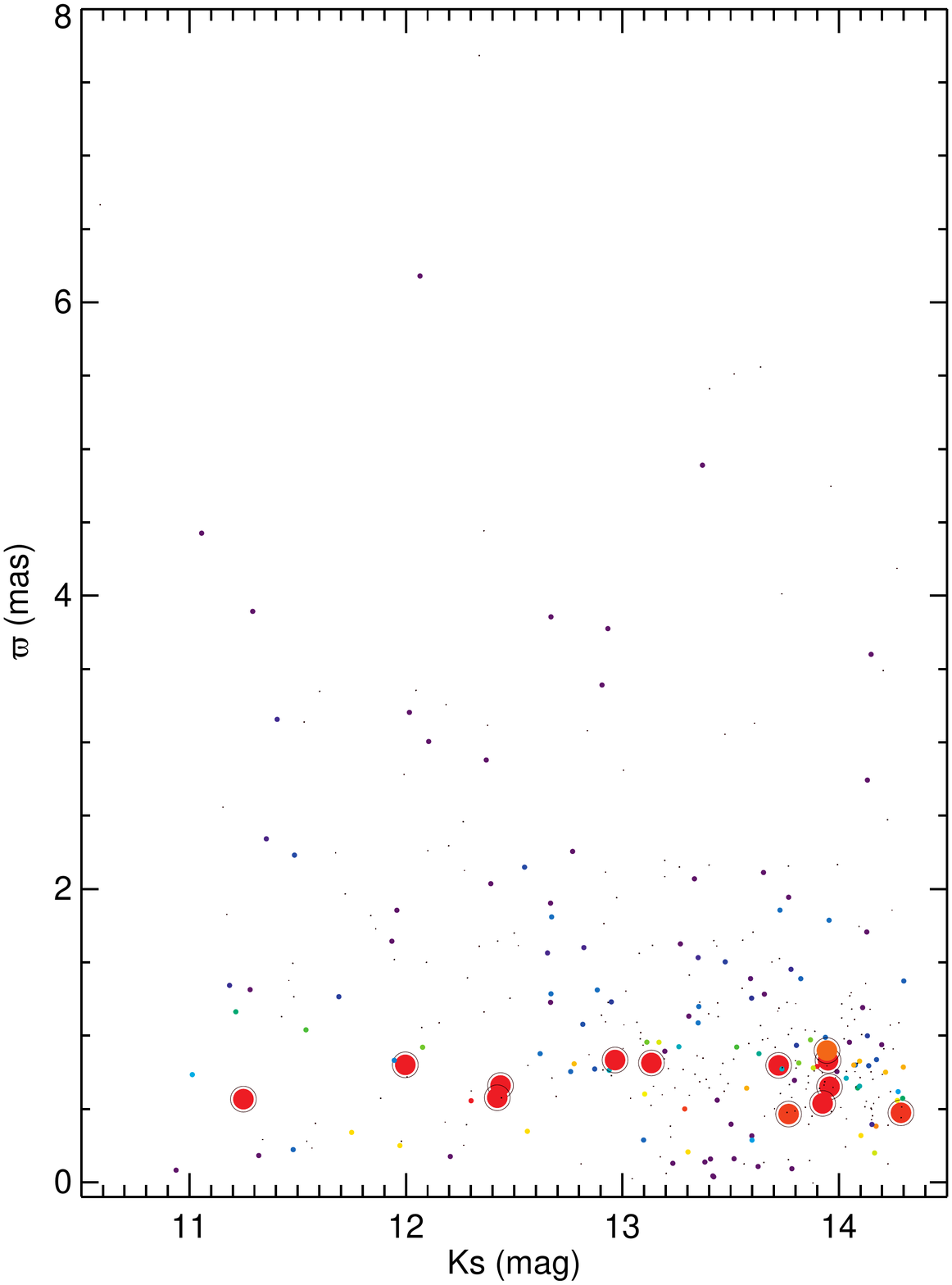}
        \includegraphics[width=0.33333\textwidth]{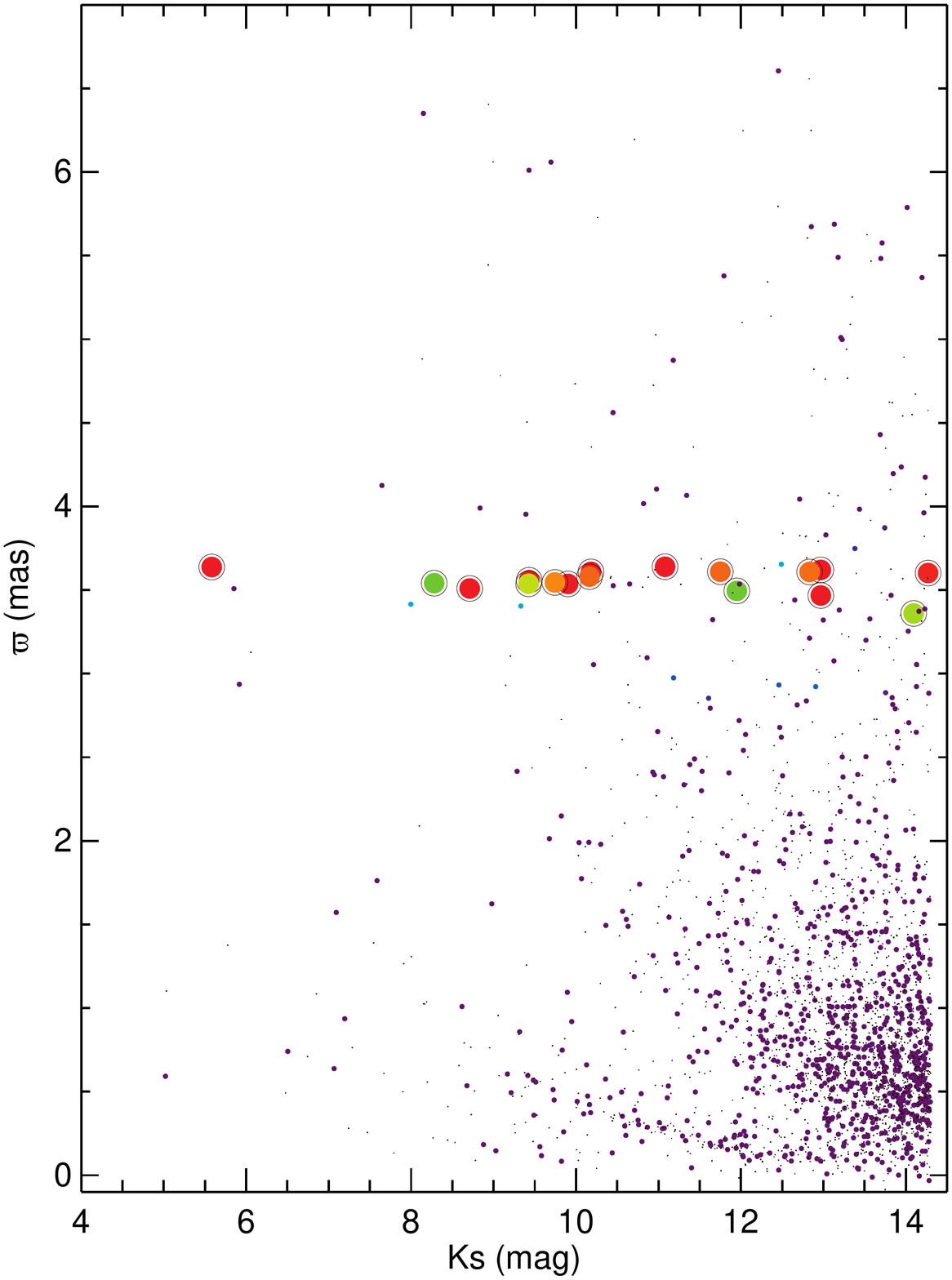}
        \includegraphics[width=0.33333\textwidth]{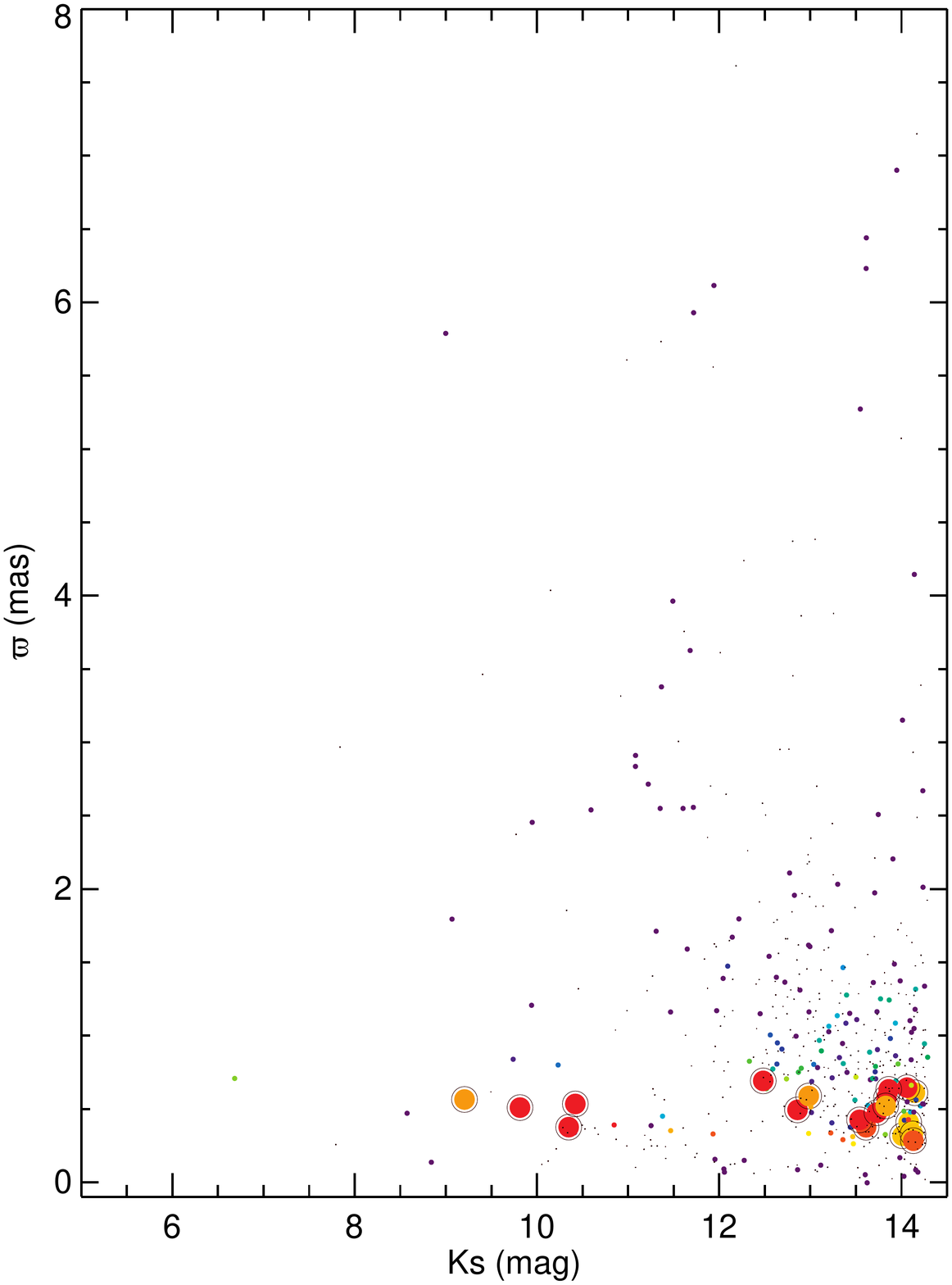}
        
  }
\caption{ Same as Figure \ref{VPDs_plx_Ks_NGC752_NGC188}, but for the OCRs NGC\,7036, Alessi\,3, and ESO\,426-SC26. }

\label{VPDs_plx_Ks_parte2}
\end{center}
\end{figure*}


\subsection{Mass functions}
\label{mass_functions}

\begin{figure*}
\begin{center}
\parbox[c]{0.81\textwidth}
  {
   
   \includegraphics[width=0.27\textwidth]{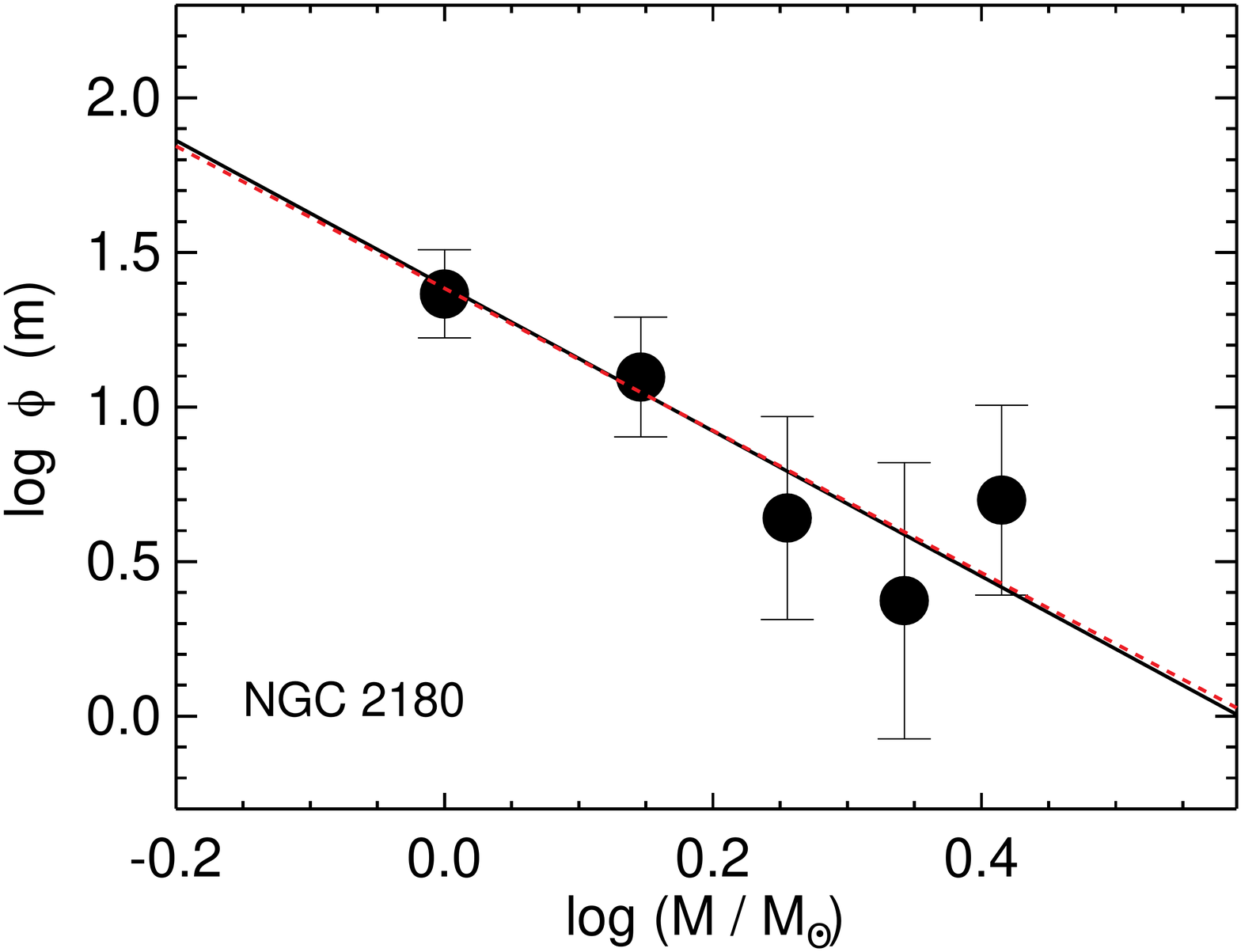}
   \includegraphics[width=0.27\textwidth]{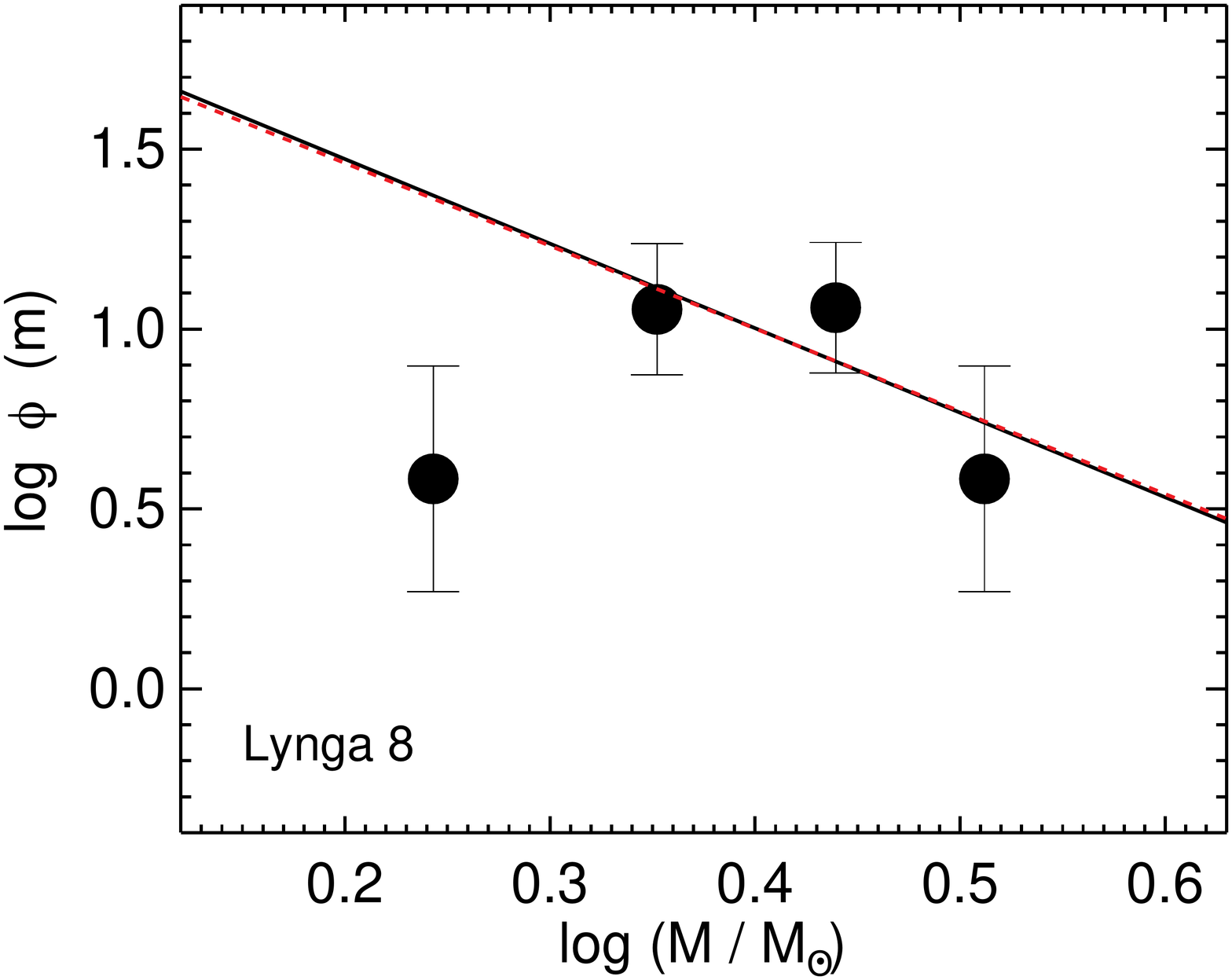}   
   \includegraphics[width=0.27\textwidth]{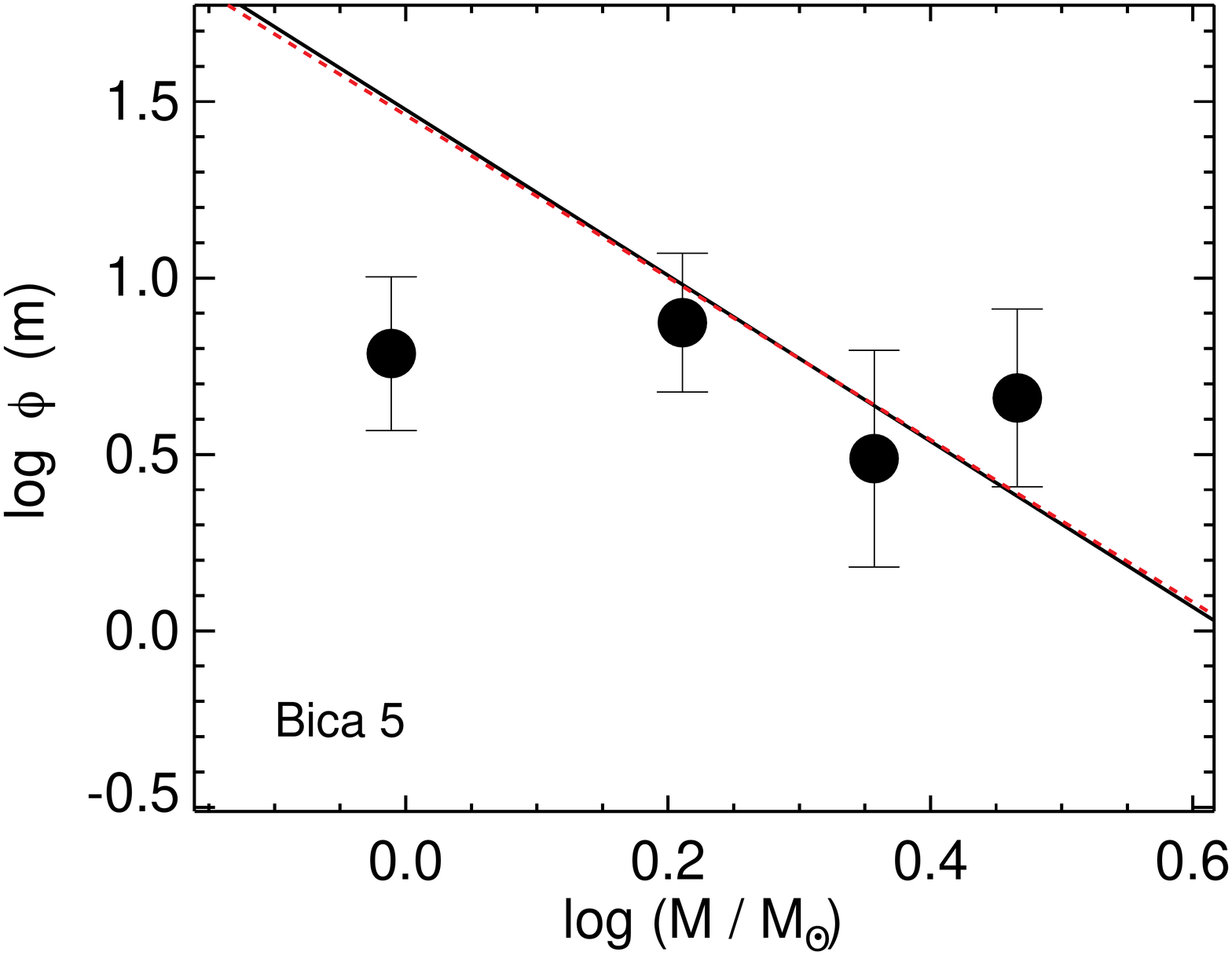}
   \includegraphics[width=0.27\textwidth]{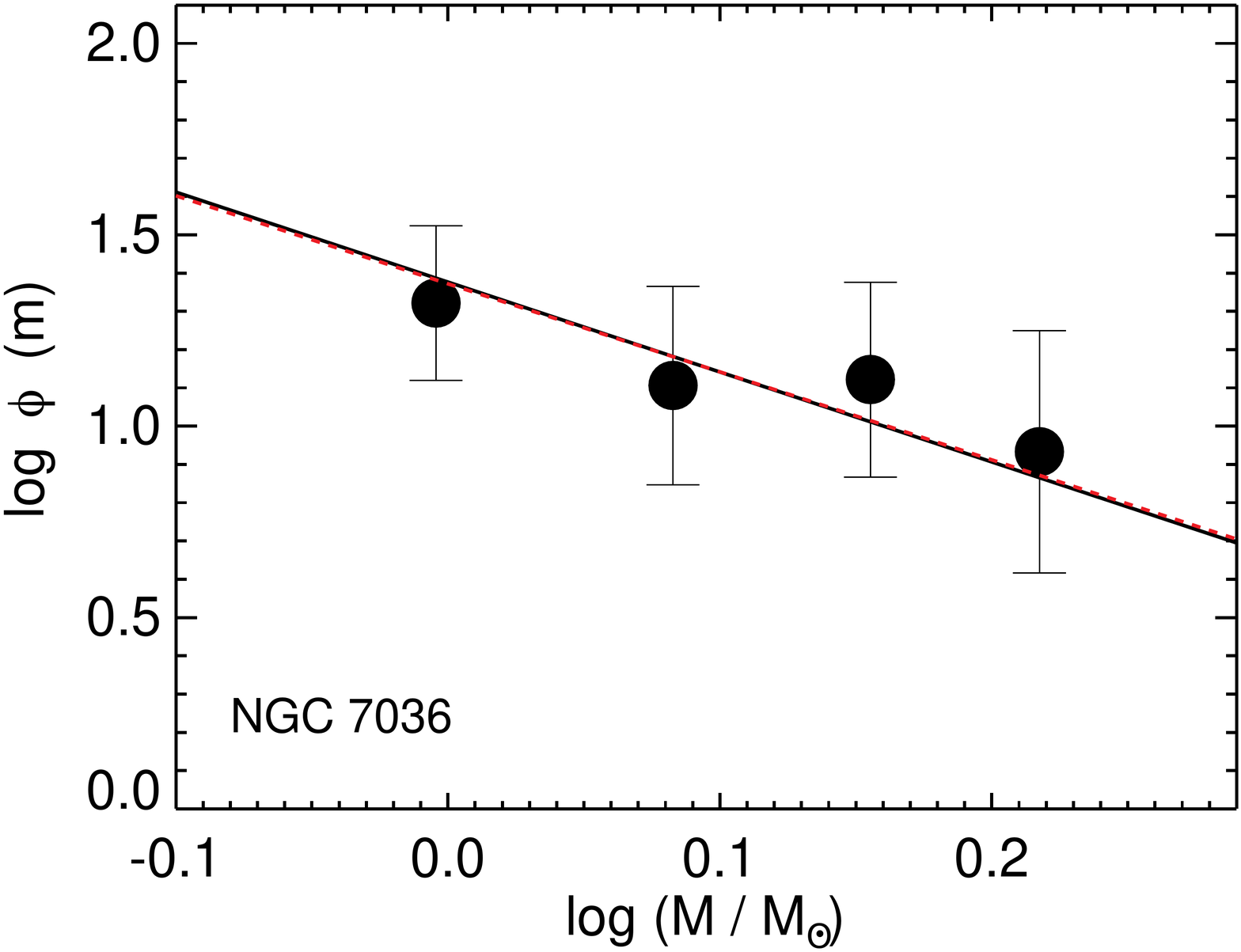}
   \includegraphics[width=0.27\textwidth]{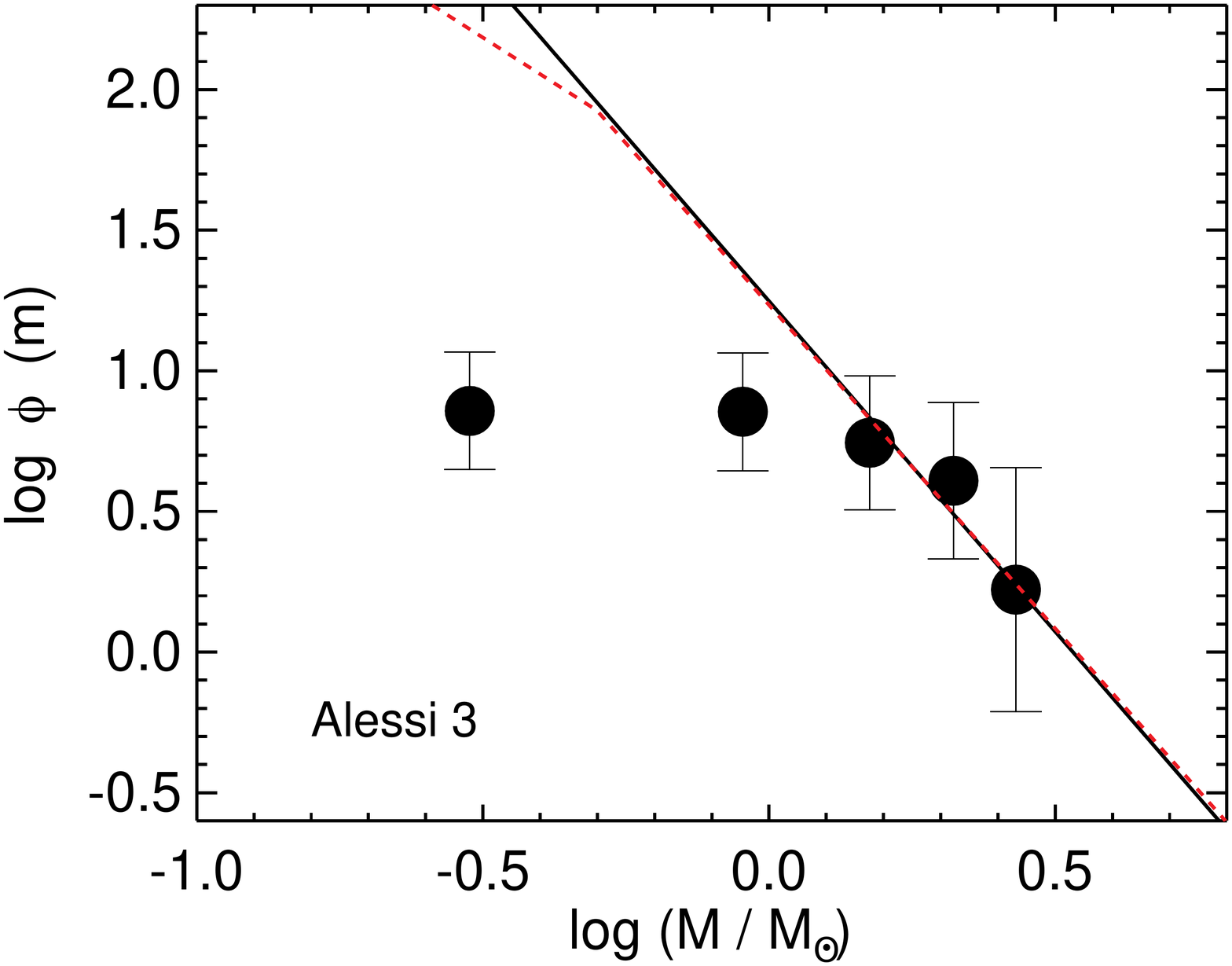}
   \includegraphics[width=0.27\textwidth]{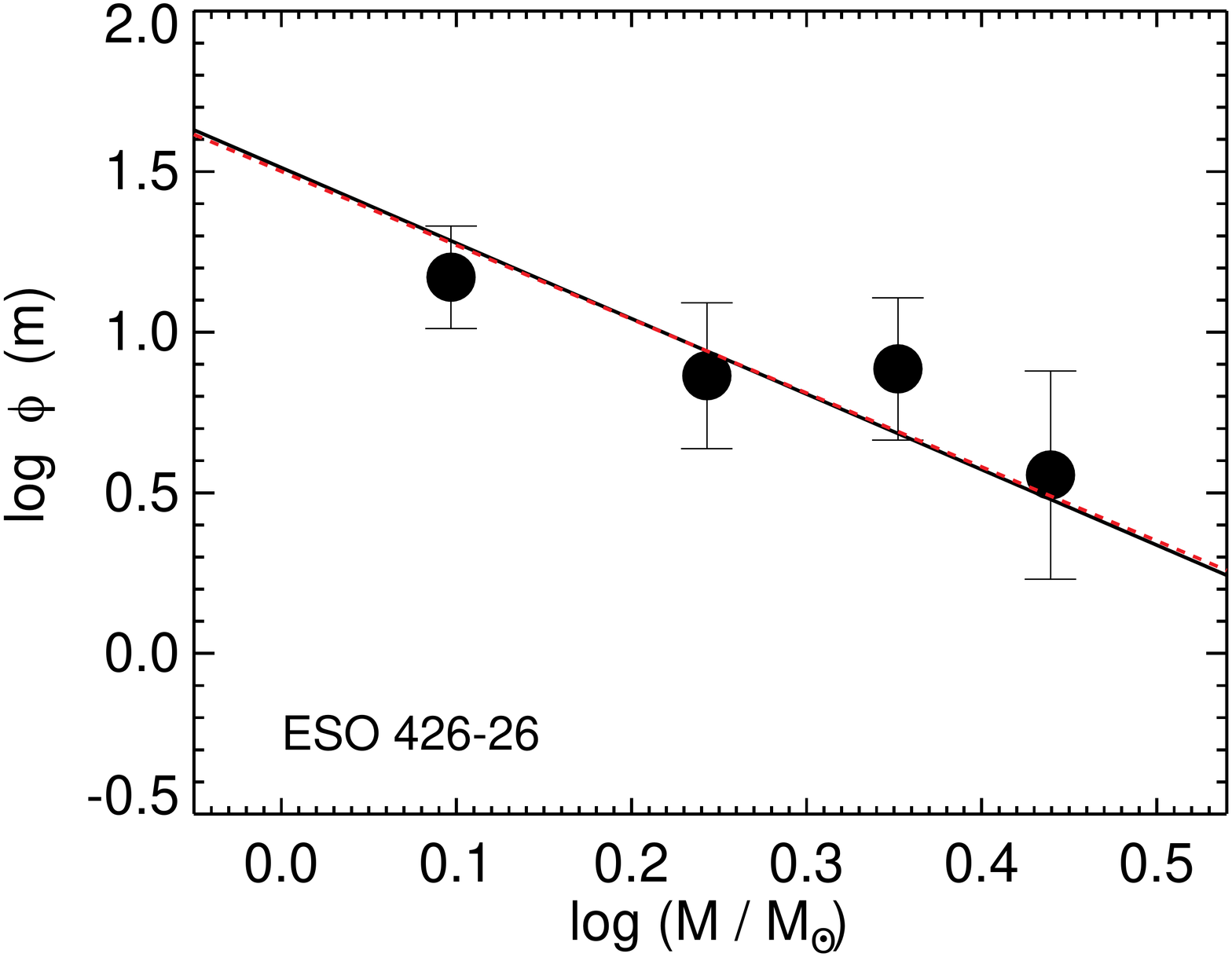}

  }

\caption{ Observed mass functions for (from top left to lower right): NGC\,2180, Lynga\,8, Bica\,5, NGC\,7036, Alessi\,3, and ESO\,426-SC26. For comparison, the IMFs of \citeauthor{Salpeter:1955}\,\,(\citeyear{Salpeter:1955}, continuous black line) and \citeauthor{Kroupa:2001}\,\,(\citeyear{Kroupa:2001}, dashed red line) are shown. }
\label{mass_func_parte1}
\end{center}
\end{figure*}

Individual masses for member stars were derived by interpolation in the theoretical isochrone, properly corrected for reddening and distance modulus (Figure \ref{Ks_JKs_clusters_decontaminated_parte1}; see the appendix for other clusters), from the star $K_{\textrm{s}}$ magnitude. Each cluster mass function was constructed by counting the number of stars inside linear mass bins $[\phi(m)=dN/dm]$ and then converted into the logarithmic scale. Star counts inside each bin were weighted by the  membership likelihoods (Section \ref{memberships}), and uncertainties come from Poisson statistics. The results are shown in Figure \ref{mass_func_parte1}. The initial mass functions (IMF) of \cite{Salpeter:1955} and \cite{Kroupa:2001} are shown, for comparison purposes. Both IMFs were scaled according to the total photometric cluster mass.

In some cases (e.g. Lynga\,8, Bica\,5, Alessi\,3, and Ruprecht\,31) the mass functions show depletion of main-sequence stars, especially in fainter magnitude bins. This result should not be attributed to photometric incompleteness because we have restricted our data to stars for which $J, H,$ and $K_{\textrm{s}}$ magnitudes are brighter than the 2MASS completeness limits. This depletion may be a  consequence of preferential low-mass stars evaporation, which is a signature of dynamically evolved OCs (\citeauthor{de-La-Fuente-Marcos:1997}\,\,\citeyear{de-La-Fuente-Marcos:1997}, hereafter M97; \citeauthor{Portegies-Zwart:2001}\,\,\citeyear{Portegies-Zwart:2001}). Other clusters (e.g. NGC\,2180, NGC\,7036, ESO\,426-SC26, and ESO\,425-SC15) seem less severely depleted because their mass functions are nearly compatible with Salpeter and Kroupa mass functions, considering the 2MASS limiting magnitudes.

\subsection{Non-existent clusters}
\label{asterisms}

Our original sample contained the OCR candidates NGC\,1252, NGC\,7772, ESO\,570-12, ESO\,383-SC10, and ESO\,065-SC03. We found that these sparse stellar concentrations are not clusters but are instead chance projections on the sky. Below we present previous information found in the literature about these objects.

\begin{itemize}
\item  NGC\,1252 was studied by \citeauthor{de-la-Fuente-Marcos:2013}\,\,(\citeyear{de-la-Fuente-Marcos:2013}, hereafter MMM13) by means of an investigation that involved $UBVI$ photometry, proper motions from the Yale/San Juan SPM 4.0 catalogue \citep{Girard:2011}, and high-resolution spectroscopy of 13 stars. It was considered a high-altitude metal-poor $\sim3\,$Gyr cluster, located at $\sim1\,$kpc away from the Sun. \\

\item \cite{Carraro:2002} investigated the object NGC\,7772 by employing multi-colour CMDs and colour-colour diagrams from $UBVI$ photometry. The author proposed that NGC\,7772 is a strong low-mass star depleted cluster located at 1.5\,kpc away from the Sun and its age is 1.5\,Gyr.\\

\item \cite{Pavani:2007} employed 2MASS photometry and proper motions from the UCAC4 \citep{Zacharias:2013} catalogue in the study of the object ESO\,383-SC10; it was classified as a sparse OCR. They obtained 2.0\,Gyr and 1.0\,kpc for its age and heliocentric distance. In contrast to their conclusions, \cite{Piatti:2017} employed multi-band $UBVRI$ and $CT_{1}T_{2}$ photometry, complemented with astrometric data from the GAIA DR1 \citep{Gaia-Collaboration:2016a} catalogue, to conclude that ESO\,383-SC10 is not a real stellar concentration. Similarly to ESO\,383-SC10, in the same paper \cite{Pavani:2007} evaluated the physical nature of the OCR candidate ESO\,570-SC12; the authors considered it a 800\,Myr loose remnant, located at 640\,pc away from the Sun.\\

\item \cite{Joshi:2015} investigated the case of the OCR candidated ESO\,065-SC03; they employed multi-band CMDs and CCDs based on photometric data from the WISE \citep{Cutri:2012} and 2MASS catalogues and proper motion data from the SPM\,4.0 and UCAC4 catalogues. The authors estimated 562\,Myr and 2.3\,kpc for the object age and heliocentric distance.\\

\end{itemize}

\begin{figure*}
\begin{center}
\parbox[c]{1.0\textwidth}
  {
   
        \includegraphics[width=0.33333\textwidth]{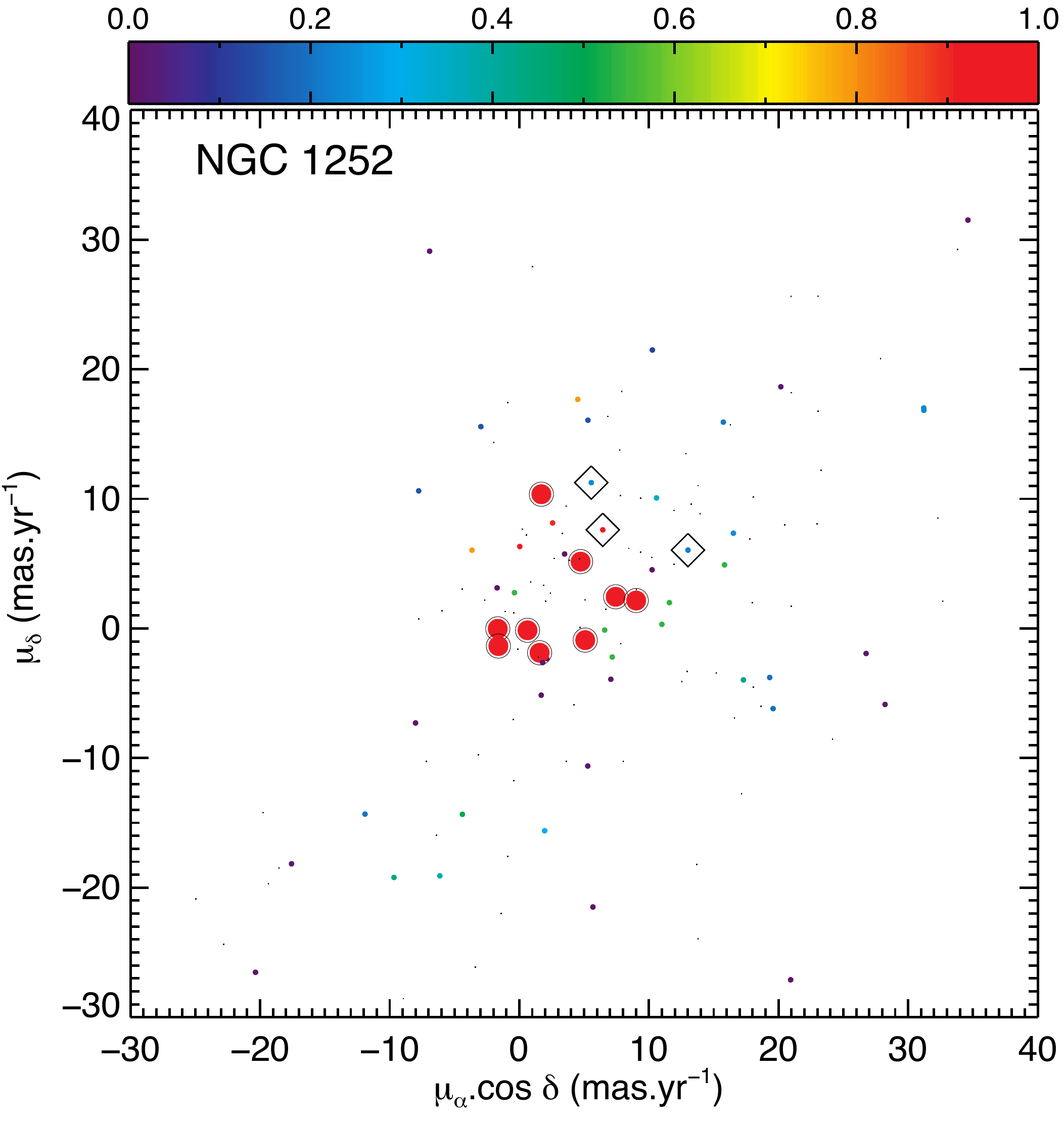}
        \includegraphics[width=0.33333\textwidth]{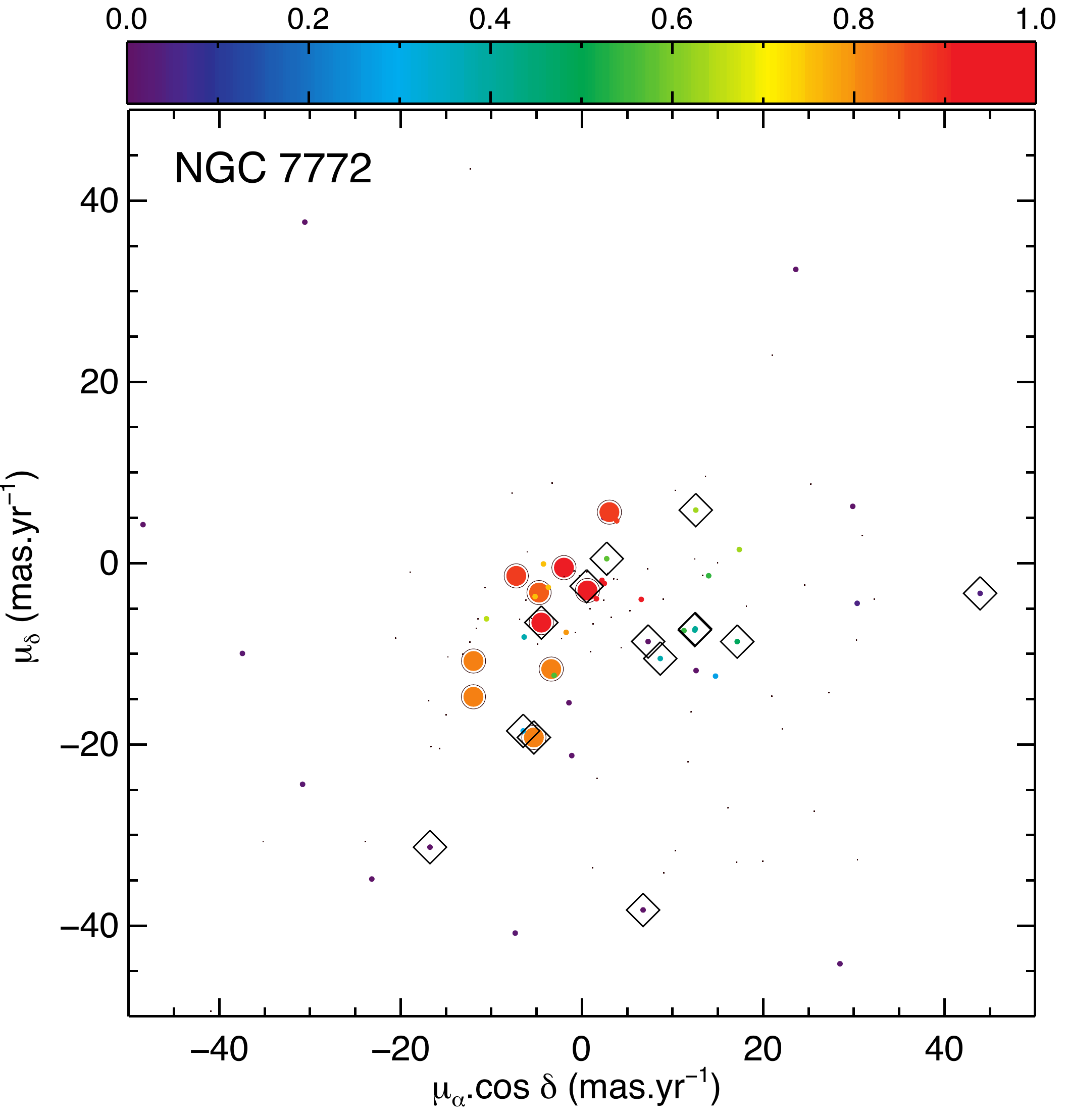}
        \includegraphics[width=0.33333\textwidth]{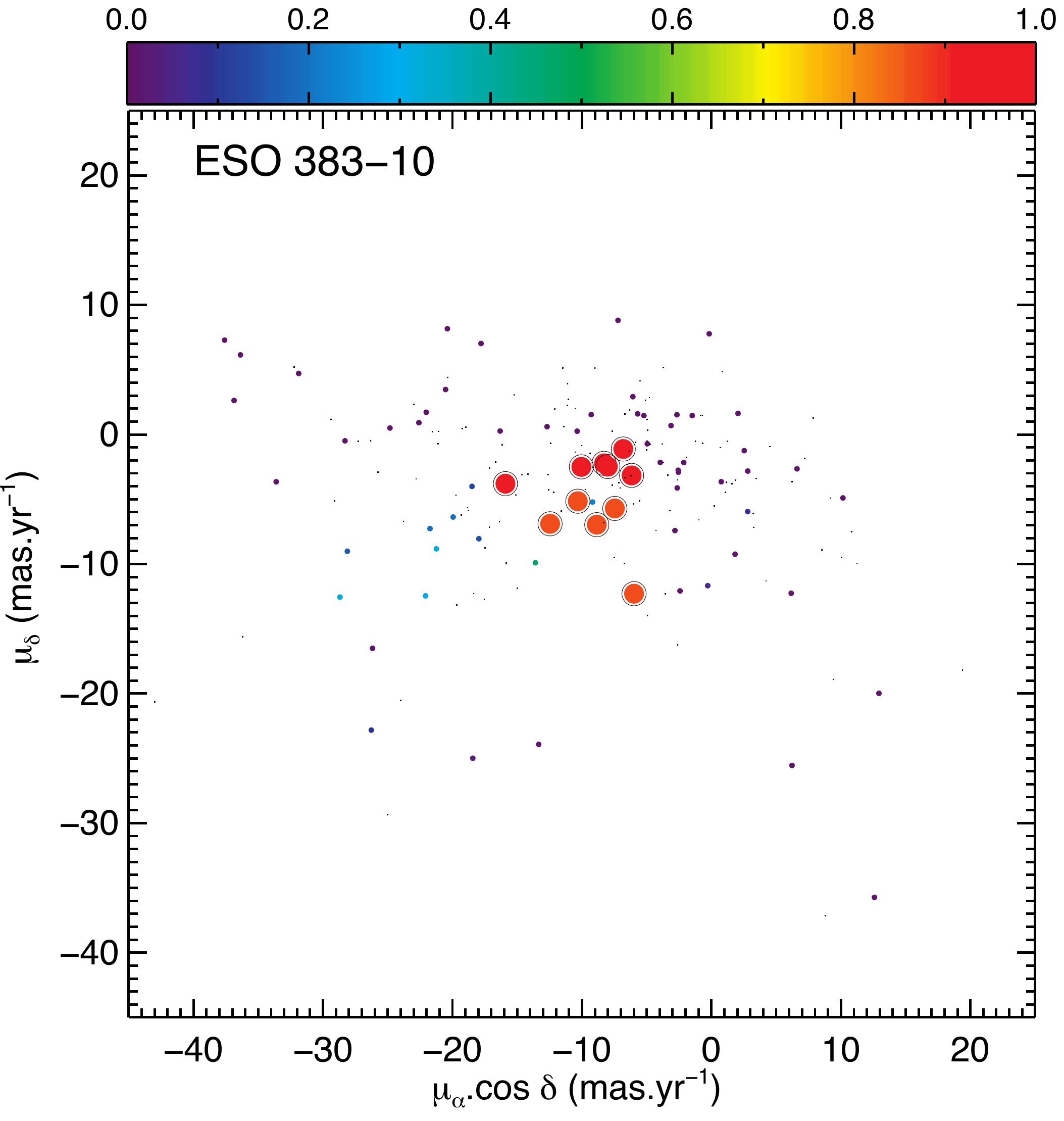}
        
        \begin{center}
        \includegraphics[width=0.33333\textwidth]{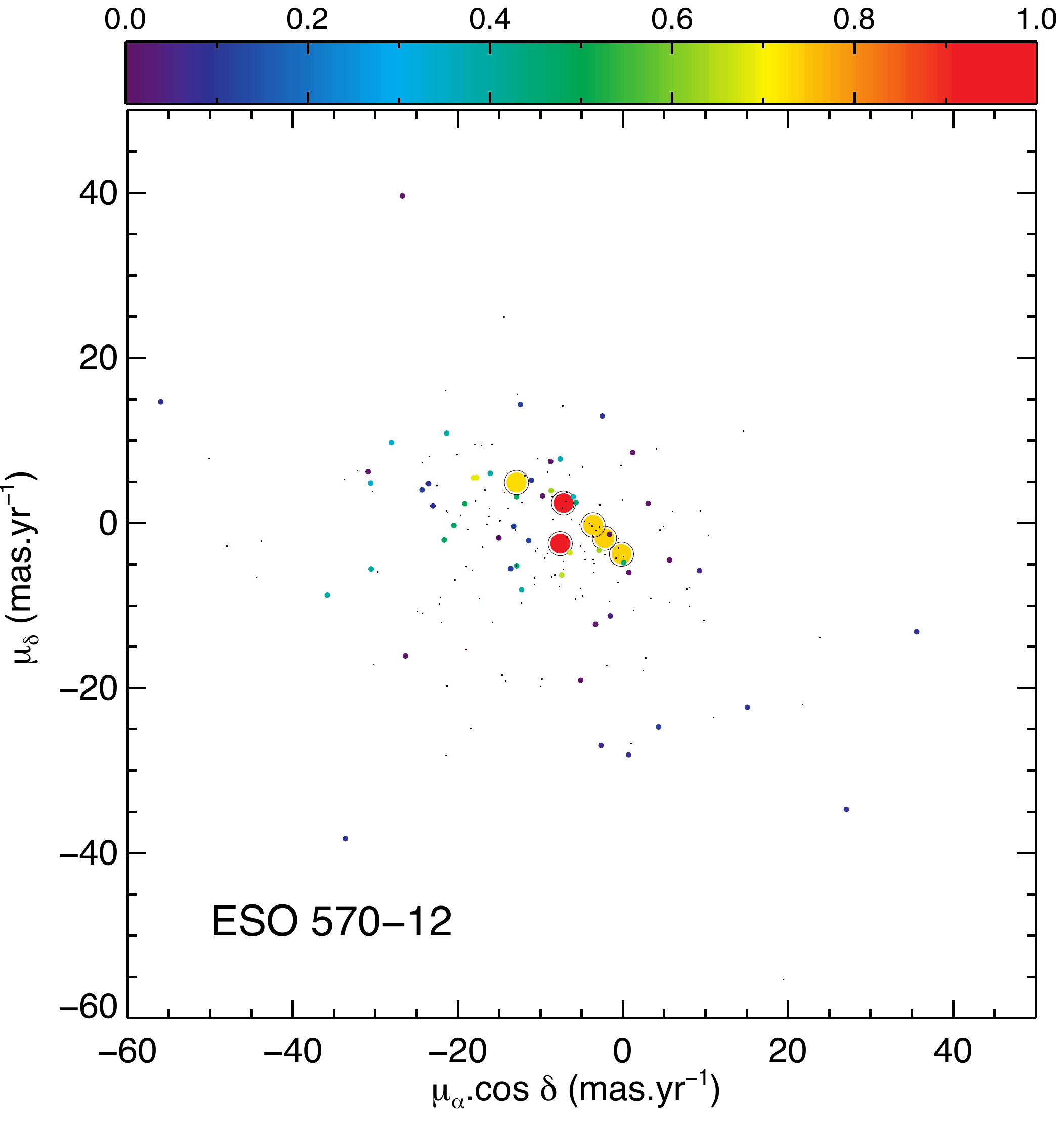}
        \includegraphics[width=0.33333\textwidth]{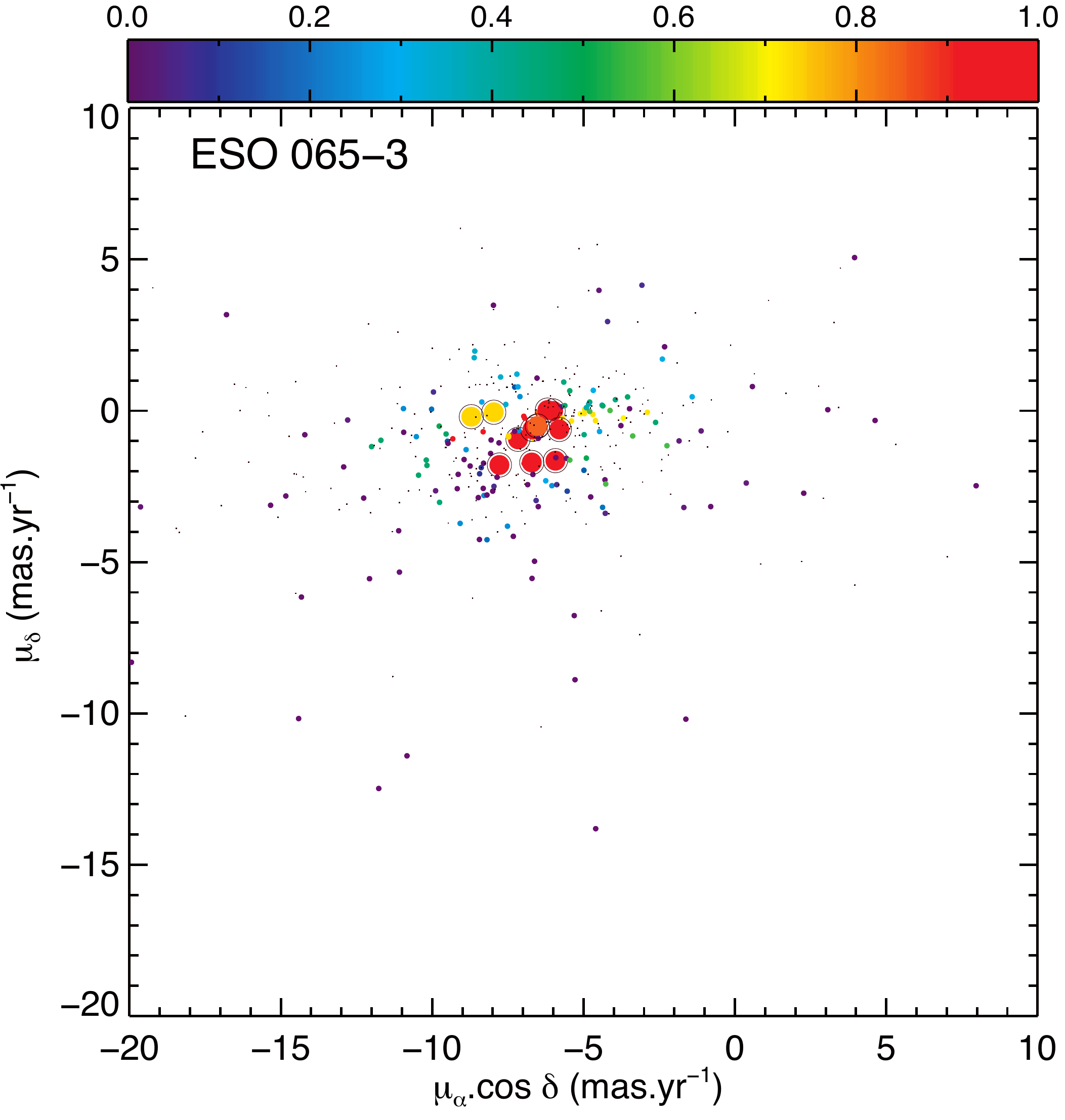}
        \end{center}
        
  }
\caption{ Same as Figure \ref{VPDs_plx_Ks_NGC752_NGC188}, but for the asterisms NGC\,1252, NGC\,7772, ESO\,383-SC10, ESO\,570-SC12, and ESO\,065-SC03. The diamonds represent member stars taken from the literature (see text for details). }

\label{VPDs_asterisms}
\end{center}
\end{figure*}

\begin{figure}
\begin{center}

     \begin{center}
        \includegraphics[width=0.3\textwidth]{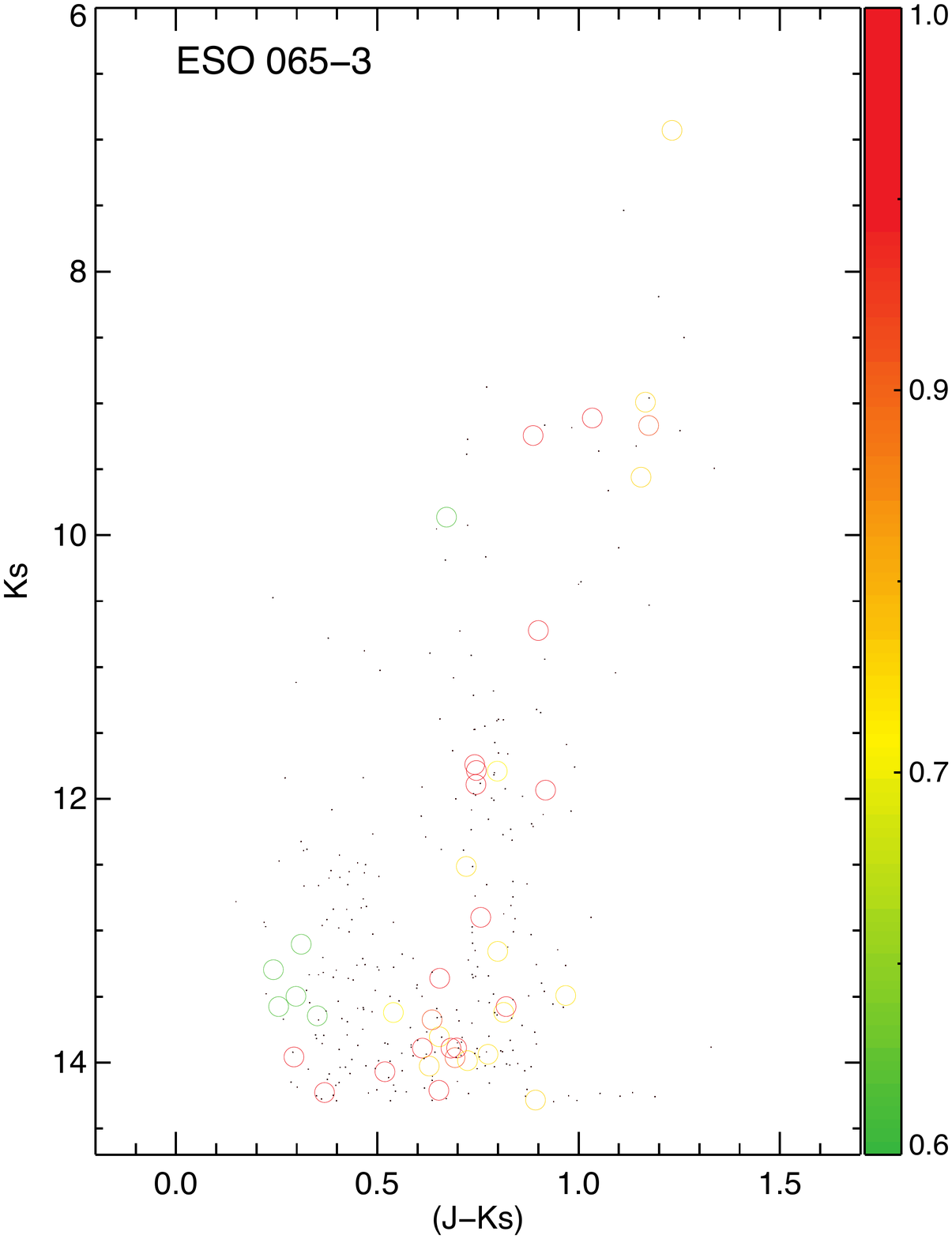}
     \end{center}

\caption{ CMD $K_{\textrm{s}}\times(J-K_{\textrm{s}})$ of the asterism ESO\,065-SC03. }

\label{Ks_JKs_ESO6503}
\end{center}
\end{figure}

As shown in Figure \ref{VPDs_asterisms}, the high dispersion of the kinematic data in these objects' VPDs makes the hypothesis of gravitationally bound systems unlikely, even considering stars with higher membership likelihoods, after employing our decontamination method. Using the method described in Section \ref{velocity_dispersions} to determine velocity dispersions ($\sigma_v$), we obtained $\sigma_{v}$ in the range 13\,$-$\,34\,km.s$^{-1}$ (the lower limit for ESO\,065-SC03 and the upper limit for NGC\,1252). These values of $\sigma_v$ are considerably higher than those determined for the sample of OCs and OCRs ($\sigma_v$\,$\sim$\,1\,$-$\,7\,km.s$^{-1}$; see Figure \ref{compara_densidade_rlim_sigma_v}, left panel).

In the case of ESO\,065-SC03, despite the apparent clustering of data around ($\mu_{\alpha},\mu_{\delta}$)\,$\sim$\,(-7.0\,,\,-1.0)\,km.s$^{-1}$, even the more probable members do not seem to form recognisable evolutionary sequences in the CMD (Figure \ref{Ks_JKs_ESO6503}), therefore we were unable to fit a single isochrone to this object's photometric data.

The three stars marked with open diamonds in the NGC\,1252 VPDs are member candidates identified by MMM13 (their table 4), based on spectroscopic and proper motions analysis. Their astrometric parameters $\pi$\,(mas), $\mu_{\alpha}$\,(mas.yr$^{-1}$) and $\mu_{\delta}$(mas.yr$^{-1}$) are (1.4967$\pm$0.0238, 6.441$\pm$0.045, 7.607$\pm$0.044), (1.0936$\pm$0.0207, 5.553$\pm$0.035, 11.251$\pm$0.033) and (2.2619$\pm$0.0170, 13.013$\pm$0.032, 6.045$\pm$0.032). These stars seem to follow the same general scatter of data as shown by the higher membership stars according to our method, therefore they do not exhibit any particular concentration in the VPD. In addition, their parallaxes are incompatible. 

In the case of NGC\,7772, stars marked with open diamonds are member stars according to \cite{Carraro:2002}. These stars present an even greater scatter in the VPD than our higher membership stars. In addition, the parallaxes of the stars in Carraro's (2002) sample range from $\sim$0.5$-$$\sim$5.5\,mas, which precludes the hypothesis that these stars form a physical system. ESO\,383-SC10, ESO\,570-SC12, and ESO\,065-SC03 are not in any published lists of member stars in the literature.

The scattering of data in the astrometric space led to the conclusion that the five objects present in this section are chance alignment of stars along the line of sight. This is the same conclusion as reported by \cite{Kos:2018} for the cases of NGC\,1252 and NGC\,7772; ESO\,383-SC10 was also considered as asterism by \cite{Piatti:2017}. The set of astrometric and photometric data for the highlighted stars in Figure \ref{VPDs_asterisms} is shown in the appendix.
\section{Evaluating an evolutionary connection between OCs and OCRs}
\label{discussion}

\subsection{Mass, age, and limiting radius}
\label{mass_age_rlim}

Comparisons between evolved OCs and remnants are useful to establish possible evolutionary connections (\citeauthor{Bonatto:2004a}\,\,\citeyear{Bonatto:2004a},\,hereafter BBP04). In this sense, the parameters shown in Table \ref{info_sample_OCs_OCRs} were employed to compare the groups of ($i$) dynamically evolved OCs and ($ii$) sparsely populated OCRs. The dynamical state of each object is characterised by the parameters age, limiting radius, and stellar mass. Velocity dispersions are analysed in Sect. \ref{velocity_dispersions}.

\begin{figure*}
 \centering
 \includegraphics[width=16cm]{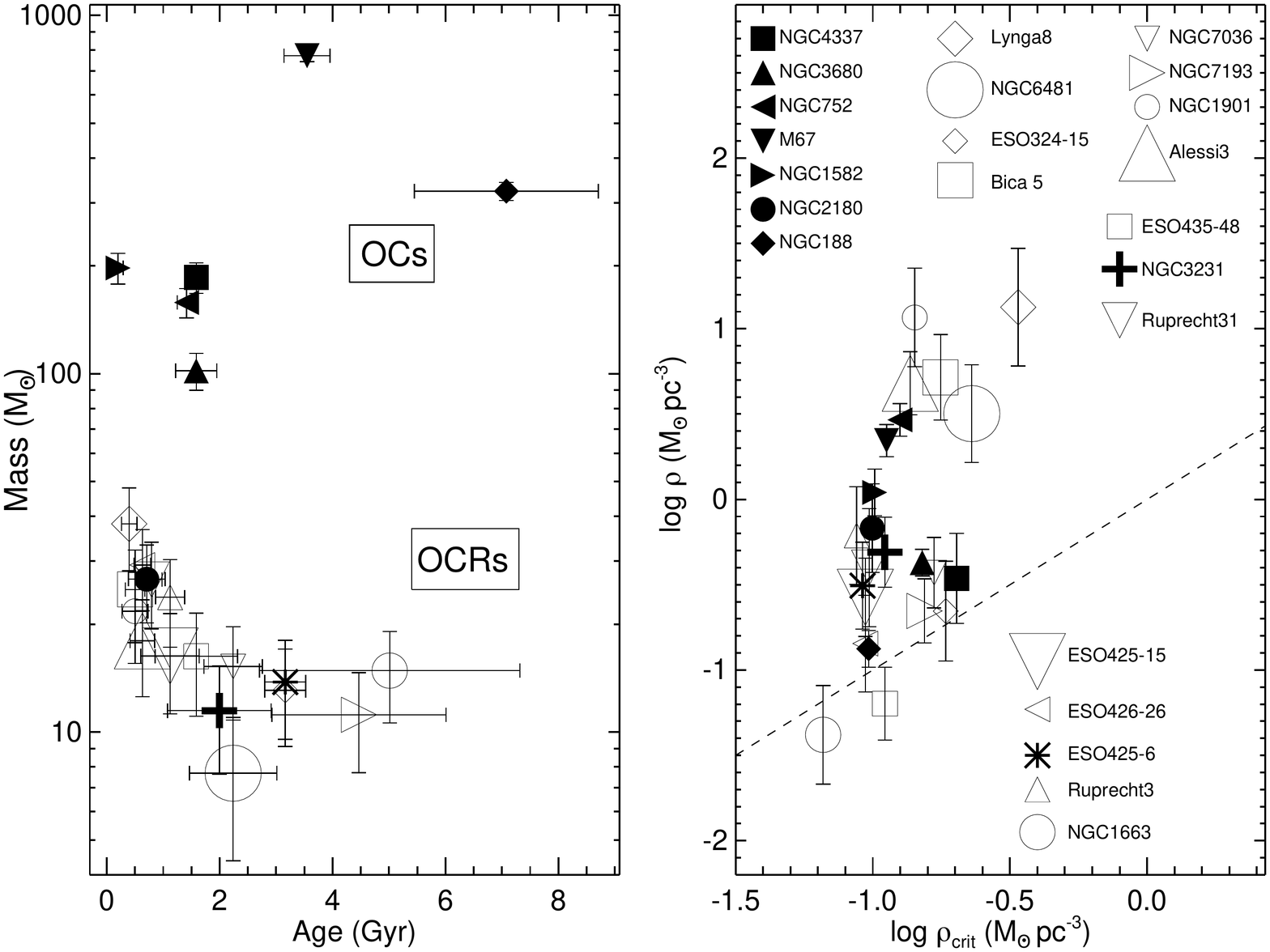}

\caption{  Mass vs. age (left panel) and density vs. critical density (right panel) for a set of 6 OCs, 16 OCRs and the transition object NGC\,2180. The dashed line shows the identity $\rho=\rho_{\textrm{crit}}$. The set of OCs, OCRs and their respective symbols are shown in the right panel. }


   \label{plots_rlim_age_mass_POCRs_forpaper_parte1}
\end{figure*}

\begin{figure*}
 \centering
 \includegraphics[width=11cm]{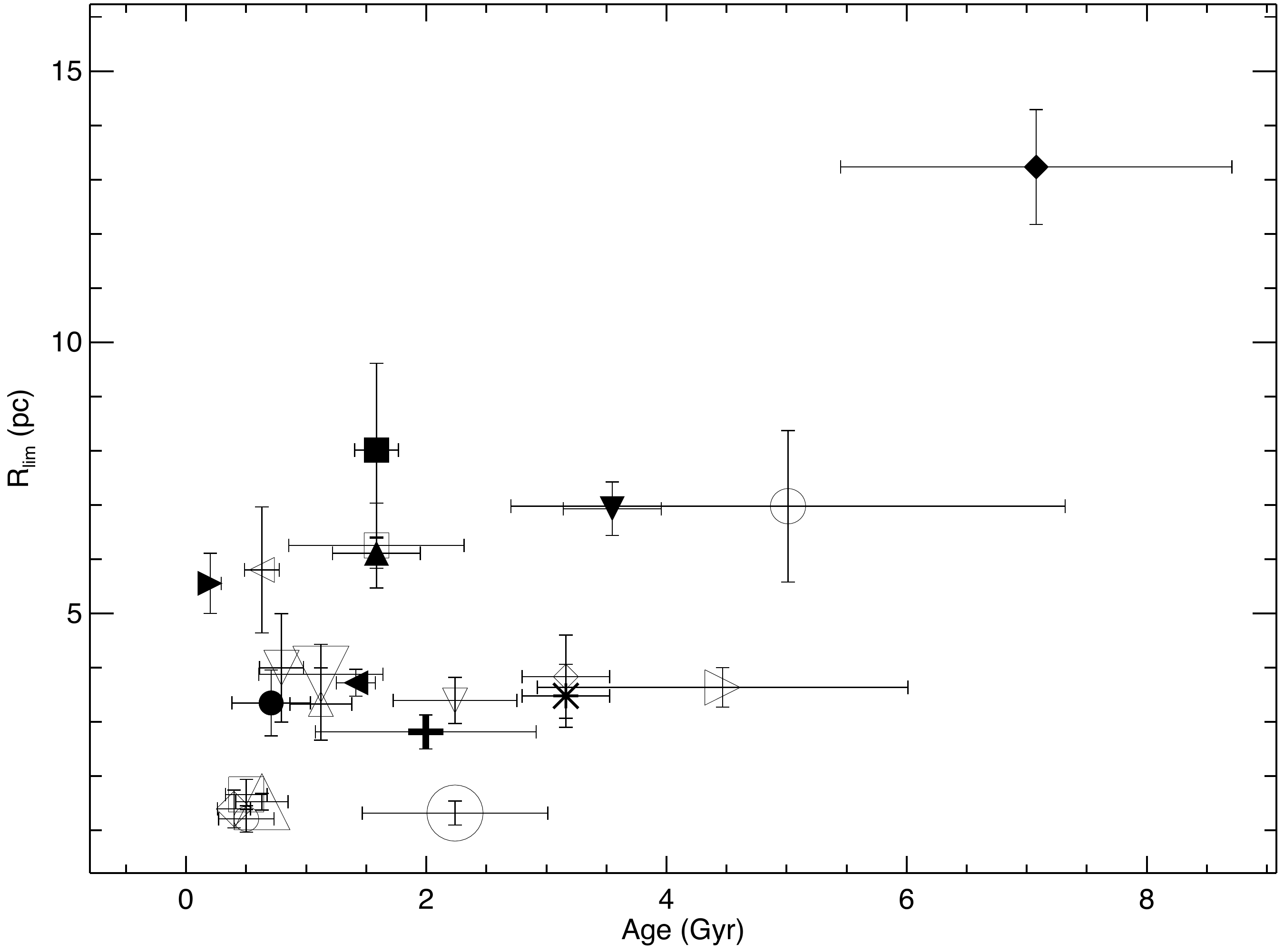}
 \caption{Limiting radius ($R_{\textrm{lim}}$) as a function of age for the 23 studied OCs (filled symbols) and OCRs (open symbols).}
   \label{plots_rlim_age_mass_POCRs_forpaper_parte2}
\end{figure*}

Fig. \ref{plots_rlim_age_mass_POCRs_forpaper_parte1} shows a plot of mass versus age (left panel) and the density  versus critical density ($\rho\times\rho_{\textrm{crit}}$; right panel) for our sample. We performed half-mass density estimations ($\rho$\,=\,$3M/(8\pi R_{\textrm{hm}}^3)$) by assuming an approximate relation $R_{\textrm{hm}}\sim0.5\,R_{\textrm{lim}}$. Filled symbols represent OCs and open ones (including crosses and asterisks) represent OCRs. 



Older clusters among the dynamically evolved OCs tend to present higher masses. Massive OCs tend to survive longer against destructive effects such as Galactic tidal stripping and evaporation due to dynamical internal relaxation. According to $N-$body simulations (e.g. M97; \citeauthor{Lamers:2005}\,\,\citeyear{Lamers:2005}), there is a strong dependence between cluster disruption time and initial population size for a given Galactocentric distance. OCR masses are confined to the interval $\sim$10$-$40\,M$_{\odot}$ (Fig. \ref{plots_rlim_age_mass_POCRs_forpaper_parte1}, left panel). 

Most OCRs present density estimates that are greater than $\rho_{\textrm{crit}}$ (Figure \ref{plots_rlim_age_mass_POCRs_forpaper_parte1}, right panel), the critical density for survival against Galactic tidal disruption,

\begin{equation}
   \rho_{\textrm{crit}}\sim\frac{M_{\textrm{clu}}}{(4/3)\pi R_\textrm{J}^3}
   \label{equation_rho_crit}
\end{equation}

\noindent
for a cluster completely filling its Roche lobe; we employed the formulation of \cite{von-Hoerner:1957} for the Jacobi radius ($R_{J}$) determination

\begin{equation}
   R_J=\left(\frac{M_{\textrm{clu}}}{3\,M_{\textrm{G}}}\right)^{1/3}\times R_{\textrm{G}}
   \label{equation_Rj}
,\end{equation}

\noindent
which assumes a circular orbit around a point mass galaxy ($M_{\textrm{G}}$$\sim$1.0$\times$$10^{11}$\,M$_{\odot}$; \citeauthor{Carraro:1994}\,\,\citeyear{Carraro:1994}; \citeauthor{Bonatto:2005}\,\,\citeyear{Bonatto:2005}; \citeauthor{Taylor:2016}\,\,\citeyear{Taylor:2016}). Equations \ref{equation_rho_crit} and \ref{equation_Rj} combined result in

\begin{equation}
   \rho_{\textrm{crit}}=\left(\frac{9}{4\pi}\right)\frac{M_{\textrm{G}}}{R_{\textrm{G}}^3}
,\end{equation}

\noindent
for which we employed the $R_{\textrm{G}}$ values of Table \ref{info_sample_OCs_OCRs}. NGC\,1663 and ESO\,435-SC48 present density estimates that are marginally compatible with $\rho_{\textrm{crit}}$, considering the uncertainties. Both OCRs are currently far away from the Galactic plane ($\vert Z\vert\sim1\,$kpc, see Table \ref{info_sample_OCs_OCRs}). Despite their low densities, it is known that moving in an inclined orbit increases the survival chances of a star cluster (e.g. MMM13). Far away from the Galactic plane during most of their dynamical evolution, NGC\,1663 and ESO\,435-SC48 may have been able to relax their internal mass distribution in larger limiting radii without completely dispersing their stellar content.

The evolved OCs present a lower limit of R$_{\textrm{lim}}\sim3$\,pc (Figure \ref{plots_rlim_age_mass_POCRs_forpaper_parte2}), which may be a consequence of the Galactic tidal pull and internal dynamical heating. The limiting radii of the OCRs do not exhibit a trend with age. OCRs are divided into two groups (see also Fig. \ref{compara_densidade_rlim_sigma_v}, right panel): (a) the more compact OCRs, for which $R_{\textrm{lim}}\leq2\,$pc, and (b) the less compact OCRs, for which $2< R_{\textrm{lim}}\leq7\,$pc. Values for group (a) are compatible with the core radii ($r_\textrm{c}$) obtained by M98 through $N-$body simulations for clusters with initial populations of $N_{0}\sim10^{3}-10^{4}$ stars and containing primordial binaries; the simulated clusters present $r_\textrm{c}\lesssim2\,$pc for log($t.$yr$^{-1}$)$\gtrsim8.5$ (his figures 2b and 2c). Considering the uncertainties, values for group (b) are comparable with the half-mass radii ($r_{\textrm{hm}}$) obtained by M98 for the $N_{0}\sim10^4$ simulated clusters, for which $3\lesssim r_{\textrm{hm}}\lesssim5\,$pc for all ages (figure 2c of M98).  

Based on this comparison, it may be suggested that initially massive clusters tend to leave only a compact structure as a fossil residue of their evolution, with most of their stellar content dispersed into the field (BBP04 and references therein). Although sparsely populated, the higher densities of these OCRs apparently allowed them to keep their residual stellar content throughout their dynamical evolution. As stated by M98, when considering the solar neighbourhood, OCRs of initially massive ($N_{0}\sim10^{4}$) simulated clusters have stellar densities higher than the local Galactic value (0.044\,$M_{\odot}\,$pc$^{-3}$) and higher than 0.08\,$M_{\odot}\,$pc$^{-3}$, the critical density value for a cluster to be stable against tidal disruption in the solar neighbourhood, even when the residual population is as low as 30 stars.

The cluster NGC\,2180, previously catalogued as an OC (\citeauthor{Dias:2002}\,\,\citeyear{Dias:2002} and WEBDA\footnote[3]{https://www.univie.ac.at/webda/}), may represent intermediate evolutionary states between OCs and OCRs. It is  consistently closer to the loci occupied by remnants in Figs. \ref{plots_rlim_age_mass_POCRs_forpaper_parte1} (left panel) and \ref{compara_densidade_rlim_sigma_v} (right panel, Section \ref{analysing_OCRs_evolut_stages}). This was also verified by BBP04. It is enlightening to make a comparison between NGC\,2180 and NGC\,1582. Both objects have marginally comparable ages, considering the uncertainties, and are located at the same Galactocentric distance (Table \ref{info_sample_OCs_OCRs}). That is, both clusters have been subjected to nearly the same external effects during their evolution. Despite this, NGC\,1582 is much more massive. This comparison suggests that the differences in their dynamical states may be traced back to their progenitor OCs. As stated by BBP04, the lower-mass nature of NGC\,2180 probably accelerated its evolutionary timescale, setting it in a more dynamically advanced state than NGC\,1582.

Similar observations can be drawn for the pairs NGC\,4337\,-\,NGC\,6481, NGC\,3680\,-\,NGC\,7036, and NGC\,752\,-\,NGC\,3231. For each of these three pairs, the OC and the OCR have compatible ages and Galactocentric distances, considering the uncertainties, but very different observed masses ($m_{\textrm{NGC\,4337}}/m_{\textrm{NGC\,6481}}\cong24$, $m_{\textrm{NGC\,3680}}/m_{\textrm{NGC\,7036}}\cong7$, $m_{\textrm{NGC\,752}}/m_{\textrm{NGC\,3231}}\cong14$). Again, these differences may be attributed to their parent OCs. 


\subsection{Comparison with $N-$body simulation results}
\label{comparison_with_simulations}

Based on simulations for clusters in an external tidal field \citep{Baumgardt:2003}, \cite{Lamers:2005} found an approximate dependence (their eq. 8) between the disruption time ($t_{\textrm{dis}}$), the initial cluster stellar mass ($M_{\textrm{cl}}$), and the local ambient density ($\rho_{\textrm{amb}}$): 

\begin{equation}
   t_{\textrm{dis}}\,\simeq\,810\,\textrm{Myr}\,\left(\frac{M_{\textrm{cl}}}{10^4\,M_{\odot}}\right)^{0.62}\,\left(\frac{\rho_{\textrm{amb}}}{M_{\odot}\,\textrm{pc}^{-3}}\right)^{-1/2} 
.\end{equation}

\noindent
Assuming an initial mean stellar mass of $\langle m\rangle$\,=\,$0.54\,M_{\odot}$ \citep{Lamers:2005a}, this equation converts to 

\begin{equation}
  t_{\textrm{dis}}\,\simeq\,553\,\textrm{Myr}\,\left(\frac{N_{0}}{10^4}\right)^{0.62}\,\left(\frac{\rho_{\textrm{amb}}}{M_{\odot}\,\textrm{pc}^{-3}}\right)^{-1/2}
,\end{equation}

\noindent
where $N_{0}$ is the initial number of stars. Therefore, a cluster with an initial number of stars $N_{0}$$\,\sim\,$$10^{5}$ and located in the solar neighbourhood (heliocentric distance $d_{\odot}\lesssim\,1\,$kpc; mean gas and stellar density $\rho_{\textrm{amb}}=0.10\pm0.01\,M_{\odot}\,$pc$^{-3}$; \citeauthor{Holmberg:2000}\,\,\citeyear{Holmberg:2000}) is expected to survive for about 10\,Gyr, in agreement with \cite{Portegies-Zwart:2010}. 


 Based on this scaling, 
 
\begin{equation}
   t_{\textrm{dis}}\,\sim\,10\,\textrm{Gyr}\,\left(\frac{N_{0}}{10^5}\right)^{0.62}
,\end{equation} 
 
\noindent 
and assuming that our OCRs are indeed on the verge of complete dissolution within the general Galactic field (i.e., age$_{\textrm{OCRs}}\,\sim t_{\textrm{dis}}$), we find that 8 OCRs (i.e., NGC\,6481, ESO\,324-SC15, NGC\,7036, NGC\,7193, ESO\,435-SC48, NGC\,3231, ESO\,425-SC06 and NGC\,1663) may have been as rich as $N_{0}\sim10^4$ stars in the past. This result is in agreement with M98, whose simulations with $N_{0}\sim10^4$ stars are able to reproduce observable quantities of open clusters (see his figure 2). We estimate that the other 9 OCRs (i.e., Lynga\,8, Bica\,5, NGC\,1901, Alessi\,3, Ruprecht\,31, ESO\,425-SC15, ESO\,426-SC26, Ruprecht\,3, and NGC\,2180; this last one included here as an OCR) may have been as rich as $N_{0}\sim10^3$ when they were formed. These OCRs have estimated ages in the range $0.4-1.1\,$Gyr. This is consistent with the results of M97, whose expected disruption times for simulated clusters with an initial population of about 1\,000 stars (including primordial binaries) are in the range $\sim0.5-1.1\,$Gyr, depending on the assumed IMF (see his table 2). The evaporation times predicted by M97 compare well with those obtained by \cite{Portegies-Zwart:2001}, whose simulated clusters with $N_{0}=3\,000$ stars dissolve in $\sim1-1.6\,$Gyr, depending on the initial adopted density profile (see their table 10).              




 \section{Comparison of OCR dynamical properties}    
 \label{comparison_OCR_dyn_properties} 
 \subsection{Velocity dispersions}
 \label{velocity_dispersions}     

\begin{figure*}
\begin{center}
\parbox[c]{1.0\textwidth}
  {

   \includegraphics[width=0.33333\textwidth]{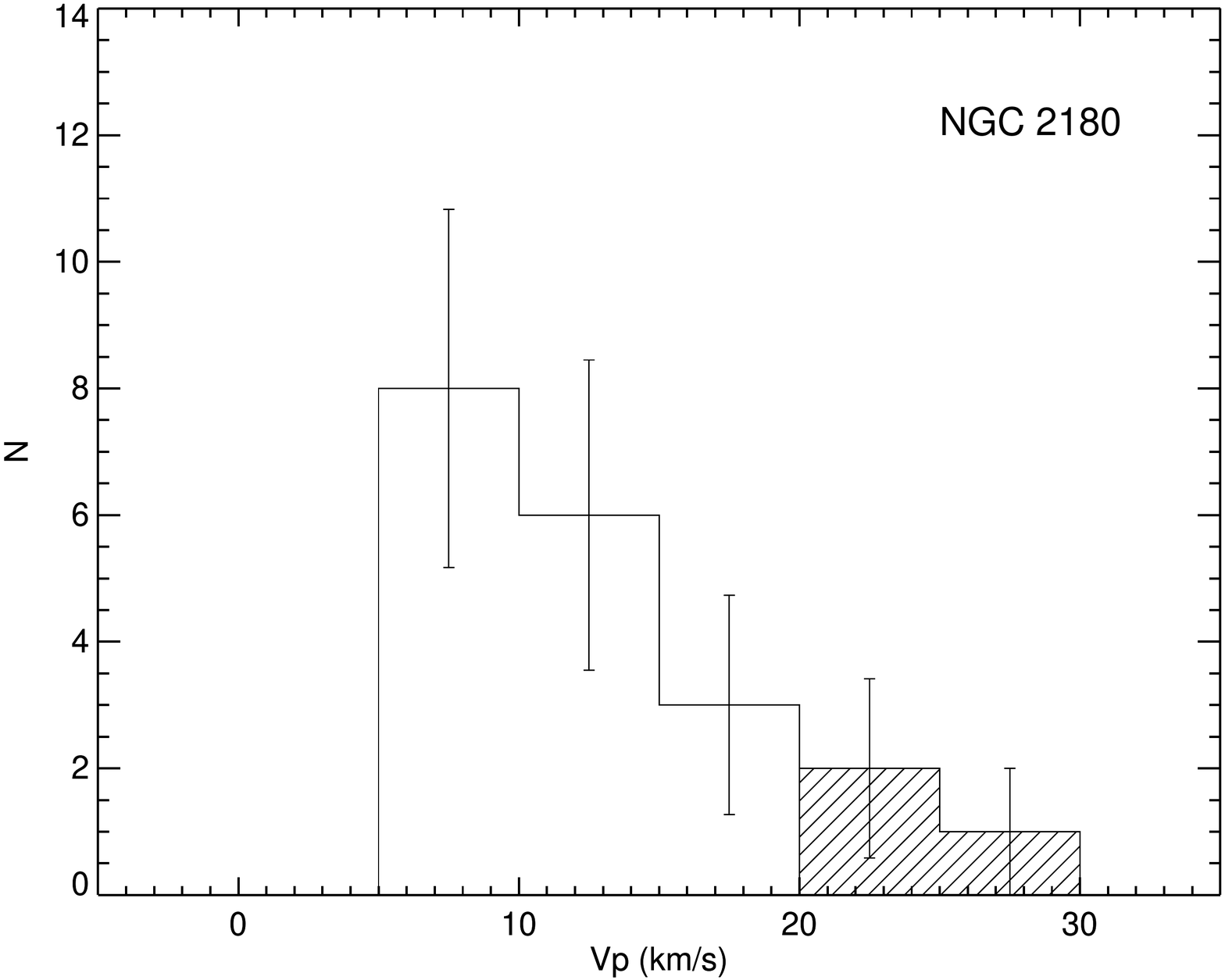}   
   \includegraphics[width=0.33333\textwidth]{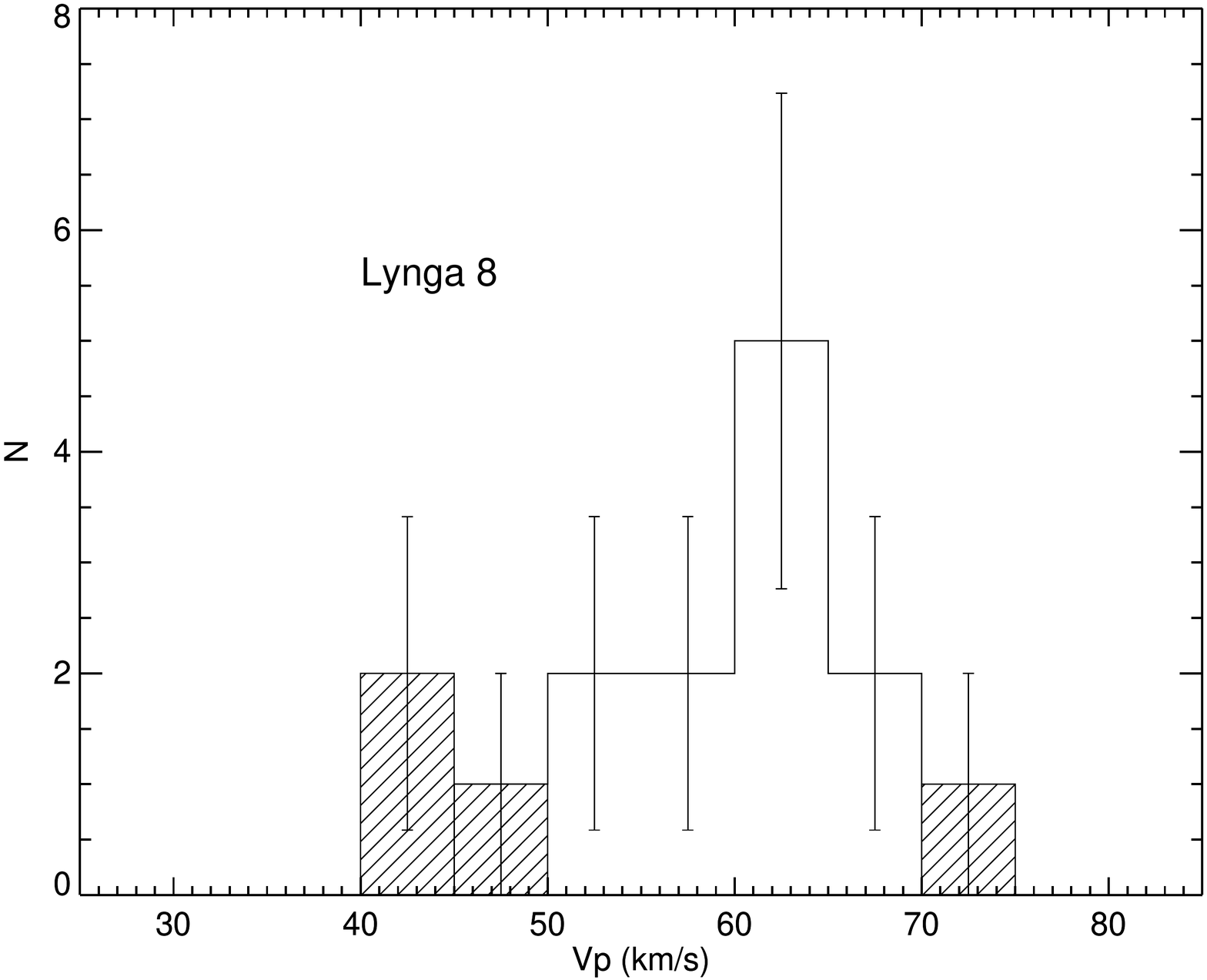}
   \includegraphics[width=0.33333\textwidth]{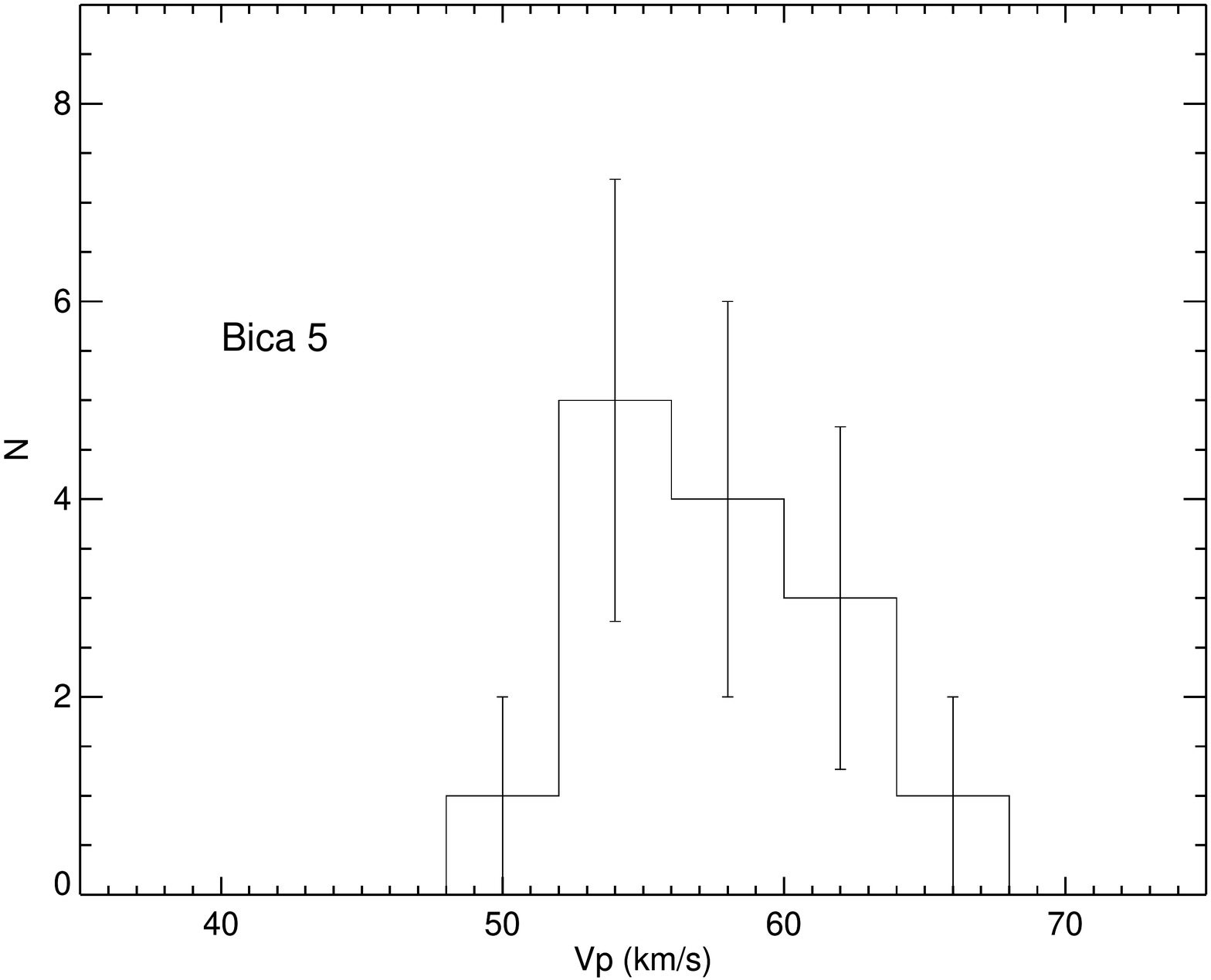}
   \includegraphics[width=0.33333\textwidth]{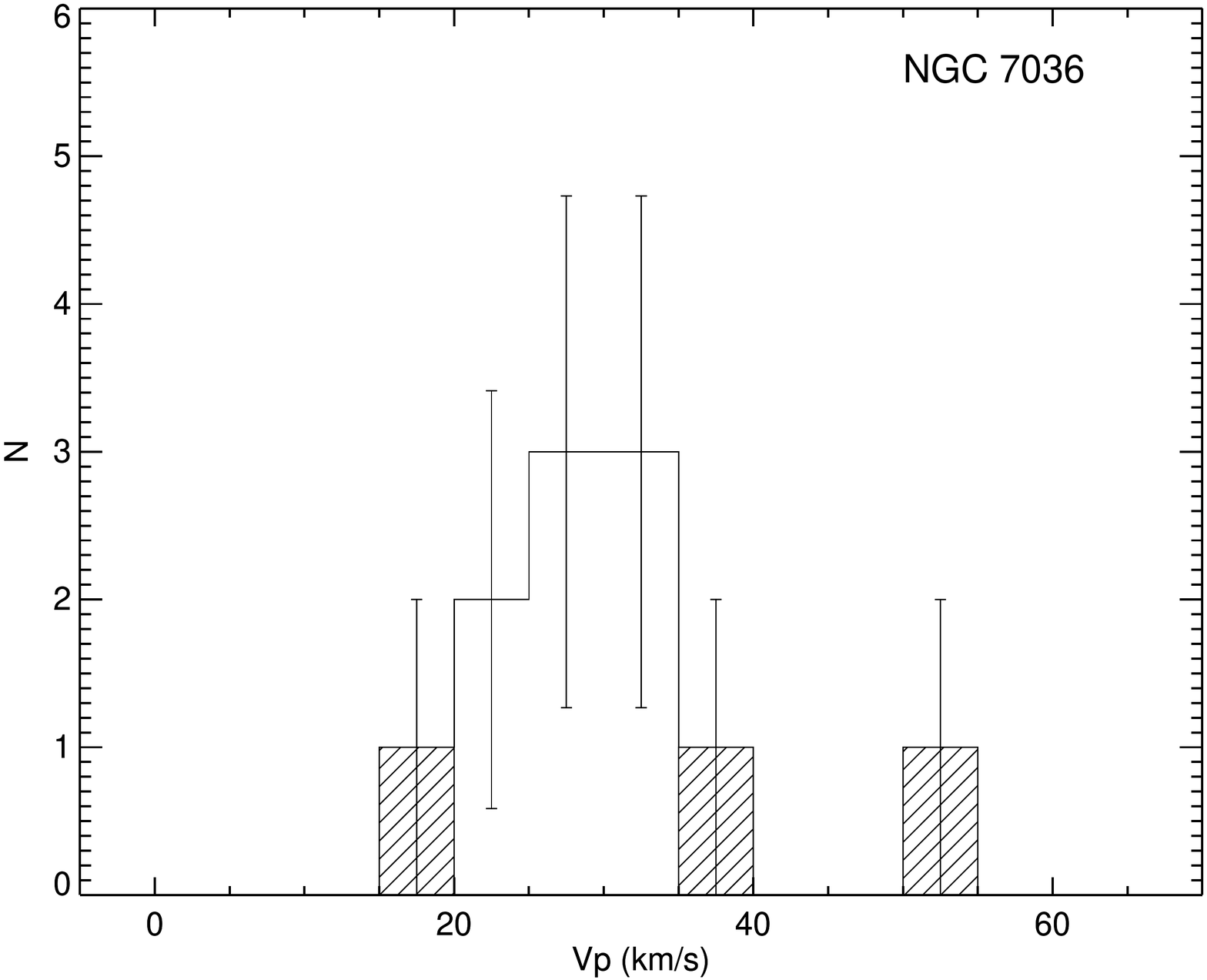}
   \includegraphics[width=0.33333\textwidth]{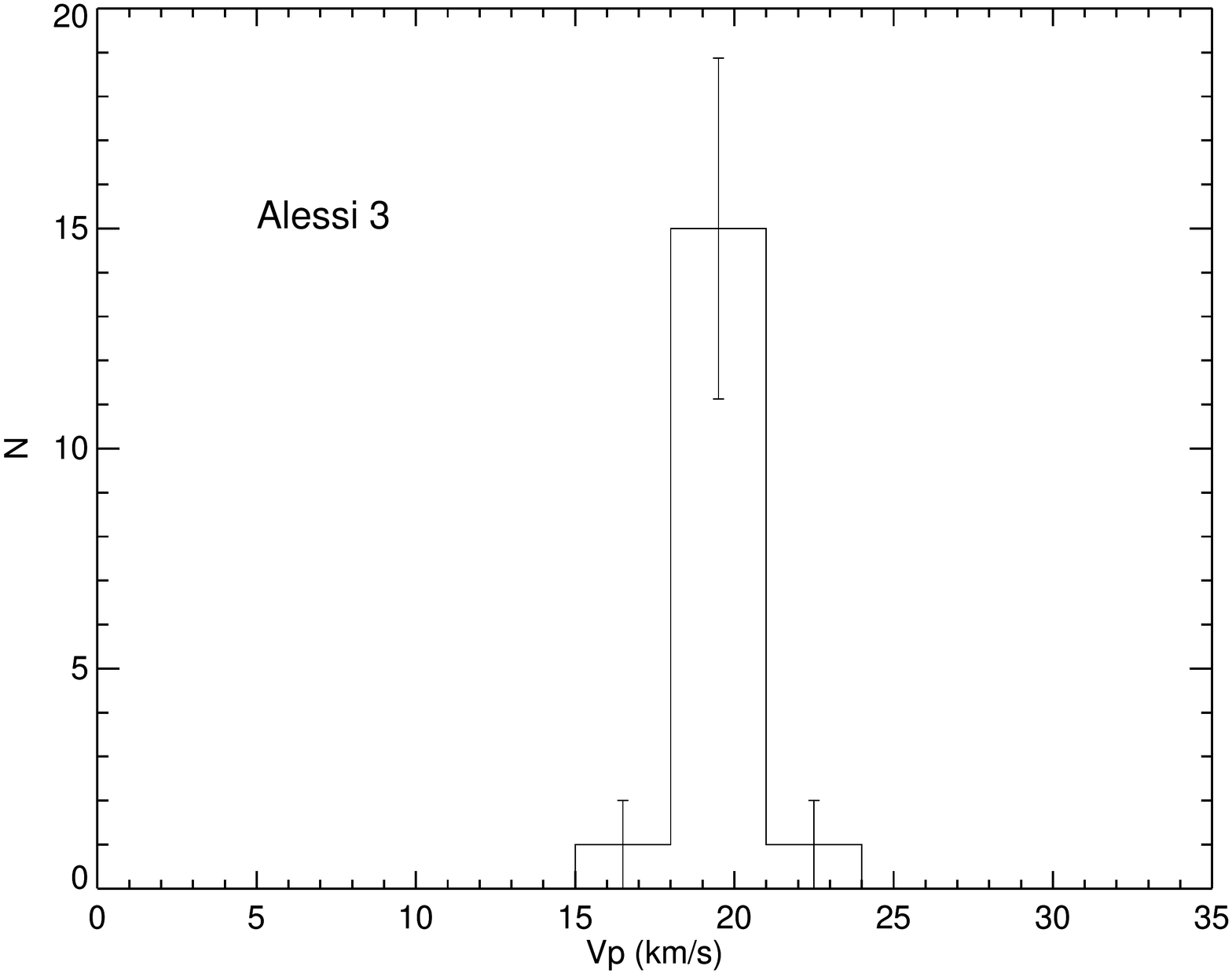}
   \includegraphics[width=0.33333\textwidth]{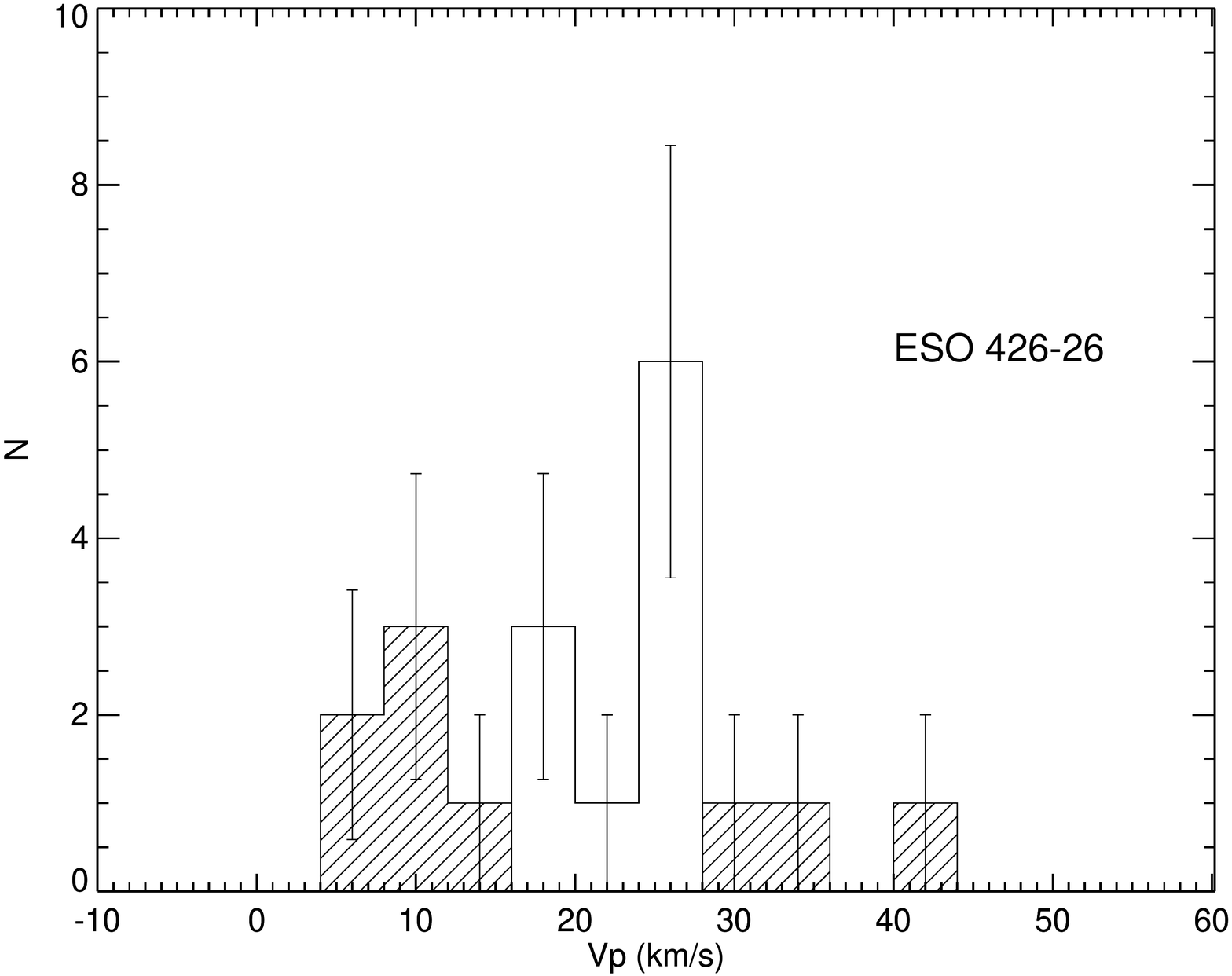}   

  }

\caption{Projected velocity distributions of member stars (Sect. \ref{results}). The hatched bins show stars that highly deviate from the samples means after applying a sigma-clipping routine. Open histograms were employed to calculate $\sigma_{v_{p}}$ and $\sigma_{v}$. Poisson uncertainties are also shown. Histograms are shown for NGC\,2180, Lynga\,8, Bica\,5, NGC\,7036, Alessi\,3, and ESO\,426-SC26.}
\label{hist_vp_members_parte1}
\end{center}
\end{figure*}

 
Proper motions from the GAIA DR2 catalogue were extracted for each object member star. Then the distribution of projected angular velocities ($V_{p}\,=\,\sqrt{  \mu_{\alpha}^2\,\mathrm{cos}^2\delta + \mu_{\delta}^2  }   $) was constructed as shown in Fig. \ref{hist_vp_members_parte1}. Angular velocities (expressed in mas\,yr$^{-1}$) were converted into km\,s$^{-1}$ using the distances in Table \ref{info_sample_OCs_OCRs}. 



Firstly, following a procedure analogous to that of \cite{Cantat-Gaudin:2018}, an iterative 2$\sigma$ clipping exclusion routine was applied to the data in Fig. \ref{hist_vp_members_parte1} in order to identify peaks that highly deviate from the sample means (hatched bins). As stated by \cite{Bica:2005}, highly discrepant peaks may be mostly produced by unresolved multiple systems. Then the dispersion of projected velocities ($\sigma_{v_{p}}$) of each ``cleaned"\, histogram was determined. The uncertainties in the proper motions were properly taken into account in the calculation of $\sigma_{v_{p}}$ by means of the procedure described in section 5 of \cite{van-Altena:2013} and also in \cite{Sagar:1989}. 3D velocity dispersions ($\sigma_{v}$) were then obtained from $\sigma_{v_{p}}$ assuming that velocity components of stars relative to each object centre are isotropically distributed. With this approximation, $\sigma_{v}\,=\,\sqrt{3/2}\,\sigma_{v_{p}}$. The $\sigma_{v}$ values for our objects are listed in the last column of Table \ref{info_sample_OCs_OCRs}. Obtaining sample standard deviations $\sigma_{v_{p}}$ was preferred rather than fitting Gaussians to the histogram peaks (as performed by \citeauthor{Bica:2005}\,\,\citeyear{Bica:2005}) because of the scarcity of stars in the case of the OCRs, which precludes statistically significant fits.

\subsection{Analysing the OCR evolutionary stages}
\label{analysing_OCRs_evolut_stages}

 \begin{figure*}
 \centering
 \includegraphics[width=16cm]{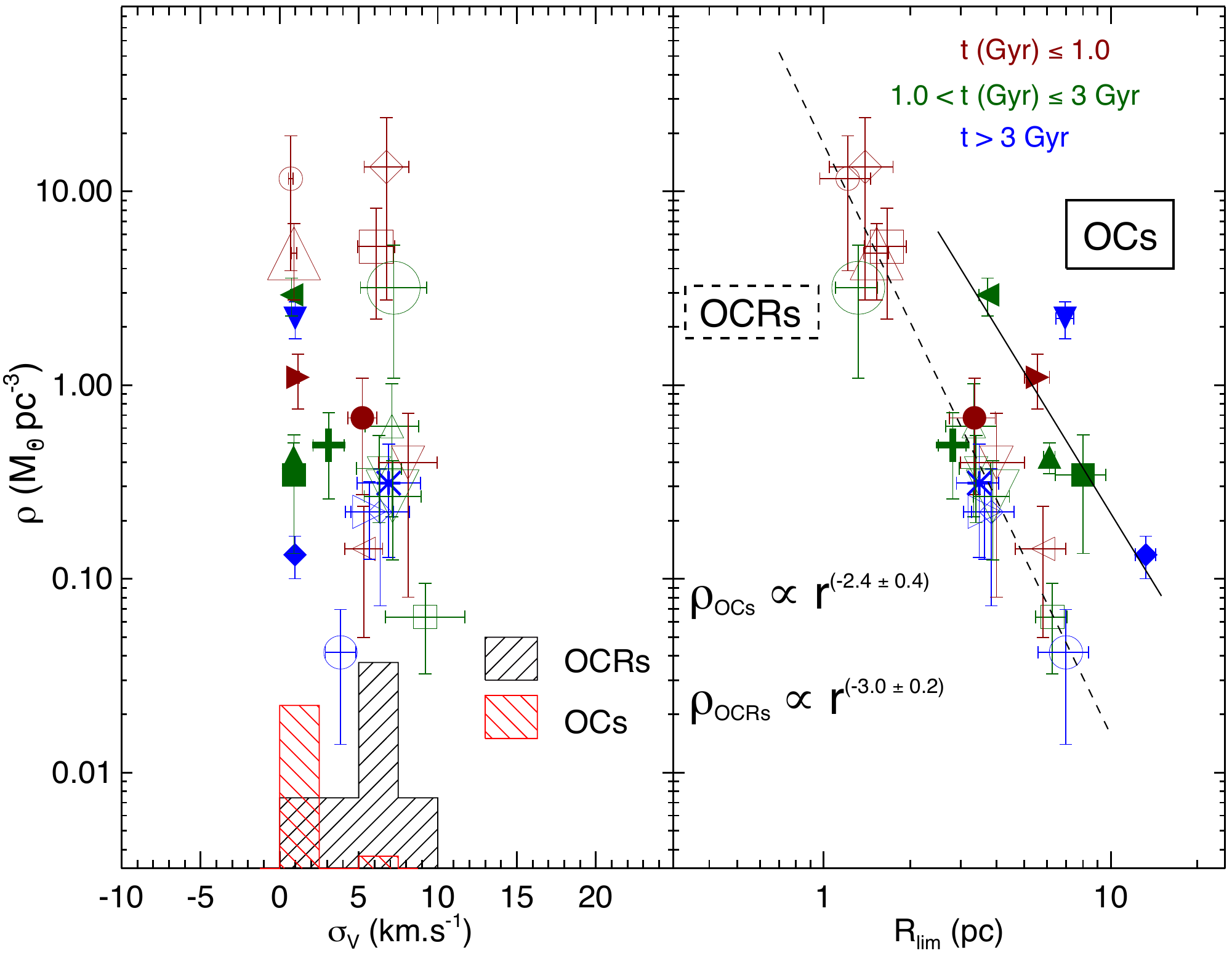}
 \caption{Left: Mean stellar densities as function of the velocity dispersions $\sigma_{v}$ for the 23 OCs and OCRs in our sample. Symbols are the same as those in Figs. \ref{plots_rlim_age_mass_POCRs_forpaper_parte1} and \ref{plots_rlim_age_mass_POCRs_forpaper_parte2}. A normalized 3D velocity dispersion distribution of OCs (red) and OCRs (black) is represented by the hatched histogram. Right: Mean stellar densities as a function of the limiting radii ($R_{\textrm{lim}}$). The dashed and continuous lines are linear fittings to the loci of points defined by the OCRs and OCs, respectively. Clusters were divided into three age bins, as indicated.}
   \label{compara_densidade_rlim_sigma_v}
\end{figure*}

In the left panel of Fig. \ref{compara_densidade_rlim_sigma_v}, the derived dispersions $\sigma_{v}$ are plotted as a function of the stellar densities. The sample was divided into three coloured age bins: $t$(Gyr)$\,\leq\,$1.0 (red), 1.0$\,<t$(Gyr)$\,\leq\,3$ (green), and $t>3$\,Gyr (blue). Typical $\sigma_{v}$ values for the OCRs are between $\sim1-7\,$km\,s$^{-1}$, as shown in the hatched histograms in the bottom of the figure. This dispersion range is consistent with the data presented by \cite{Cantat-Gaudin:2018}, who obtained mean astrometric parameters for 128 clusters closer than about 2 kpc. By inverting the mean parallaxes and taking the dispersions $\sigma_{\mu_{\alpha}}$ and $\sigma_{\mu_{\delta}}$ listed in their tables 1 and 2 for clusters with $n_{\textrm{members}}$ $\leq$ 20 (to be consistent with the typical number of member stars found in our OCRs sample), we derived velocity dispersions (assuming isotropy) in the range $\sim0.5-7\,$km.s$^{-1}$ for 53 clusters of their sample. Numerical simulations for open clusters containing a fraction of primordial binaries show that OCRs are expected to be highly biased towards stable dynamical configurations, containing binaries and long-lived triples, which increases the cluster velocity dispersions. The total binary fraction increases with time when the cluster population reaches values lower than one-third of the initial (M97; M98).

The results shown in Figs. \ref{plots_rlim_age_mass_POCRs_forpaper_parte1}, \ref{plots_rlim_age_mass_POCRs_forpaper_parte2}, and \ref{compara_densidade_rlim_sigma_v}  suggest that regardless of their densities and ages, clusters that are close to complete disruption tend to present internal velocity dispersions, masses, and limiting radii within reasonably well-defined ranges ($\sigma_{v}$ between $\sim1-7\,$km\,s$^{-1}$, $M$ between $\sim10-40\,$M$_{\odot}$, $R_{\textrm{lim}}\lesssim2\,$pc for the compact OCRs, and $R_{\textrm{lim}}$ between $\sim2-7\,$pc for the less compact OCRs), that is, they are dynamically similar. These results suggest that in the case of initially populous clusters ($N_{0}\sim10^3-10^4$), the long-term evolution causes the clusters to lose the memory of their initial forming conditions and reach final (remnant) dynamical states that tend to be compatible with each other.    


In the right panel of Fig. \ref{compara_densidade_rlim_sigma_v}, stellar densities are plotted as a function of limiting radii. Because the OCRs have similar masses, their $\rho$ values follow a $R^{-3}$ relation, as expected. The dashed line represents a linear fitting for which 

\begin{equation}
   \rho_{_{\textrm{OCRs}}}\,\propto\,R\,^{(-3.0\,\pm\,0.2)} 
   \label{rho_versus_R_OCRs}   
.\end{equation}

\noindent
As we showed in Fig. \ref{plots_rlim_age_mass_POCRs_forpaper_parte2}, the $R_{\textrm{lim}}$ of these objects do not show any trend along the time, so that the data points are not age segregated along the OCRs locus. The OCs occupy a  different locus, for which a tentative fitting (continuous line) resulted in a slightly different slope: 

\begin{equation}
  \rho_{_{\textrm{OCs}}}\,\propto\,R\,^{(-2.4\,\pm\,0.4)}
  \label{rho_versus_R_OCs}
,\end{equation}

\noindent
which still compatible within the uncertainties with the characteristic slope defined by the OCRs locus, however. A possible speculation regarding the reason of this similarity is that although hugely different in terms of stellar content, the physical processes (internal relaxation and Galactic tidal pull) that have driven the dynamical evolution of both OCs and OCRs are similar, which reinforces the evolutionary connection between both populations. Despite this, the dependences expressed in eqs. \ref{rho_versus_R_OCRs} and \ref{rho_versus_R_OCs} will be better constrained when the number of investigated objects is larger.

It could be argued that sparsely populated objects such as OCRs should evaporate within a relatively short timescale ($t_{\textrm{evap}}$), proportional to the relaxation time 

\begin{equation}
  t_{\textrm{relax}}\,\approx\,\left(\frac{0.1\,N}{\ln\,N}\right)\,t_{\textrm{cross}}
,\end{equation}

\noindent
where $N$ is the number of stars and $t_{\textrm{cross}}$ the time required for a star to cross the cluster diameter ($t_{\textrm{cross}}$\,$\approx\,$2\,$R_{\textrm{lim}}/\sigma_{v}$) and $t_{\textrm{evap}}$\,$\approx$\,$140\,t_{\textrm{relax}}$ \citep{Binney:2008}. Taking into account the number of members estimated for each of our OCRs together with their $R_{\textrm{lim}}$ and $\sigma_{v}$ values, the resulting $t_{\textrm{evap}}$ are in the range $\sim$\,$20-300\,$Myr. With such short timescales, it would be unlikely that we observe such  objects on their way to disruption. Although this reasoning is quite correct in the case of OCs born with just a few tens of stars (MMM13), currently observed OCRs are the descendants of much more massive progenitors, therefore such analysis based on the $t_{\textrm{evap}}$ cannot be applied to them. As described previously, clusters with initial populations $N_{0}$ in the range $\sim$$10^3-10^4$ stars are expected to survive for much longer timescales, as shown by $N$-body simulations. For instance, a particular simulation run by M98 starting with a cluster containing 10010 stars evolved to an OCR with 32 stars at an age of 5.4\,Gyr. 

\section{Summary and concluding remarks}
\label{conclusions}

We have analysed the evolutionary stages of a set of 16 objects previously catalogued as OCRs or POCRs. We have employed parameters that are directly associated with dynamical evolution: age, limiting radius, stellar mass, and velocity dispersion. A sample of 6 dynamically evolved OCs was also employed for comparison purposes. Here we suggest that the cluster NGC\,2180, previously classified as an OC, is better classified as an OCR.

Limiting radii were derived by means of the construction of RDPs. Then we evaluated the contrast between each object and its surrounding field regarding the number of stars counted in the object inner area and within circular field samples that were  chosen randomly. In all cases there is at least one $R_{\textrm{lim}}$ value for which the associated percentile is above 90\% (Fig. \ref{percentile_vs_radius_parte1}), which reveals a significant contrast with respect to the surrounding field.  

Membership likelihoods were assigned to stars by means of an algorithm that evaluates the overdensity of stars in each part of the 3D astrometric parameters space ($\mu_{\alpha}, \mu_{\delta}, \varpi$) relative to the real dispersion of data of a comparison field. We identified evolutionary sequences in the cluster $K_{\textrm{s}}\times{(J-K_{\textrm{s}})}$ CMDs by taking the more probable member star and fitted isochrones in order to derive fundamental parameters. Photometric data for member stars were employed to build mass functions and to determine total masses ($M$). We used proper motion data for these members and derived velocity dispersions ($\sigma_{v}$), after applying a proper iterative exclusion routine. 

The mass functions for many of our OCs and OCRs (Fig. \ref{mass_func_parte1}) show signals of depletion of low main-sequence stars, which may be a consequence of preferential low-mass star evaporation. Other objects seem less severely depleted because their mass functions present smaller deviations with respect to \cite{Kroupa:2001} and \cite{Salpeter:1955} scaled IMFs. Differently from the OCs, OCR masses do not exhibit a clear trend with age. Instead, their masses are confined to the interval $\sim10-40\,M_{\odot}$. The limiting radii for our compact OCR sample are $\lesssim2\,$pc, which are similar to the core radii obtained by M98 through $N-$body simulations for clusters with initial populations $N_{0}\sim10^3-10^4$ stars and containing primordial binaries. The less compact OCRs present $R_{\textrm{lim}}$ between $\sim2-7\,$pc, which is comparable with the OCs and with the half-mass radii ($r_{\textrm{hm}}$) obtained by M98 for a $N_{0}\sim10^4$ simulated cluster. The estimated stellar densities of our OCRs are typically greater than the critical density for a cluster to be stable against Galactic tidal disruption. 

Some of our OCRs present ages and Galactocentric distances that are similar to some of the OCs, as we showed for the pairs NGC\,4337$-$NGC\,6481, NGC\,3680$-$NGC\,7036, NGC\,752$-$NGC\,3231, and NGC\,2180$-$NGC\,1582. Although these objects have been subject to nearly the same external tidal fields, the OCRs are in a much more advanced dynamical stage than the corresponding OC. This difference may be traced back to their progenitor open clusters. We also verified that NGC\,2180 is entering in a remnant evolutionary stage, consistently with the interpretation of BBP04. 

From the results of $N-$body simulations, we used an approximate scaling between disruption time and initial number of stars. We suggest that currently observed OCRs are in fact descendants of initially populous clusters containing $N_{0}\sim10^3-10^4$ stars. The velocity dispersions derived for our OCRs are typically in the range $\sim1-7\,$km\,s$^{-1}$. This $\sigma_{v}$ range is consistent with that found in the literature \citep{Cantat-Gaudin:2018} for other objects of this same nature. As a consequence of the internal relaxation, the fraction of binaries (either primordial or dynamically formed) in the central parts of evolved clusters tend to increase with time; indeed, binaries act as internal heating sources, thus contributing to the increase in the velocity dispersions.


We conclude that clusters that are close to complete disruption tend to present masses, limiting radii, and internal velocity dispersions within reasonably well-defined ranges. We may suggest that in the case of initially populous clusters (which survive the destructive effects of the initial evolutionary stages), the outcome of the long-term evolution (which includes internal relaxation, external interactions due to the Galactic tidal field and stochastic tidal effects, such as encounters with dense clouds and disc shocking) is to bring the final residues of the OCs to  dynamical states that are similar to each other, thus masking out the memory of star clusters initial formation conditions.

Observational constraints to evolutionary models will be improved as we increase the number of investigated OCRs, allowing us to make more general conclusions about the dynamical properties of these challenging objects. The study of the OCRs is a subject of great interest because these objects are important for our understanding of the formation and evolution of the Galactic disc.

\begin{acknowledgements}
     We thank the anonymous referee for helpful suggestions. We thank the Brazilian financial agencies CNPq (grant 304654/2017-5) and Fapemig. F.~Maia acknowledges FAPESP funding through the fellowship n$^o$ 2018/05535-3. This study was also partially funded by the Coordenação de Aperfeiçoamento de Pessoal de Nível Superior - Brasil (CAPES) - Finance Code 001. This publication makes use of data products from the Two Micron All Sky Survey, which is a joint project of the University of Massachusetts and the Infrared Processing and Analysis Center/California Institute of Technology, funded by the National Aeronautics and Space Administration and the National Science Foundation. This research has made use of the WEBDA database, operated at the Department of Theoretical Physics and Astrophysics of the Masaryk University, and of the SIMBAD database, operated at CDS, Strasbourg, France. This research has made use of Aladin sky Atlas. This work presents results from the European Space Agency (ESA) space mission Gaia. Gaia data are being processed by the Gaia Data Processing and Analysis Consortium (DPAC). Funding for the DPAC is provided by national institutions, in particular the institutions participating in the Gaia MultiLateral Agreement (MLA). The Gaia mission website is https://www.cosmos.esa.int/gaia. The Gaia archive website is https://archives.esac.esa.int/gaia.     
      
\end{acknowledgements}

%
%

{\footnotesize
\bibliographystyle{aa}
\bibliography{referencias}}

\begin{appendix} 
\section{Supplementary material}

\clearpage















\clearpage


\end{appendix}

\end{document}